\documentclass[12pt]{article}%
\usepackage{amsmath}
\usepackage{amsfonts}
\usepackage{amssymb}
\usepackage[small,compact]{titlesec}
\usepackage{geometry}
\usepackage{enumerate}
\usepackage{graphicx}
\usepackage{dcolumn}
\usepackage{booktabs}%
\providecommand{\U}[1]{\protect\rule{.1in}{.1in}}
\newtheorem{theorem}{Theorem}

\newtheorem{definition}[theorem]{Definition}

\renewcommand{\mathbf}{\boldsymbol}
\expandafter
\def \expandafter \normalsize \expandafter{\normalsize \setlength \abovedisplayskip{10pt plus 2pt minus 7pt}}
\expandafter
\def \expandafter \normalsize \expandafter{\normalsize \setlength \abovedisplayshortskip{0pt plus 2pt}}
\expandafter
\def \expandafter \normalsize \expandafter{\normalsize \setlength \belowdisplayskip{10pt plus 2pt minus 7pt}}
\expandafter
\def \expandafter \normalsize \expandafter{\normalsize \setlength \belowdisplayshortskip{5pt plus 2pt minus 3pt}}
\newcolumntype{d}[1]{D{.}{.}{#1}}
\begin{document}

\title{Fractional integration and cointegration\thanks{We thank Giuseppe Cavaliere
for helpful discussions. Javier Hualde's research is supported by the Spanish
Ministerio de Ciencia e Innovaci\'{o}n through project PGC2018-093542-B-I00.}}
\author{Javier Hualde\thanks{Corresponding author.}\\Universidad P\'{u}blica de Navarra\\\texttt{javier.hualde@unavarra.es}
\and Morten \O rregaard Nielsen\\Aarhus University\\\texttt{mon@econ.au.dk}}
\maketitle

\begin{abstract}
In this chapter we present an overview of the main ideas and methods in the
fractional integration and cointegration literature. We do not attempt to give
a complete survey of this enormous literature, but rather a more introductory
treatment suitable for a researcher or graduate student wishing to learn about
this exciting field of research. With this aim, we have surely overlooked many
relevant references for which we apologize in advance. Knowledge of standard
time series methods, and in particular methods related to nonstationary time
series, at the level of a standard graduate course or advanced undergraduate
course is assumed.



\end{abstract}

\section{Introduction}

Fractional time series constitute a general class of models which are able to
capture a wide range of stationary and nonstationary behaviours and can
display the so-called long memory property, where a single parameter (known as
memory) characterizes a fundamental part of the persistence of the time
series. In the classical sense, a purely stochastic scalar time series
$\zeta_{t}$, $t\in\mathbb{Z}=\{t:t=0,\pm1,\ldots\}$, is integrated of order
$d$, denoted $\zeta_{t}\in I(d)$, if it can be represented as a stationary and
invertible autoregressive-moving average (ARMA) process after differencing it
$d$ times. Usually, the parameter $d$ has been assumed to be either 0, 1,
or~2, but the class of fractional time series models is characterized by a
non-integer value of $d$. Earlier surveys of this phenomenon include Ballie
(1996), Robinson (2003), and Gil-Alana and Hualde (2009), and a recent
monograph treatment is given by Hassler (2018).

Undoubtedly, the key aspect of the definition of (fractional) integration is
the concept of an $I(0)$ process, which in popular terms has been referred to
as a \textquotedblleft short memory\textquotedblright, \textquotedblleft
weakly dependent\textquotedblright, \textquotedblleft short-range
dependent\textquotedblright,\ or \textquotedblleft weakly
autocorrelated\textquotedblright\ process. The $I(0)$ concept has been given
different, although relatively closely related, meanings in the literature.
Define the spectral density of $\zeta_{t}$ as
\[
f_{\zeta}(\lambda)=\frac{1}{2\pi}\sum_{j=-\infty}^{\infty}\gamma_{\zeta
}(j)e^{ij\lambda},
\]
where $\gamma_{\zeta}(j)$ represents the lag $j$ autocovariance of the process
$\zeta_{t}$. Given the spectral density $f_{\zeta}(\lambda)$, one definition
of $I(0)$ is the following. A zero-mean scalar covariance stationary process
$\zeta_{t}$, $t\in\mathbb{Z}$,\ with spectral density $f_{\zeta}(\lambda)$\ is
integrated of order zero, denoted $\zeta_{t}\in I(0)$,\ if
\begin{equation}
0<f_{\zeta}(\lambda)<\infty\text{ for all }\lambda\in(-\pi,\pi].
\label{def zero}%
\end{equation}
\smallskip This definition strengthens the original idea of an $I(0)$ process
(which was based on stationarity and invertibility and related only to
frequency zero behaviour), and avoids complications due to poles or zeros of
the spectral density outside frequency zero which may have important consequences.

An alternative definition of $I(0)$ is to include all processes $\zeta_{t}$
that satisfy a functional central limit theorem (FCLT) with a Brownian motion
limit process; that is, with $\left\lfloor \cdot\right\rfloor $ denoting
integer part and $T$ the sample size, for $r\in\lbrack0,1]$,%
\[
T^{-1/2}\sum_{t=1}^{\left\lfloor Tr\right\rfloor }\zeta_{t}\Rightarrow
W(r)\text{\ as }T\rightarrow\infty,
\]
where \textquotedblleft$\Rightarrow$\textquotedblright\ denotes weak
convergence of the associated probability measures and $W(r)$ denotes a
Brownian motion with variance $\sigma^{2}=\lim_{T\rightarrow\infty}%
T^{-1}E((\sum_{t=1}^{T}\zeta_{t})^{2})>0$. In any case, the distinction
between these different definitions of $I(0)$ is not that relevant because
proper additional conditions are usually imposed so that an invariance
principle holds.

Letting $u_{t}\in I(0)$, we can define an $I(1)$ process $X_{t}$ as%
\begin{equation}
\Delta X_{t}=u_{t},\quad t\in\mathbb{Z}, \label{a1}%
\end{equation}
where $\Delta=1-L$ and $L$ are the difference and the lag operators,
respectively. Inverting the $\Delta$ operator, it is immediate to show that%
\begin{equation}
X_{t}=X_{0}+\sum_{j=1}^{t}u_{j}=X_{0}+\sum_{j=0}^{t-1}u_{t-j},\quad t\geq1,
\label{integration}%
\end{equation}
where $\sum_{j=0}^{t-1}L^{j}$ is an integration (or cumulation or partial
summation) operator. Note that the integration operator in (\ref{integration})
ends at $j=t-1$ because the summation $\sum_{j=0}^{\infty}u_{t-j}$ is not
well-defined in the mean-square sense. The choice of starting point, $t=0$, in
(\ref{integration}) is arbitrary, and as such nothing is said about $X_{t}$
for $t<0$, which may be taken to be zero. It can be shown that the stochastic
difference equation (\ref{a1}) does not have a stationary solution
(e.g.,\ Brockwell and Davis, 1991, p.\ 86), so, in this sense, we say that
process $X_{t}$ is nonstationary. Similar arguments and conclusions can be
made for $\Delta^{2}X_{t}=u_{t}$, etc.

In a multivariate context, the concepts of nonstationarity and integration
lead naturally to that of cointegration. An early definition of cointegration
from Engle and Granger (1987), see also Granger (1981), is that the
$p$-dimensional vector $X_{t}$ is cointegrated if all the components of
$X_{t}$\ are $I(d)$ and there exists a $\beta\neq0$\ such that $\beta^{\prime
}X_{t}\in I(d-b)$. This definition has commonly been applied, especially in
regression setups with $d=b=1$, such that the components of $X_{t}$ are $I(1)$
but $\beta^{\prime}X_{t}\in I(0)$. A problem with this simple definition is
that it is not invariant to nonsingular linear transformations (for example a
transformation involving $\beta^{\prime}X_{t}$ will no longer be $I(1)$ by
this definition). Thus, more general definitions have been provided, starting
with a definition of $I(0)$ for multivariate processes. We next give a general
definition of $I(0)$ and $I(d)$ inspired by that of Johansen (1995).

\begin{definition}
[Integration order]\label{def vector zero}Let $X_{t}$ be a $p$-dimensional
process such that%
\[
X_{t}-E(X_{t})=\sum_{j=0}^{\infty}C_{j}\varepsilon_{t-j},
\]
where $\varepsilon_{t}$ is a $p$-dimensional zero mean independent and
identically distributed (i.i.d.) sequence with $\operatorname{Var}%
(\varepsilon_{t})=\Omega$, and $C(z)=\sum_{j=0}^{\infty}C_{j}z^{j}$ is
convergent for $|z|\leq1+\delta$ for some $\delta>0$. We say that $X_{t}\in
I(0)$ if $C(1)\neq0$. Furthermore, we say that $X_{t}\in I(d)$ if $\Delta
^{d}X_{t}\in I(0)$.
\end{definition}

For example, for $d=1$ the definition of a multivariate $I(1)$ process in
Definition~\ref{def vector zero} coincides with that in (\ref{integration}).
Unlike the simple definition from Engle and Granger (1987),
Definition~\ref{def vector zero} implies that the integration order of an
$I(d)$ vector ($d=0,1$) is invariant to nonsingular linear transformations;
that is, if the $p$-dimensional vector $X_{t}\in I(d)$, then $AX_{t}\in I(d)$
for any nonsingular $p\times p$ matrix $A$.

\begin{definition}
[Cointegration]\label{def cointegration}The $p$-dimensional process $X_{t}$ is
cointegrated if $X_{t}\in I(d)$ and there exists a $\beta\neq0$\ such that
$\beta^{\prime}X_{t}\in I(d-b)$ for $b>0$.
\end{definition}

The concept of cointegration captures the idea of long-run equilibrium among
the observables. When $d=b=1$, this is understood in the sense that these are
themselves nonstationary, or $I(1)$, so that the variables have no natural
\textquotedblleft level\textquotedblright. However, there exists an (economic)
equilibrium relation given by $\beta$, such that the equilibrium relation
$\beta^{\prime}X_{t}$ is stationary, or $I(0)$, possibly around a non-zero
mean/level. In other words, while shocks to the variables $X_{t}$ themselves
have a permanent effect, any deviation from the equilibrium $\beta^{\prime
}X_{t}$ has only a transitory effect. The success of nonstationary time series
analysis and the concept of cointegration seems largely due to this
interpretation of cointegration as a long-run equilibrium among the (economic) variables.

The definitions of integration and cointegration given above are valid not
only for integer values of $d$ and $b$, but also for real values. Non-integer
values of $d$ and/or $b$ lead to the concepts of fractional integration and
fractional cointegration. Importantly, the interpretation of cointegration as
an (economic) equilibrium concept still applies in a more general form in the
fractional setting.

In this chapter we present an overview of the concepts discussed above when
$d$ and/or $b$ are not integers. In particular, we present and discuss the
main ideas and methods in the fractional integration and cointegration
literature, as related to definitions, estimation, and inference. We
deliberately do not attempt to give a complete survey of this enormous
literature, but rather a more introductory treatment. The aim is to provide a
treatment that is useful for a researcher or graduate student wishing to learn
about this exciting field of research or possibly to apply some of the methods
in practice. For these reasons, we have surely overlooked many relevant
references for which we apologize in advance. We will assume knowledge of
standard time series methods, and in particular methods related to
nonstationary time series, at the level of a standard graduate course or
advanced undergraduate course.

The remainder of this chapter is laid out as follows. In the next section we
present and discuss definitions and general issues related to fractional
integration. In Section~\ref{sec:inference} we discuss inference in fractional
models, and in Section~\ref{sec:cointegration} we discuss fractional
cointegration. In both sections we present both semiparametric and parametric
methods as well as hypothesis testing approaches.
Section~\ref{sec:applications} discusses applications of fractional
integration and cointegration. However, we keep this section relatively brief
in view of the survey of applications by Henry and Zaffaroni (2003). Finally,
in Section~\ref{sec:concluding} we conclude the chapter by briefly mentioning
some additional related topics with further references to relevant work.

\section{Fractional integration}

\label{sec:FI}

For a number of years, increased interest has developed in a wider framework
which takes into account that $I\left(  0\right)  $, and also $I\left(
1\right)  $, $I\left(  2\right)  $,..., are very specific types of stationary
and nonstationary processes, respectively. Thus, as a direct consequence of
the definition of integrated process given above, one could think about a
process which is $I(0)$ after $d$-differencing, where $d$ is not necessarily
an integer. Following early contributions by Granger and Joyeux (1980) and
Hosking (1981), we can define this operation by means of the binomial
expansion%
\begin{equation}
\Delta^{-d}=(1-L)^{-d}=\sum\limits_{j=0}^{\infty}\pi_{j}(d)L^{j}, \label{a4}%
\end{equation}
where $\pi_{0}(d)=1$ for all $d$,
\[
\pi_{j}(d)=\frac{\Gamma(j+d)}{\Gamma(d)\Gamma(j+1)},\quad j>0,
\]
and $\Gamma(\cdot)$ denotes the Gamma function with the convention
$\Gamma(d)=\infty$ for $d=0,-1,-2,\ldots$.

Letting $u_{t}\in I(0)$ with $E(u_{t})=0$, we can say that $X_{t}\in I(d)$ or
$X_{t}\in F(d)$ (that is, $X_{t}$ is a fractional process of order $d$) if%
\begin{equation}
\Delta^{d}X_{t}=u_{t},\quad t\in\mathbb{Z}.\label{a2}%
\end{equation}
The summation on the left-hand side of (\ref{a2}) is well-defined in the
mean-square sense if $d>-1/2$. This property is sometimes referred to as
invertibility. Furthermore, if $d<1/2$, it can be shown that (\ref{a2}) has a
unique stationary solution given by%
\begin{equation}
X_{t}=\Delta^{-d}u_{t}=\sum_{j=0}^{\infty}\pi_{j}(d)u_{t-j},\label{a3}%
\end{equation}
where the summation is well defined in this case. Thus, for (\ref{a2}) and
(\ref{a3}) to make sense, it would usually be assumed that $d\in(-1/2,1/2)$.
Note that, by Stirling's approximation,%
\begin{equation}
\pi_{j}(d)\sim\frac{1}{\Gamma(d)}j^{d-1}\text{ as }j\rightarrow\infty
,\label{a6}%
\end{equation}
where \textquotedblleft$\sim$\textquotedblright\ means that the ratio of the
left- and right-hand sides tends to one. This shows how (\ref{a2}) and
(\ref{a3}) are well-defined in the mean-square sense when $d>-1/2$ and
$d<1/2$, respectively. For example, (\ref{a3}) is well-defined when $d<1/2$
because%
\[
\sum\limits_{j=0}^{\infty}\pi_{j}^{2}\left(  d\right)  \leq K\sum
\limits_{j=1}^{\infty}j^{2\left(  d-1\right)  }<\infty,
\]
where throughout $K$ represents an arbitrarily large, finite constant.

In the particular case where $u_{t}$ is a white noise process with
$\operatorname{Var}(u_{t})=\sigma^{2}$, we say that $X_{t}=\Delta^{-d}u_{t}$
is a fractionally integrated noise. The impulse response function then takes
the simple form%
\[
\frac{\partial X_{t}}{\partial u_{t-j}}=\pi_{j}(d),\quad j=1,2,\ldots.
\]
In view of (\ref{a6}) and $d\in(-1/2,1/2)$, it follows that shocks are
transitory and their impact show a slow hyperbolic decay, which stands in
contrast with the much faster exponential decay in stationary and invertible
(finite-order) ARMA settings. Furthermore, it can be shown that the lag $j$
autocovariance of the process $X_{t}$ is given by%
\[
\gamma_{X}(j)=\frac{\sigma^{2}\Gamma(1-2d)\Gamma(j+d)}{\Gamma(d)\Gamma
(1-d)\Gamma(1+j-d)},\quad j\geq0,
\]
where by Stirling's approximation,%
\begin{equation}
\gamma_{X}(j)\sim\frac{\sigma^{2}\Gamma(1-2d)}{\Gamma(d)\Gamma(1-d)}%
j^{2d-1}\text{ as }j\rightarrow\infty. \label{autocorr}%
\end{equation}
Thus, $\gamma_{X}(j)$ also shows slow hyperbolic decay and, in fact, is not
summable when $d>0$. The spectral density of $X_{t}$ in this case is%
\begin{equation}
f_{X}(\lambda)=\frac{\sigma^{2}}{2\pi}\left\vert 1-e^{i\lambda}\right\vert
^{-2d}=\frac{\sigma^{2}}{2\pi}\left(  2\sin\left(  \lambda/2\right)  \right)
^{-2d}\sim\frac{\sigma^{2}}{2\pi}\lambda^{-2d}\text{ as }\lambda\rightarrow0,
\label{a7}%
\end{equation}
which is unbounded at the origin when $d>0$.

The behavior of the impulse response, autocovariance, and spectral density
functions have given rise to the (now standard) terminology that $X_{t}$ has
long memory, short memory, or negative memory depending on whether $d>0$,
$d=0$, or $d<0$, respectively.

For a general $I(0)$ process $u_{t}$, Granger and Joyeux (1980) and Hosking
(1981) showed that, under certain additional regularity conditions,%
\begin{equation}
\gamma_{X}(j)\sim Kj^{2d-1}\text{ as\ }j\rightarrow\infty,\label{Iaa}%
\end{equation}
where $K$ is a constant depending only on the parameters, and
\begin{equation}
f_{X}(\lambda)=\left\vert 1-e^{i\lambda}\right\vert ^{-2d}f_{u}(\lambda)\sim
f_{u}(0)\lambda^{-2d}\text{ as }\lambda\rightarrow0.\label{a40}%
\end{equation}
Thus, essentially the same results given for the fractionally integrated noise
apply. The behavior of the spectral density also relates directly to the
concept of an $I(0)$ process as implied by Robinson's (1993) definition of a
covariance stationary $I(d)$ process, which he defined as one with spectral
density
\[
g(\lambda)=\left\vert 1-e^{i\lambda}\right\vert ^{-2d}\overline{g}(\lambda),
\]
where $0<\overline{g}(0)<\infty$. This implied definition of an $I(0)$ process
also appears in Robinson (1994a), Marinucci and Robinson (2001), and Robinson
and Yajima (2002).

When $d>0$ we note that
\[
\sum\limits_{j=0}^{\infty}\pi_{j}(-d)=0.
\]
This implies that the process $X_{t}$ is covariance stationary, but with zero
spectrum at the origin, see also (\ref{a7}). It also implies that, for any
constant $\mu$ and $d>0$, the fractional difference $\Delta^{d}\mu=\mu
\sum_{j=0}^{\infty}\pi_{j}(-d)=0$, thus generalizing in a natural way the fact
that $\Delta\mu=0$.

\subsection{Nonstationary fractional integration}

The above discussion focused on the stationary case. However, it is both
empirically and theoretically relevant to consider nonstationary situations,
preferably including the celebrated $I(1)$ process as a special case. In such
situations, the treatment of initial conditions (pre-sample observations) may
be critical and, in fact, affects not only the choice of a convention
regarding initial observations from which to obtain parameter estimates, but
also the underlying process under consideration.

Two different definitions have been applied in the literature. The first of
these (Type~I) uses the idea in (\ref{integration}), while the second
(Type~II) uses truncation. These definitions mirror different definitions of
fractional Brownian motions (also denoted Type~I and Type~II) to which the
suitably normalized fractionally integrated processes converge; see Marinucci
and Robinson (1999) for a very detailed analysis of the different types of
convergence and Section~\ref{sec:fclt} below for an overview.

\begin{definition}
[Type I fractionally integrated process]\label{def type I}Let $k$ be an
integer and $d$ a real number such that $-1/2<d-k<1/2$. Given a scalar
$I(0)$\ process $u_{t}$ with mean zero, the process $X_{t}$ is Type~I
fractionally integrated of order $d$, denoted $X_{t}\in F_{1}(d)$, if%
\[
\Delta^{k}X_{t}=\Delta^{-(d-k)}u_{t}.
\]
\smallskip
\end{definition}

We note that the differencing operators in Definition~\ref{def type I} cannot
be combined. Specifically, the integer-order difference $\Delta^{k}$ involves
only a finite summation, and hence always exists, whereas the fractional
difference $\Delta^{-(d-k)}$ exists because it is assumed that $-1/2<d-k<1/2$.
Thus, the integer-differenced series can be defined as a fractional process,
which is then cumulated back. This interpretation has given rise to the notion
of `difference-and-cumulate-back'.

As an example, when $1/2<d<3/2$, the process $\Delta X_{t}$ is an $I\left(
d-1\right)  $ covariance stationary process, but $X_{t}$ itself is
nonstationary. Inverting the $\Delta$-operator we find the following
generalization of (\ref{integration}),
\begin{equation}
X_{t}=X_{0}+\sum\limits_{j=1}^{t}\Delta^{-(d-k)}u_{j},\quad t\geq1,
\label{integration 2}%
\end{equation}
so that when $d=1$ Definition~\ref{def type I} implies the standard definition
of an $I(1)$ process known from the integer-integration framework. The same
considerations apply for larger integer orders. In particular, the choice of
starting point $t=0$ in (\ref{integration 2}) is arbitrary, as in
(\ref{integration}), and for larger values of $k$ the choice $t=1-k$ would
result in a very similar equation and interpretation. As in (\ref{integration}%
), the values of $X_{t}$ for $t<1-k$ may then be taken to be zero.

Some early works using Type~I fractionally integrated processes include Sowell
(1990), Hurvich and Ray (1995), Chan and Terrin (1995), Jeganathan (1999),
Velasco (1999a,b), among others.

An important limitation of the Type~I process in Definition~\ref{def type I}
is the (necessary) requirement that $-1/2<d-k<1/2$. This implies that, in
principle, a different definition is given for each value of the integer $k$.
Perhaps more importantly, the definition does not cover the values
$d=\pm1/2,\pm3/2$, etc. To make the parameter space for $d$ compact in an
estimation setting, it would then be necessary to remove open neighborhoods of
these values from the parameter space. This is not only mathematically
displeasing, but has important implications for hypothesis testing. The value
$d=1/2$, for example, is exactly the boundary between stationarity and
nonstationarity and is therefore an obvious value to impose under the null
hypothesis so it is unfortunate if it is not in the parameter space.

\begin{definition}
[Type II\ fractionally integrated process]\label{def type II}Let $d$ be any
real number. Given a scalar $I(0)$\ process $u_{t}$ with mean zero, the
process $X_{t}$ is Type~II fractionally integrated of order $d$, denoted
$X_{t}\in F_{2}(d)$, if
\begin{equation}
X_{t}=\sum\limits_{j=0}^{t-1}\pi_{j}\left(  d\right)  u_{t-j}=\Delta_{+}%
^{-d}u_{t} \label{Ia6}%
\end{equation}
with the obvious definition of the truncated operator $\Delta_{+}^{-d}$.
\end{definition}

This definition has different implications from those of
Definition~\ref{def type I}. For example, in the case $d<1/2$ and $d\neq0$,
the Type~II process\ is nonstationary while the Type~I process is stationary.
However, as shown in Lemma~3.4 of Robinson and Marinucci (2001), under
relatively mild conditions,%
\begin{equation}
\lim_{t\rightarrow\infty}\left\vert \operatorname{Cov}(\Delta^{-k}%
\Delta^{-(d-k)}u_{t},\Delta^{-k}\Delta^{-(d-k)}u_{t+k})-\operatorname{Cov}%
(\Delta_{+}^{-d}u_{t},\Delta_{+}^{-d}u_{t+k})\right\vert =0\text{ for all
}k\geq0.\label{asy stat}%
\end{equation}
Thus, in this case, the Type~II process $\Delta_{+}^{-d}u_{t}$ could be
considered \textquotedblleft asymptotically stationary\textquotedblright%
\ because the nonstationarity is due only to the truncation on the right-hand
side of (\ref{Ia6}). For $d\geq1/2$, $\Delta_{+}^{-d}u_{t}$ is purely
nonstationary and the truncation in (\ref{Ia6}) ensures that the Type~II
process is well-defined in the mean-square sense. Robinson (2005a) provides
additional bounds for differences between the two fractionally integrated
processes. Finally, we note that the linear convolution in (\ref{Ia6}) can be
computed very efficiently by the use of the fast Fourier transform; see Jensen
and Nielsen (2014).

The Type~II definition has been applied by Marinucci and Robinson (2000,
2001), Robinson and Marinucci (2001), Robinson and Hualde (2003), Shimotsu and
Phillips (2005), Johansen and Nielsen (2010, 2012a), among others.

Note that both Definitions~\ref{def type I} and~\ref{def type II} imply that
$E(X_{t})=0$. However, it is straighforward to extend the definitions to cover
processes with a nonzero deterministic component by defining $Y_{t}=\mu
_{t}+X_{t}$, where $X_{t}$ is defined in either of
Definitions~\ref{def type I} or~\ref{def type II} and $\mu_{t}$ is an
arbitrary deterministic term.

In the remainder of this chapter, the type of fractional integration will in
most cases either be clear from the context or the distinction is not
relevant, and we will then use the generic notation $I(d)$. In cases where the
distinction is important we will clearly indicate either Type~I or Type~II in
the text.

Note that Definitions~\ref{def type I} and~\ref{def type II} are identical for
$d=0$ and for positive integers. Also note that they both impose zero initial
conditions for the nonstationary range of the memory parameter~$d$. However,
this can easily be relaxed because, focusing on (\ref{a2}),%
\begin{equation}
\Delta^{d}X_{t}=\sum\limits_{j=0}^{\infty}\pi_{j}(-d)X_{t-j}=\sum
\limits_{j=0}^{t-1}\pi_{j}(-d)X_{t-j}+\sum\limits_{j=t}^{\infty}\pi
_{j}(-d)X_{t-j}=\Delta_{+}^{d}X_{t}+\Delta_{-}^{d}X_{t}, \label{trunc}%
\end{equation}
using the obvious definitions for the operators $\Delta_{+}^{d}$
and~$\Delta_{-}^{d}$. Re-arranging and using (\ref{a2}),%
\[
\Delta_{+}^{d}X_{t}=u_{t}-\Delta_{-}^{d}X_{t}.
\]
As justified by Johansen (2008) and Johansen and Nielsen (2010, 2012a),
$\Delta_{+}^{d}$ is an invertible operator for any $d$ (because it involves
only a finite summation). Consequently,%
\begin{equation}
X_{t}=\Delta_{+}^{-d}u_{t}-\Delta_{+}^{-d}\Delta_{-}^{d}X_{t}=\sum
\limits_{j=0}^{t-1}\pi_{j}(d)u_{t-j}+\mu_{t}, \label{a5}%
\end{equation}
where $\mu_{t}$ is an initial condition term which depends only on $X_{-j}$,
$j\geq0$.

Clearly, (\ref{a5}) generalizes (\ref{integration}), (\ref{integration 2}),
and (\ref{Ia6}) to allow for a more general initial condition if conditions
are imposed such that $\mu_{t}$ exists. In the special case where $d=1$ it
holds that $\mu_{t}=X_{0}$ which results in (\ref{integration}). The most
common assumption in the literature (either implicitly or explicitly) is that
$X_{t}=0$ for $t\leq0$, which results in $\mu_{t}=0$ and hence the Type~II
definition with zero initial condition. More generally, Johansen and Nielsen
(2010, 2012a, 2016) give several different conditions on $X_{-j}$, $j\geq0$,
that ensure existence of $\mu_{t}$. For example, setting $X_{-j}=0$ but only
for all $j\geq N_{0}$, where $N_{0}$ could be an arbitrarily large but finite integer.

\subsection{ARFIMA processes}

The autoregressive fractionally integrated moving-average (ARFIMA, or
sometimes FARIMA) class is the most popular family of parametric models
displaying fractional integration.

\begin{definition}
[ARFIMA process]\label{Def arfima}Let $d$ be such that $-1/2<d<1/2$ and let
$\varepsilon_{t}$ be a zero mean white noise process with $\operatorname{Var}%
(\varepsilon_{t})=\sigma^{2}$. Then $X_{t}$, $t\in\mathbb{Z}$, is an
ARFIMA($p,d,q$) process if%
\begin{equation}
\phi(L)\Delta^{d}X_{t}=\theta(L)\varepsilon_{t}, \label{a8}%
\end{equation}
where%
\begin{align*}
\phi(L)  &  =1-\phi_{1}L-\phi_{2}L^{2}-\dots-\phi_{p}L^{p},\\
\theta(L)  &  =1+\theta_{1}L+\theta_{2}L^{2}+\dots+\theta_{q}L^{q},
\end{align*}
and $\phi(z)$, $\theta(z)$ have no common zeroes and satisfy $\phi
(z)\theta(z)\neq0$ for $|z|\leq1$.
\end{definition}

Under these conditions, the AR and MA polynomials, $\phi(z)$ and $\theta(z)$,
imply both stationarity and invertibility. Hence, (\ref{a8}) has the unique
stationary solution
\begin{equation}
X_{t}=\phi^{-1}(L)\theta(L)\Delta^{-d}\varepsilon_{t}, \label{a9}%
\end{equation}
and $X_{t}$ is a stationary and invertible process. Furthermore, $X_{t}$ has
mean zero, spectral density%
\[
f_{X}(\lambda)=\frac{\sigma^{2}}{2\pi}\left\vert \phi(e^{i\lambda})\right\vert
^{-2}\left\vert \theta(e^{i\lambda})\right\vert ^{2}\left\vert 1-e^{i\lambda
}\right\vert ^{-2d}\sim\frac{\sigma^{2}\theta^{2}(1)}{2\pi\phi^{2}(1)}%
\lambda^{-2d}\text{ as }\lambda\rightarrow0,
\]
and autocovariances that satisfy (\ref{Iaa}). Interestingly, for this process
the memory parameter $d$ drives the long-run behavior, whereas the ARMA
parameters in $\phi(z)$ and $\theta(z)$ control the short-run dynamics.

For $d\leq-1/2$, the ARFIMA($p,d,q$) process $X_{t}$, $t\in\mathbb{Z}$, can be
defined as%
\begin{equation}
\phi(L)X_{t}=\theta(L)\Delta^{-d}\varepsilon_{t}, \label{a10}%
\end{equation}
imposing that $\phi(z)$, $\theta(z)$ have no common zeroes and that
$\phi(z)\theta(z)\neq0$ for $|z|\leq1$. In this case, the unique stationary
solution to (\ref{a10}) is given by (\ref{a9}), although this solution is not invertible.

Nonstationarity can be accommodated within the ARFIMA framework using two
different strategies. One possibility in the spirit of the Type~I process is
to keep (\ref{a8}), but relax the conditions on $\phi(z)$ so as to allow for
unit roots in the AR polynomial. This alternative was pursued by, e.g.,
Jeganathan (1999), and it implies that, depending on the number of unit roots,
$X_{t}$ could be $I(d+1)$, $I(d+2)$, etc. Another possibility is to define the
observable process as a Type~II fractionally integrated process and maintain
the conditions on $\phi(z)$. That is, instead of (\ref{a8}), defining%
\[
\phi(L)\Delta_{+}^{d}X_{t}=\theta(L)\varepsilon_{t},
\]
using the truncated filter in (\ref{Ia6}).

\subsection{Fractional Brownian motion and FCLT}

\label{sec:fclt}

Marinucci and Robinson (1999) describe the two alternative forms of fractional
Brownian motion that have been employed in the literature. These two forms are
linked to the different types of fractionally integrated processes described
before, in the sense that the two types of fractional Brownian motions are the
limits of such processes in corresponding functional limit theorems. First,
the Type~I fractional Brownian motion was introduced by Mandelbrot and Van
Ness (1968). Letting $B(r)$ be a standard Brownian motion, for $-1/2<d<1/2$,
one version of this process is%
\[
B_{d}(r)=\left\{
\begin{array}
[c]{cc}%
\frac{1}{A(d)}\left(  \int_{-\infty}^{0}\left(  (r-s)^{d}-(-s)^{d}\right)
\mathsf{d}B(s)+\int_{0}^{r}(r-s)^{d}\mathsf{d}B(s)\right)  & \text{for }%
r\geq0,\\
\frac{1}{A(d)}\left(  \int_{-\infty}^{r}\left(  (r-s)^{d}-(-s)^{d}\right)
\mathsf{d}B(s)-\int_{r}^{0}(-s)^{d}\mathsf{d}B(s)\right)  & \text{for }r<0,
\end{array}
\right.
\]
where%
\[
A(d)=\left(  \frac{1}{2d+1}+\int_{0}^{\infty}\left(  (1+s)^{d}-s^{d}\right)
^{2}\mathsf{d}s\right)  ^{1/2}.
\]

It can be shown that $E(B_{d}(r))=0$ for $r\in%
\mathbb{R}
$,
\begin{equation}
E(B_{d}(r_{1})B_{d}(r_{2}))=\frac{1}{2}\left(  \left\vert r_{1}\right\vert
^{2d+1}+\left\vert r_{2}\right\vert ^{2d+1}-\left\vert r_{2}-r_{1}\right\vert
^{2d+1}\right)  \text{ for\ }r_{1},r_{2}\in%
\mathbb{R}
, \label{a12}%
\end{equation}
and also that, for $j\in%
\mathbb{Z}
$, the sequence of increments $B_{d}(j+1)-B_{d}(j)$ is identically $N(0,1)$
distributed. However, unlike in the short memory case, where $d=0$ (and
$B_{0}(r)=B(r)$), this sequence is correlated if $d\neq0$ with autocovariances
decaying at rate $j^{2d-1}$. It can be shown that if $X_{t}$ is a mean zero
Type~I fractionally integrated process with $-1/2<d<12$, then%
\begin{equation}
\frac{1}{T^{d+1/2}}\sum_{t=1}^{\left\lfloor Tr\right\rfloor }X_{t}\Rightarrow
c_{d}B_{d}(r)\text{ for\ }r\in\lbrack0,1], \label{a13}%
\end{equation}
where $c_{d}$ is a constant that is proportional to the long-run variance of
$\Delta^{d}X_{t}$.

Alternatively, for any $d>-1/2$, the Type~II fractional Brownian motion
$W_{d}(r)$ can be defined as%
\begin{equation}
W_{d}(r)=\left\{
\begin{array}
[c]{cc}%
(2d+1)^{1/2}\int_{0}^{r}(r-s)^{d}\mathsf{d}B(s) & \text{for }r\geq0,\\
-(2d+1)^{1/2}\int_{r}^{0}(s-r)^{d}\mathsf{d}B(s) & \text{for }r<0,
\end{array}
\right.  \label{a33}%
\end{equation}
where, as for $B_{d}(r)$, $W_{0}(r)=B(r)$. It can be shown that $E(W_{d}%
(r))=0$ for $r\in%
\mathbb{R}
$ and
\[
E(W_{d}^{2}(r))=\left\vert r\right\vert ^{2d+1},
\]
so that $E(W_{d}^{2}(r))=E(B_{d}^{2}(r))$. However, as explained in Marinucci
and Robinson (1999), the covariance structure of $W_{d}(r)$ is substantially
different from (\ref{a12}), noting that the increments $W_{d}(r_{2}%
)-W_{d}(r_{1})$ at equally-spaced intervals are nonstationary (although
asymptotically stationary in a sense similar to (\ref{asy stat})).
Additionally, for $j=0,1,\dots$, the increments $W_{d}(j+1)-W_{d}(j)$ are also
nonstationary, having zero mean but a non-constant variance and time-dependent
covariances. However, $W_{d}\left(  r\right)  $ is self-similar of degree
$d+1/2$, meaning that the distribution of $W_{d}(kr)$ is identical to that of
$k^{d+1/2}W_{d}(r)$. This self-similarity property is also shared by
$B_{d}(r)$. As in (\ref{a13}), it can be shown that if $X_{t}$ is a mean zero
Type~II fractionally integrated process of order $d$ with $d>-1/2$, then%
\begin{equation}
\frac{1}{T^{d+1/2}}%
{\displaystyle\sum\limits_{t=1}^{\left\lfloor Tr\right\rfloor }}
X_{t}\Rightarrow k_{d}W_{d}(r)\text{ for\ }r\in\lbrack0,1], \label{a14}%
\end{equation}
where, again, $k_{d}$ is a constant that is proportional to the long-run
variance of $\Delta_{+}^{d}X_{t}$.

The result (\ref{a13}) has been established under various conditions; see
Davydov (1970), Taqqu (1975), and others. The result (\ref{a14}) has been
established by Akonom and Gourieroux (1987) and Marinucci and Robinson (2000).
Assuming that $u_{t}$ in (\ref{Ia6}) is a linear processes generated by a
zero-mean white noise process $\varepsilon_{t}$, the most stringent condition
is the requirement $E\left\vert \varepsilon_{t}\right\vert ^{q}<\infty$ for
some $q>\max\{2,1/(d+1/2)\}$, so that a large number of moments might be
required if $d$ is close to $-1/2$ (of course, Gaussianity suffices). Johansen
and Nielsen (2012b) showed that this moment condition is necessary for the
functional central limit theorem.

\subsection{Sources of long memory}

\label{sec:sources}

Several (economic) mechanisms have been proposed that generate long memory in
observed data. These include aggregation of heterogeneous units, the error
duration model, models of learning with rational agents, and marginalization
in large systems. The earliest and most well-known mechanism seems to be that
the aggregation of heterogeneous units can produce processes which display
long memory properties, so we will discuss this in more detail.

The conceptual framework for the aggregation idea is the random coefficients
AR(1) process,%
\begin{equation}
X_{i,t}=\alpha_{i}X_{i,t-1}+\varepsilon_{i,t},\quad t\in\mathbb{Z},\quad
i=1,\dots,N, \label{a11}%
\end{equation}
where $\varepsilon_{i,t}$ is a zero mean white noise process independent of
the random coefficients $\alpha_{i}$, which are i.i.d.\ Beta with shape
parameters $p>1$ and $q>1$. Robinson (1978) explored the properties of this
process for a more general distribution function for $\alpha_{i}$,
establishing necessary and sufficient conditions for the continuity of the
spectral density of $X_{i,t}$, examplifying also his results to the Beta
distribution case.

Granger (1980) aggregated $X_{i,t}$ in (\ref{a11}) over units, defining
$X_{t}=\sum_{i=1}^{N}X_{i,t}$, and then showed that, as $N\rightarrow\infty$,
the aggregated series $X_{t}$ has autocovariances that decay at a hyperbolic
rate like a long memory process. More specifically, the autocovariances of
$X_{t}$ behave like (\ref{Iaa}) with memory parameter $d=1-q/2$. Hence, the
parameter $q$, which determines the slope of the density function of
$\alpha_{i}$ near one, determines the amount of memory in $X_{t}$. Similar
ideas have been explored by Zaffaroni (2004), Davidson and Sibbertsen (2005),
and others, and aggregation as a source of long memory has been widely used in
empirical works; e.g., Gadea and Mayoral (2006) for inflation series or Byers,
Davidson, and Peel (2007) for series of political support.

Haldrup and Vera Vald\'{e}s (2017) justified that the aggregated series,
although displaying the usual long memory properties, does not satisfy an
ARFIMA specification. In particular, they derived the properties of a
fractionally differenced (with the adequate memory parameter) cross-sectional
aggregated series and show that its autocovariances, while summable, do not
correspond to those of an ARMA. This result introduces a word of caution when
modelling aggregated series with a parametric model.

An alternative source of long memory is the error duration model proposed by
Parke (1999). In this model, the i.i.d.\ shock $\varepsilon_{t}$ has a
stochastic duration $n_{s}$, after which it disappears from the observed
process. Letting $g_{s,t}=\mathbb{I}(t\leq s+n_{s})$, where $\mathbb{I}(A)$
denotes the indicator function of the event $A$, the observed process is
$y_{t}=\sum_{s=-\infty}^{t}g_{s,t}\varepsilon_{s}$. Parke (1999) showed that,
depending on the values of the survival parameters $p_{k}=P(g_{s,s+k}=1)$,
$y_{t}$ can exhibit both long memory and fractional integration. For example,
if $p_{k}=\Gamma(k+d)\Gamma(2-d)/(\Gamma(k+2-d)\Gamma(d))$ for $d\in
\lbrack0,1]$ then $y_{t}$ is fractionally integrated noise with memory
parameter $d$.

More recently, Chevillon and Mavroeidis (2017) have proposed a model of
learning that can generate long memory endogenously, without any persistence
in the exogenous shocks. The learning mechanism is embedded in a prototypical
representative-agent forward-looking model in which agents' beliefs are
updated using linear learning algorithms. Depending on the weights that agents
place on past observations when they update their beliefs, and on the
magnitude of the feedback from expectations to the endogenous variable, this
model and learning mechanism can generate long memory in the observed series.
This is distinctly different from the case of rational expectations, where the
memory of the endogenous variable is determined exogenously.

Also very recently, Chevillon, Hecq, and Laurent (2018) have shown how a
high-dimensional vector autoregressive (VAR) model of finite order can
generate fractional integration in the marginalized univariate series. That
is, when an observed univariate series is in fact generated by a
high-dimensional VAR with cross-section dependence, this leads to long memory
behavior in the univariate series. An example is given where the final
equation representation of a VAR(1) leads to univariate fractional noises.

\subsection{Long memory and structural breaks}

\label{sec:breaks}

An important early literature showed that structural change and unit roots can
be confused (e.g.\ Perron, 1989, and Stock, 1994). This phenomenon generalizes
to long memory in the sense that processes like regime-switching can display
typical long memory (or fractionally integrated) behaviour. This issue was
anticipated in the pioneering work of Klemes (1974) who justified that the
Hurst phenomenon could be caused by two features which arise in hydrologic
process: nonstationarity in the mean and random walks with one absorbing
barrier. In a similar vein, Lobato and Savin (1997) pointed out
nonstationarity in mean as a reason for rejecting the weak dependence
hypothesis in favor of the long memory alternative. Granger and Ter\"{a}svirta
(1999) showed that a similar behaviour could occur within a stationary
setting. Specifically, they analyzed the nonlinear autoregressive model%
\[
y_{t}=\operatorname{sign}(y_{t-1})+\varepsilon_{t},
\]
where $\operatorname{sign}(\cdot)=\pm1$ with the same sign as the argument and
$\varepsilon_{t}$ is i.i.d.$N(0,\sigma^{2})$. This process has zero mean and
it is stationary with $k$'th order autocorrelation given by $(1-2p)^{k}$,
where $p=\Pr(\varepsilon_{t}>1)$. Although the autocorrelation does not
exhibit hyperbolic decay as in (\ref{autocorr}), it is clear that, as $p$
decreases, switches are more infrequent so the process has more memory.
Looking at the linear properties of the data through estimated correlations or
the periodogram, it can well be concluded that the behavior of $y_{t}$ is
closer to that of an $I\left(  d\right)  $ process rather than short memory.

Diebold and Inoue (2001) justified how stochastic regime-switching models can
display typical long memory behaviour. More specifically, they first studied
the properties of a mixture model $y_{t}$ given by%
\[
y_{t}=\left\{
\begin{array}
[c]{l}%
0\text{ with probability }1-p_{T},\\
\varepsilon_{t}\text{ with probability }p_{T},
\end{array}
\right.
\]
where $\varepsilon_{t}$ is i.i.d.$N(0,\sigma^{2})$. Here, it is easily shown
that, letting the mixture probability $p_{T}$ tend to zero at an appropriate
rate, the variance of the partial sum of $y_{t}$ behaves like that of an
$I(d)$ process. Extensions of this simple mixture model present very similar
patterns. Additionally, they studied the behaviour of a particular stochastic
permanent break model given by%
\begin{align*}
y_{t}  &  =\mu_{t}+\varepsilon_{t},\\
\mu_{t}  &  =\mu_{t-1}+\frac{\varepsilon_{t-1}^{2}}{\gamma_{T}+\varepsilon
_{t-1}^{2}}\varepsilon_{t-1},
\end{align*}
where $\varepsilon_{t}$ is i.i.d.$N(0,\sigma^{2})$ and $\gamma_{T}>0$ is
varying with $T$. This model is an approximation of a mean-plus-noise
extension of the mixture model presented before, so it is expected that a
similar long memory pattern appears, which is also the case when $\gamma_{T}$
grows with the sample size appropriately. Finally, they discussed the
properties of the Markov-switching model of Hamilton (1989). Again,the
variance of the partial sum of this process matches the behaviour of an $I(d)$
process when transition probabilities tend to one at an appropriate rate. In
any case, Diebold and Inoue (2001) warn against the temptation to draw
potentially na\"{\i}ve conclusions about the specific character of $I(d)$
processes. In their view, at least in the settings described in their paper,
structural change and $I(d)$ are two different labels for the same phenomenon,
so there is little value in assigning labels like \textquotedblleft
true\textquotedblright\ or \textquotedblleft spurious\textquotedblright\ to
one or the other.

Similar works include Gourieroux and Jasiak (2001), who provided further
evidence on the connection between infrequent breaks and long memory by
analyzing the estimated correlogram of a regime-switching model, and Granger
and Hyung (2004), who (in contemporaneous and independent work from Diebold
and Inoue, 2001), studied a similar occasional break process. Specifically,
they proposed
\begin{align*}
y_{t}  &  =m_{t}+\varepsilon_{t},\\
m_{t}  &  =m_{t-1}+q_{t}\eta_{t},
\end{align*}
where $\varepsilon_{t}$ and $\eta_{t}$ are white noises,%
\[
q_{t}=\left\{
\begin{array}
[c]{l}%
0\text{ with probability }1-p_{T},\\
1\text{ with probability }p_{T},
\end{array}
\right.
\]
and $p_{T}$ converges slowly to zero with $T$ such that $Tp_{T}$ tends to a
non-zero finite constant. In this setting, $y_{t}$ displays some of the
properties of an $I(d)$ process.

\section{Inference in (univariate) fractionally integrated models}

\label{sec:inference}

\subsection{Semiparametric estimation of fractional integration}

Semiparametric estimation of the memory parameter has a long tradition in the
time series literature. These methods are based on the following local
approximation of the spectral density of a stationary long memory
process$\ X_{t}$,%
\begin{equation}
f_{X}(\lambda)\sim G\lambda^{-2d}\text{ as }\lambda\rightarrow\infty,
\label{a15}%
\end{equation}
where $G$ is a positive constant and $-1/2<d<1/2$; see also (\ref{a7}). Here,
two classes of frequency domain estimators have become very popular: the
log-periodogram approach proposed by Geweke and Porter-Hudak
(1983)\footnote{On the occasion of its 35th anniversary, this famous article
was celebrated with a special issue of the \emph{Journal of Time Series
Analysis}; see Nielsen and Hualde (2019).} whose theoretical properties were
first analyzed by Robinson (1995a), and the local Whittle (also known as
Gaussian semiparametric) estimator, proposed by K\"{u}nsch (1987) and analyzed
by Robinson (1995b). See Velasco (2006) for a very detailed review on these
inference methods and their extensions.

The main reason why these methods have become so popular is that, unlike fully
parametric approaches, semiparametric methods remain agnostic about the
short-run structure of the model. The obvious advantage of this strategy is
that it limits considerably the risk of misspecification. Additionally, given
that these procedures rely on conditions of the spectral density
$f_{X}(\lambda)$ near zero frequency, the estimation methods are robust to a
nonstandard behaviour of $f_{X}(\lambda)$ at higher frequencies, like that
implied by poles or zeros. Therefore, the methods are valid for a very wide
range of processes. As will be seen, most of these methods are invariant to
the presence of a non-zero mean. This is a very attractive feature of the
semiparametric approach because in a stationary long memory context the mean
is estimated potentially with a very slow rate (this is particularly the case
if $d$ is close to 1/2); see (\ref{a13}). Of course, this advantage is not
exclusive of semiparametric methods, but, given that these procedures are
nearly always expressed in the frequency domain, dealing with deterministic
components is typically easier in this context.

The main price to pay for the greater generality that semiparametric
estimation allows is a slower rate of convergence compared with that achieved
by parametric alternatives. In particular, given a user-chosen parameter $m$
(known as bandwidth) such that $m/T\rightarrow0$ as $T\rightarrow\infty$, the
typical rate achieved by semiparametric methods is $\sqrt{m}$ which is slower
than the usual $\sqrt{T}$ parametric rate. Methods to improve this rate in a
semiparametric framework have been proposed, but these always rely on imposing
stronger smoothness conditions on $f_{X}(\lambda)$, so there is a trade-off
between generality and rate of convergence. This discussion prompts another
important issue which always arises in semiparametric estimation, namely
bandwidth choice.

The log-periodogram estimator is based on taking logs in (\ref{a15}) and
approximating the log-spectral density by the log-periodogram. Define the
Fourier frequencies as $\lambda_{j}=2\pi j/T$ for $j=0,1,\dots,T$, the
discrete Fourier transform of an arbitrary vector process $\zeta_{t}$ as%
\[
w_{\zeta}(\lambda)=\frac{1}{\sqrt{2\pi T}}\sum_{t=1}^{T}\zeta_{t}e^{it\lambda
},
\]
and for another vector process $\xi_{t}$ (possibly the same one), the
(cross-)periodogram as%
\begin{equation}
I_{\zeta\xi}(\lambda)=w_{\zeta}(\lambda)w_{\xi}^{\prime}(-\lambda),\quad
I_{\zeta}(\lambda)=I_{\zeta\zeta}(\lambda), \label{a23}%
\end{equation}
with prime denoting transposition. Then, for an observable scalar process
$X_{t}$ with spectral density satisfying (\ref{a15}), the following regression
model is proposed%
\begin{equation}
\log I_{X}(\lambda_{j})=\log G-2d\log\lambda_{j}+v_{j}, \label{a16}%
\end{equation}
where $v_{j}$ is an error term. Because (\ref{a15}) holds for $\lambda
\rightarrow0$, the approximation (\ref{a16}) is expected to work well for low
frequencies $\lambda_{j}$, and the log-periodogram estimator of $d$, denoted
$\widehat{d}_{LP}$, is derived from estimating (\ref{a16}) by ordinary least
squares (OLS) using observations corresponding to $j=l,\dots,m$. Here, $l$ is
a trimming parameter, so the lowest Fourier frequencies $\lambda_{0}%
,\dots,\lambda_{l-1}$ are not considered in the estimation. By the properties
of the complex exponential, removal of the components associated with
$\lambda_{0}$ makes the estimation procedure invariant to a non-zero mean.

The limiting properties of $\widehat{d}_{LP}$ were justified by Robinson
(1995a) (actually, his estimator was more general, including possible pooling
of adjacent Fourier frequencies). The assumptions included $X_{t}$ being
Gaussian and a smoothness condition for $f_{X}(\lambda)$, namely%
\begin{equation}
f_{X}(\lambda)=G\lambda^{-2d}(1+O(\lambda^{\beta}))\text{ as }\lambda
\rightarrow\infty,\quad\beta\in(0,2]. \label{a19}%
\end{equation}
Thus, the larger $\beta$, the closer is $f\left(  \lambda\right)  $ to
$G\lambda^{-2d}$ locally (for an ARFIMA process $\beta=2$). Imposing the
bandwidth condition%
\begin{equation}
\frac{\sqrt{m}\log m}{l}+\frac{l\left(  \log T\right)  ^{2}}{m}+\frac
{m^{1+2\beta}}{T^{2\beta}}\rightarrow0\text{ as }T\rightarrow\infty,
\label{a18}%
\end{equation}
and using very technical arguments to deal with the nonlinearity in $\log
I_{X}(\lambda_{j})$, Robinson (1995a) justified that%
\begin{equation}
\sqrt{m}(\widehat{d}_{LP}-d)\rightarrow_{d}N(0,\pi^{2}/24)\text{ as
}T\rightarrow\infty. \label{a17}%
\end{equation}
Nicely, as long as (\ref{a18}) is satisfied, (\ref{a17}) does not depend on
the trimming parameter $l$, which was introduced because removal of the
frequencies closest to zero allowed for controlling adequately an inherent
bias problem in the estimation. From the theoretical viewpoint, some important
improvements over Robinson's (1995a) results have been introduced, like
relaxing the Gaussian assumption (Velasco, 2000) and avoiding trimming and
using frequencies $\lambda_{1},\dots,\lambda_{m}$ in the estimation (Hurvich,
Deo and Brodsky, 1998).

An alternative semiparametric approach is based on maximizing a local
approximation to the frequency domain Whittle likelihood. The parametric
Whittle log-likelihood is given by%
\[
Q_{T}(\theta)=\frac{1}{T}\sum_{j=1}^{T}\left(  \log f_{X}(\lambda_{j}%
;\theta)+\frac{I_{X}(\lambda_{j})}{f_{X}(\lambda_{j};\theta)}\right)  ,
\]
where the spectral density of $X_{t}$ is considered to be a known function,
$f_{X}(\lambda_{j};\theta)$, up to a vector of unknown parameters, $\theta$.
K\"{u}nsch (1987) suggested to use just $m$ frequencies, with $m/T\rightarrow
0$ as $T\rightarrow\infty$, replacing also $f_{X}(\lambda_{j};\theta)$ by the
local approximation (\ref{a15}), to obtain the local Whittle log-likelihood%
\begin{equation}
Q_{m}(G,d)=\frac{1}{m}\sum_{j=1}^{m}\left(  \log(G\lambda_{j}^{-2d}%
)+\frac{I_{X}(\lambda_{j})}{G\lambda_{j}^{-2d}}\right)  . \label{a46}%
\end{equation}
Concentrating $Q_{m}(G,d)$ with respect to $G$, we define the local Whittle
estimator of $d$ as%
\[
\widehat{d}_{LW}=\arg\min_{d\in D}R_{m}(d),
\]
where%
\[
R_{m}(d)=\log\widehat{G}(d)-\frac{2d}{m}\sum_{j=1}^{m}\log\lambda_{j}%
,\quad\widehat{G}(d)=\frac{1}{m}\sum_{j=1}^{m}\lambda_{j}^{2d}I_{X}%
(\lambda_{j}),
\]
and $D$ is a compact subset of $(-1/2,1/2)$. The local Whittle estimator is
attractive because of the likelihood interpretation and because it is more
efficient than $\widehat{d}_{LP}$ under weaker conditions. On the other hand,
while the log-periodogram estimator has an explicit form, the local Whittle is
defined only implicitly and hence requires numerical optimization. Robinson
(1995b) justified the limiting properties of $\widehat{d}_{LW}$ under certain
conditions, including the smoothness condition (\ref{a19}),
\begin{equation}
X_{t}=%
{\displaystyle\sum\limits_{j=0}^{\infty}}
c_{j}\varepsilon_{t-j},\quad%
{\displaystyle\sum\limits_{j=0}^{\infty}}
c_{j}^{2}<\infty, \label{a20}%
\end{equation}
the $\varepsilon_{t}$\ in (\ref{a20}) being stationary and ergodic with finite
fourth moment, $E(\varepsilon_{t}|\mathcal{F}_{t-1})=0$, $E(\varepsilon
_{t}^{2}|\mathcal{F}_{t-1})=1$ a.s., where $\mathcal{F}_{t}$\ is the $\sigma
$-field of events generated by $\varepsilon_{s}$, $s\leq t$, and conditional
(on $\mathcal{F}_{t-1}$) third and fourth moments of $\varepsilon_{t}$\ equal
the corresponding unconditional moments. Under the bandwidth condition%
\[
\frac{1}{m}+\frac{m^{1+2\beta}(\log m)^{2}}{T^{2\beta}}\rightarrow0\text{ as
}T\rightarrow\infty,
\]
and following the usual strategy for dealing with implicitly defined
estimators, Robinson (1995b) first proved that $\widehat{d}_{LW}$ is
consistent (in fact, under weaker conditions than those above), and then%
\begin{equation}
\sqrt{m}(\widehat{d}_{LW}-d)\rightarrow_{d}N(0,1/4)\text{ as }T\rightarrow
\infty. \label{a21}%
\end{equation}
Compared with (\ref{a17}), (\ref{a21}) implies that $\widehat{d}_{LW}$ is more
efficient than $\widehat{d}_{LP}$. Indeed, the asymptotic variance of
$\widehat{d}_{LP}$ is approximately 1.6 times that of $\widehat{d}_{LW}$.

The finite sample performance of both estimators have been thoroughly
illustrated by means of many Monte Carlo experiments. A particular phenomenon
worth mentioning is the large bias reported in finite samples; see, e.g.,
Agiakloglou, Newbold, and Wohar (1993) or Nielsen and Frederiksen (2005). The
classical example of this problem is the behaviour of the semiparametric
estimators when the data is generated by a stationary ARMA(1,0) process with
autoregressive parameter $\phi_{1}$ close to (but below)~1. In this case, the
spectral density of the process is obviously bounded, but in finite samples it
is difficult to discern from that of a long memory process (with a spectral
pole at zero frequency). As a result, semiparametric estimators typically
overestimate the true memory parameter, $d=0$ This finite-sample distortion
can be very noticeable and, in fact, this is an illustration of a well-known
identification problem: the ARFIMA(1,0,0) model with $\phi_{1}=1$ is identical
to the ARFIMA(0,1,0) model.

As with most finite-sample issues, the problem is alleviated as $T$ increases,
although in some cases very slowly. For any given sample size, there is a
trade-off between bias and variance in the choice of bandwidth, $m$, where a
higher bandwidth tends to increase bias and reduce variance. Asymptotically,
the mean-squared-error minimizing bandwidth choice is $m=O(T^{4/5})$, see
Hurvich, Deo, and Brodsky (1998), but in practice this often results in very
large bias rendering the estimator quite useless. One practical way to deal
with the bias issue and choice of bandwidth is to perform the estimation with
a range of bandwidth parameters and check sensitivity to bandwidth choice.
Excessive sensitivity of the estimates to bandwidth choice is a clear signal
that the bias problem is affecting estimation. At a theoretical level, Andrews
and Guggenberger (2003) and Andrews and Sun (2004) have extended the
log-periodogram and local Whittle approaches, respectively, by including local
polynomial approximations to the short-memory component instead of the
constant $G$. They have shown that this reduces the bias by an order of
magnitude, while only increasing variance by a multiplicative constant.

The original work on semiparametric estimation has been extended in several
directions. First, dealing with nonstationarity is a crucial issue in
practice. In the spirit of the Type~I fractionally integrated process, a
simple possibility is to apply the \textquotedblleft differencing and adding
back\textquotedblright\ strategy, although this requires some a priori
knowledge about the value of $d$. For example, if we know that $d\in
(1/2,3/2)$, we could estimate $d-1$ from $\Delta X_{t}$, obtaining
$\widehat{d-1}$ and then define $\widehat{d}=\widehat{d-1}+1$. Higher
integration orders can be dealt with by taking further integer differences. An
alternative solution was proposed by Velasco (1999a,b), who generalized the
log-periodogram and local Whittle estimators to cover nonstationarity by means
of tapering; that is, a type of periodogram smoothing. The two main drawbacks
of this strategy are the following. First, the order of the needed taper
depends on the memory of the process, hence, again, some a priori knowledge of
$d$ is needed. Second, tapering inflates the asymptotic variance of the
estimators. On the positive side, tapering is a way of dealing with the
possible presence of a polynomial deterministic structure, because an adequate
taper periodogram is invariant to such deterministic components. In a similar
vein, for a Type~II fractionally integrated process, Shimotsu and Phillips
(2005) proposed the exact local Whittle (ELW) estimator. This is a frequency
domain estimator, but instead of \textquotedblleft whitening\textquotedblright%
\ the periodogram (like previous proposals), it applies fractional differences
to the observable process. Nicely, this strategy permits consideration of any
value of $d$ as long as it is contained in a closed interval of width no
larger than 9/2 (although this requirement could possibly be relaxed by an
alternative proof method), and it retains the semiparametric efficiency result
(\ref{a21}). However, an important drawback of this estimator is in dealing
with deterministic components. Shimotsu (2010) studied the sense in which an
unknown mean could have undesirable effects on ELW\ estimation, and proposed
modified estimators to accommodate an unknown mean and a polynomial trend.
Finally, for a Type~I fractionally integrated process, Abadir, Distaso, and
Giraitis (2007) proposed the extended local Whittle estimator, which relies on
the fully extended discrete Fourier transform and periodogram as an
alternative way to deal with the nonstationarity. This estimator does not
require estimation of the mean and achieves the result (\ref{a21}) for
$d\in(-3/2,\infty)$, although excluding values $d=1/2,3/2,\dots$.

An important source of bias is structural breaks and other low-frequency
contaminations. In fact, there is a broad research topic that analyzes several
distinct issues which could fall under the general label of long memory and
structural breaks; see Section~\ref{sec:breaks}. This literature has been
developed in both parametric and semiparametric settings, but the latter
appears to be more attractive in practice since any parametric choice for a
short-memory component is complicated due to the possible presence of breaks.
We will briefly comment on some of the most significative works. Hidalgo and
Robinson (1996) proposed a test for a unique structural change at a known time
point based on OLS estimation in a regression model with a long memory error
whose autocorrelations are modeled semiparametrically. Relatedly, Ohanissian,
Russell, and Tsay (2008) introduced a testing procedure for the null of
stationary long memory against the alternative of spurious long memory
(originating, e.g., from random level shift processes or other
regime-switching-type processes). Their procedure takes advantage of the
invariance of the long memory property under temporal aggregation, and it is
based on the log-periodogram estimator. Focusing on the same problem, Qu
(2011) proposed an alternative testing strategy based on the local Whittle likelihood.

Other works attempt to analyze the properties of long memory estimators when
applied to processes with contaminations such as infrequent level shifts.
Along the evidence presented in Section~\ref{sec:breaks}, it is well known
that in this case the log-periodogram finds long memory. For example, Smith
(2005) characterized the bias of the log-periodogram estimator and proposed a
modified log-periodogram which alleviates the bias problem. Similar extensions
and modifications are proposed and analyzed by McCloskey and Perron (2013) and
Hou and Perron (2014). With a similar motivation, Iacone, Nielsen, and Taylor
(2021) extended Lobato and Robinson's (1998) semiparametric test for the null
hypothesis of $I(0)$ against fractionally integrated alternatives to test the
more general null hypothesis of $I(d)$ for $d\in(-1/2,1/2)$ while being robust
to multiple level shifts at unknown points in time. Their semiparametric
approach extends the parametric one of Iacone, Leybourne, and Taylor (2019)
who allowed for the possibility of a single break at an unknown date.

Another important research area is the extension of semiparametric methods to
deal with perturbed fractional processes, where the observable series is
composed of a long memory process contaminated by an additive noise term.
These processes are clearly related to the structural breaks contaminations,
but the perturbed fractional processes are directly motivated by long memory
stochastic volatility models, which have important applications in finance.
Works here include Breidt, Crato, and de Lima (1998), Deo and Hurvich (2001),
Hurvich and Ray (2003), Sun and Phillips (2003), Arteche (2004, 2006),
Hurvich, Moulines, and Soulier (2005), Haldrup and Nielsen (2007), Frederiksen
and Nielsen (2008), Perron and Qu (2010), Frederiksen, Nielsen, and Nielsen
(2012), McCloskey and Perron (2013), and Hou and Perron (2014).

Finally, other extensions have focused on improving the asymptotic performance
of estimators in the semiparametric setting. The standard estimators have
convergence rates no better than $T^{2/5}$ (see Giraitis, Robinson, and
Samarov, 1997), although this rate can be substantially slower if the spectral
density of the process, $f$, is not sufficiently smooth around frequency zero.
In this sense, the possible smoothness of $f$ around zero can be exploited by
means of bias-reducing techniques in order to improve the semiparametric rate
to the parametric rate, $\sqrt{T}$. As mentioned previously, Andrews and
Guggenberger (2003) and Andrews and Sun (2004) exploited smoothness by
extending the log periodogram and the local Whittle approaches, respectively,
by local polynomials. Similarly, Robinson and Henry (2003) proposed a general
M-estimator which, in particular, nests the log-periodogram and local Whittle
approaches and uses higher-order kernels. These approaches have attractive
theoretical properties, but a somewhat disappointing behaviour in finite
samples (see, e.g., Nielsen and Frederiksen, 2005, and
Garc\'{\i}a-Enr\'{\i}quez and Hualde, 2019). A similar idea is to use global
smoothness conditions on $f$ outside frequency zero to obtain improvements by
means of a broadband approach instead of focusing on a local-to-zero band of
frequencies. This latter approach is analyzed by Moulines and Soulier (1999)
and Hurvich and Brodsky (2001).

\subsection{Parametric estimation of fractional integration}

Let $X_{t}$, $t\in%
\mathbb{Z}
$, be a Gaussian covariance stationary and invertible process with
$E(X_{t})=\mu$ and spectral density $f_{X}(\lambda_{j};\theta)$ that is known
up to a vector of unknown parameters, $\theta$. Given a sample $X_{1}%
,\dots,X_{T}$, it is immediate to construct the log-likelihood. Let
$\widetilde{X}=(X_{1},\dots,X_{T})^{\prime}$, $\Sigma(\theta
)=\operatorname{Var}(\widetilde{X})$, and denote by $\widetilde{1}$ a
$T\times1$ vector of ones. Then, ignoring constants, the log-likelihood is%
\begin{equation}
Q_{1T}(\theta,\mu)=\frac{1}{T}\log|\mathbf{\Sigma}(\theta)|+\frac{1}%
{T}(\widetilde{X}-\mu\widetilde{1})^{\prime}\mathbf{\Sigma}^{-1}%
(\theta)(\widetilde{X}-\mu\widetilde{1}). \label{a24}%
\end{equation}
Letting $\overline{X}=T^{-1}\sum_{t=1}^{T}X_{t}$ and $X_{t}^{dem}%
=X_{t}-\overline{X}$, and using $\Theta$ to denote the parameter space for
$\theta$, the following estimators (at least) of $\theta$ have been considered
in the literature,%
\begin{align*}
\widehat{\theta}_{1}  &  =\arg\min_{\theta\in\Theta}Q_{1T}(\theta,\overline
{X}),\\
(\widehat{\theta}_{2},\widehat{\mu})  &  =\arg\min_{\theta\in\Theta,\mu\in%
\mathbb{R}
}Q_{1T}(\theta,\mu),\\
\widehat{\theta}_{3}  &  =\arg\min_{\theta\in\Theta}\left(  \frac{1}{2\pi}%
\int_{-\pi}^{\pi}\log f_{X}(\lambda;\theta)\mathsf{d}\lambda+\frac{1}{2\pi
}\int_{-\pi}^{\pi}\frac{I_{X^{dem}}(\lambda)}{f_{X}(\lambda;\theta)}%
\mathsf{d}\lambda\right)  ,\\
\widehat{\theta}_{4}  &  =\arg\min_{\theta\in\Theta}\left(  \frac{1}{T}%
\sum_{j=1}^{T-1}\log f_{X}(\lambda_{j};\theta)+\frac{1}{T}\sum_{j=1}%
^{T-1}\frac{I_{X}(\lambda_{j})}{f_{X}(\lambda_{j};\theta)}\right)  .
\end{align*}
Here, $\widehat{\theta}_{1}$ is a simplification over $\widehat{\theta}_{2}$,
while $\widehat{\theta}_{3}$ and $\widehat{\theta}_{4}$ optimize an
approximation to the log-likelihood function called Whittle's (1953)
approximation (in continuous and discrete versions). From a computational
point of view, the frequency domain estimators $\widehat{\theta}_{3}$ and
$\widehat{\theta}_{4}$ are simpler because there is no need to calculate
$\mathbf{\Sigma}^{-1}(\theta)$, which is complicated; see Sowell (1992) for a
discussion of exact maximum likelihood estimation of a stationary Gaussian
zero-mean ARFIMA process (so, using (\ref{a24}) with $\mu=0$), focusing on
computational issues. Also, between $\widehat{\theta}_{3}$ and
$\widehat{\theta}_{4}$, the latter seems preferable, because it is simpler
computationally (taking advantage of the fast Fourier transform), and because
it is invariant to a non-zero mean since, by the properties of the complex
exponential, $I_{X^{dem}}(\lambda_{j})=I_{X}(\lambda_{j})$ for $j=1,\dots,T-1$.

These estimators have been well studied in the time series literature; for a
comparison, see Robinson (1994a). The earliest works showed that, under
regularity conditions (requiring at least that $f(\lambda;\theta_{0})$ is
continuous in $\lambda$, where subscript zero denotes true value), all
estimators are consistent, asymptotically normal, and asymptotically efficient
(e.g.\ Hannan, 1973). Obviously, the continuity condition rules out
fractionally integrated $X_{t}$ with $d_{0}>0$. Subsequently, Fox and Taqqu
(1986) analyzed $\widehat{\theta}_{3}$ for a possibly long-range dependent
stationary Gaussian sequence $X_{t}$ and for a special parametrization of
$f_{X}(\lambda;\theta)=\sigma^{2}g_{X}(\lambda;\theta)$, where $\int_{-\pi
}^{\pi}\log g_{X}(\lambda;\theta)\mathsf{d}\lambda=0$ (so that $X_{t}/\sigma$
has one-step prediction error independent of $\theta$). They showed
consistency and asymptotic normality with rate of convergence $\sqrt{T}$
(although this was not justified for the estimator of $\sigma^{2}$). Dahlhaus
(1989) improved these results, justifying the consistency, asymptotic
normality, and asymptotic efficiency of $\widehat{\theta}_{1}$,
$\widehat{\theta}_{2}$, $\widehat{\theta}_{3}$, and suggesting it also for
$\widehat{\theta}_{4}$. Nicely, his results hold without the need of the
special parametrization imposed by Fox and Taqqu (1986). Giraitis and
Surgailis (1990) and Hosoya (1996), among others, obtained analogous results
without the need of the Gaussianity assumption.

For the nonstationary case, Velasco and Robinson (2000) proposed a tapered
version of $\widehat{\theta}_{4}$, obtaining $\sqrt{T}$-consistency and
asymptotic normality, but with an inflated asymptotic variance (due to
tapering). With the exception of this work, the previous ones consider only (a
subset of) the stationary region, which is clearly too restrictive for most
practical purposes.

An estimation method that overcomes this problem and is valid for any value of
$d$ is the conditional (or truncated) sum-of-squares method. We exemplify this
estimator for a zero-mean Type~II fractionally integrated process. Let the
observable $X_{t}$ be given by
\begin{align}
X_{t}  &  =\Delta_{+}^{-d}u_{t},\label{a26}\\
u_{t}  &  =\theta(L;\varphi)\varepsilon_{t},\qquad\theta(s;\varphi)=\sum
_{j=0}^{\infty}\theta_{j}(\varphi)s^{j}, \label{a27}%
\end{align}
where $\varepsilon_{t}$ is a zero-mean white noise with $\operatorname{Var}%
(\varepsilon_{t})=\sigma^{2}$ and $d$ is a real number lying on a closed
interval $[\bigtriangledown_{1},\bigtriangledown_{2}]$ with $\bigtriangledown
_{1}<\bigtriangledown_{2}$. Here, $\theta(s;\varphi)$ covers for example ARMA
processes or the exponential-spectrum model (Bloomfield, 1973), so $\varphi$
is in general a $p\times1$ dimensional vector of parametric short memory
parameters. The aim is to estimate $\tau_{0}=(d_{0},\varphi_{0}^{\prime
})^{\prime}$ from observables $X_{t}$, $t=1,\dots,T$. For $\tau=(d,\varphi
^{\prime})^{\prime}$, define the residual,%
\[
\varepsilon_{t}(\tau)=\Delta^{d}\theta^{-1}(L;\varphi)X_{t},
\]
and the conditional sum-of-squares estimator,%
\begin{equation}
\widetilde{\tau}=\arg\underset{\tau\in\mathcal{T}}{\min}\frac{1}{T}\sum
_{t=1}^{T}\varepsilon_{t}^{2}(\tau), \label{a25}%
\end{equation}
where $\mathcal{T=[}\bigtriangledown_{1},\bigtriangledown_{2}]\times\Psi$ and
$\Psi$ is a compact subset of $\mathbb{R}^{p}$. In (\ref{a25}) it is critical
that $X_{t}=0$ for $t\leq0$, although this could be relaxed to allow for a
finite number of non-zero (but bounded) initial values as in Johansen and
Nielsen (2010, 2012a, 2016). The loss function in (\ref{a25}) corresponds to
the conditional log-likelihood function (concentrated with respect to
$\sigma^{2}$), which has been applied for example by Box and Jenkins (1970)
for the case where $d$ is a known integer.

The conditional sum-of-squares estimator enjoys very attractive features. It
has has the same limit distribution as that derived by Fox and Taqqu (1986)
and Dahlhaus (1989) for the parametric Whittle estimator, but without assuming
Gaussianity, and it is asymptotically efficient under Gaussianity. Also, it is
computationally very simple because it does not require inversion of a large
$T\times T$ matrix. However, the main advantage over other parametric methods
is that these results hold for any value of $d$, as long as it lies on an
arbitrarily large compact interval, so that both the nonstationary and
noninvertible range is covered by the theory. The conditional sum-of-squares
estimator has been intensely studied by the literature. Li and McLeod (1986)
proposed this method for stationary ARFIMA models with $0<d<1/2$ and Robinson
(2006) analyzed it for this range of $d$. The first analysis of the estimator
in nonstationary situations was Beran (1995), who applied it to a potentially
nonstationary ARFIMA model, although the proof of consistency, which is a
necessary preliminary step for establishing the limit distribution, is not
rigorous due to a circular argument. Tanaka (1999) and Nielsen (2004a) gave
local consistency proofs, while Hualde and Robinson (2011) and Nielsen (2015)
provided rigorous global consistency proofs, so that asymptotic results for
$\widetilde{\tau}$ were finally formally established. These results were
extended by Cavaliere, Nielsen, and Taylor (2015, 2017, 2020) to cover general
forms of conditional and unconditional heteroskedasticity. The technical
difficulty with the consistency proof arises when $[\bigtriangledown
_{1},\bigtriangledown_{2}]$ is of length greater than 1/2 because the
behaviour of the loss function is completely different depending on whether
$d_{0}-d<1/2$ (where $\varepsilon_{t}(\tau)$ is asymptotically stationary) or
$d_{0}-d>1/2$ (where $\varepsilon_{t}(\tau)$ is purely nonstationary), and the
case where $d_{0}-d$ is close to 1/2 requires a very specific and detailed treatment.

From a theoretical perspective, the conditional sum-of-squares estimator
appears to be preferable. However, it can exhibit noticeable finite-sample
bias as documented by Nielsen and Frederiksen (2005). For example, this occurs
for the ARFIMA(1,$d$,0) model with a positive autoregressive parameter, which
is an issue that is also present in the semiparametric setting. Nielsen and
Frederiksen (2005) showed that the exact maximum likelihood estimator (MLE)
suffers from the same problem, but also that it is somewhat alleviated for the
Whittle estimator, although the latter covers only the stationary region.

The second limitation is dealing with deterministic components. This is a very
relevant issue because empirical evidence suggests that observed economic time
series are formed by both stochastic and deterministic components (see, e.g.,
Stock and Watson, 1988, Johansen and Juselius, 1990, Johansen, 1995). In fact,
most of the previous parametric approaches allowed for an unknown and possibly
non-zero mean, but this issue is complicated to tackle with a conditional
sum-of-squares approach when allowing also for nonstationarity. In fact, there
is no simple solution to this problem and it can be shown that apparently
sensible simple strategies like, for example, \textquotedblleft differencing
and adding back\textquotedblright\ to eliminate a drift and then estimate
$d-1$ from $\Delta X_{t}$ do not work for all values of $d$; see Hualde and
Nielsen (2021). Given (\ref{a26}) and (\ref{a27}), Hualde and Nielsen (2020)
analyze the model%
\begin{equation}
Z_{t}=\mu t_{+}^{\gamma}+X_{t}, \label{c alt}%
\end{equation}
where $t_{+}^{\gamma}=t^{\gamma}\mathbb{I}(t\geq1)$. This is related to models
of Robinson (2005b) and Robinson and Iacone (2005), although these authors
included more deterministic terms and assumed that power law parameters (like
$\gamma$) were known. Hualde and Nielsen (2020) considered conditional
sum-of-squares estimation of the parameters in (\ref{c alt}), that is
$(\mu,\gamma,d,\varphi,\sigma^{2})$, where $d$ and $\gamma$ were allowed to
belong to arbitrarily large compact sets. Their main results were that
parameter estimates related to the deterministic component are consistent and
asymptotically normal only for parts of the parameter space depending on the
relative strength of the stochastic and deterministic components, whereas
consistency, asymptotic normality, and efficiency of parameter estimates
related to the stochastic component was established for the entire parameter space.

\subsection{Testing the unit root hypothesis against fractional alternatives}

\label{sec:lm}

Testing the unit root (or $I(1)$) hypothesis has attracted great theoretical
and empirical interest. In this context, there are several different
approaches. The most common approach considers Dickey-Fuller or
Phillips-Perron-type tests, which, within an autoregressive framework, test
whether there is a unit root in the autoregressive polynomial. These tests are
designed against stationary (specifically, $I(0)$) alternatives, and have been
shown to have poor power against fractionally integrated alternatives (Diebold
and Rudebusch, 1991a). The fractionally integrated model allows for a
different strategy; that is, to test $H_{0}:d=1$ versus the alternative
$H_{1}:d\neq1$ (of course, this alternative could be one-sided). The
previously established theory in semiparametric or parametric contexts can be
used to derive Wald, likelihood ratio, or Lagrange multiplier tests with
standard asymptotic null distributions.

The main caveat here must be to ensure adequate model fit. That is, to ensure
that the estimated model fits the data prior to conducting the statistical
test, or in other words to test whether the assumptions underlying the
asymptotic theory are likely to be satisfied for the case at hand. This is
usually done by analysis of the model residuals by testing, for example,
whether these are serially uncorrelated, (conditionally) homoskedastic, etc.
This can be challenging with a purely LM-based approach, e.g.\ Robinson
(1994b), where the model has not been estimated and no residuals are available
upon which to base such model diagnostics. However, with a Wald or likelihood
ratio approach, conducting such standard model diagnostics prior to testing
the hypothesis of interest should be straightforward.

\subsection{Dickey-Fuller-type tests with fractionally integrated errors}

\label{sec:DF}

The first important contribution here is Sowell (1990), who analyzed the
asymptotic behaviour of the OLS estimator of $\phi$, say $\widehat{\phi}$, in
the model%
\[
X_{t}=\phi X_{t-1}+\Delta^{-d}u_{t},
\]
where $d$ is a known value such that $-1/2<d<1/2$, $u_{t}\in I\left(
0\right)  $, and $\phi=1$ so there is a unit root in the autoregressive
polynomial. Specifically, Sowell (1990) proved that, under mild regularity
conditions,%
\begin{equation}
T^{2d\mathbb{I}(d<0)+1}(\widehat{\phi}-1)\rightarrow_{d}A(d), \label{a28}%
\end{equation}
where $A\left(  d\right)  $ is a random variable which depends on $d$
($A\left(  0\right)  $ is the well known Dickey-Fuller distribution). Result
(\ref{a28}) is the basis to analyze the limiting distribution of the
Dickey-Fuller $t$-statistic. It is noticeable that in (\ref{a28}) the
convergence rate (and, in fact, the form of $A(d)$) depends heavily on whether
$d<0$ or $d\geq0$. In the former case, the rate can be very slow if $d$ is
close to $-1/2$. In the latter case, the usual $T$-consistency is obtained but
the form of $A(d)$ changes substantially depending on whether $d>0$ or $d=0$.

Instead of considering the OLS estimator of $\phi$, Ling and Li (2001)
considered the Gaussian MLE, which (for i.i.d.$\ u_{t}$) is the OLS estimator
of $\Delta^{d}X_{t}$ on $\Delta^{d-1}X_{t-1}$. This estimator has the usual
Dickey-Fuller distribution regardless of the value of $d$, as is seen by
(equivalently) considering $Y_{t}=\Delta^{d}X_{t}$ instead of $X_{t}$.
Relatedly, Chan and Terrin (1995) developed asymptotic theory for the OLS
estimator in an autoregressive process with fractionally integrated innovations.

In a similar fashion, Dolado, Gonzalo, and Mayoral (2002) introduced the model%
\begin{equation}
\alpha(L)\Delta X_{t}=\phi\Delta^{d_{1}}X_{t-1}+\varepsilon_{t},
\label{DGM model}%
\end{equation}
where $\varepsilon_{t}$ is white noise, $\alpha(z)$ is a polynomial with roots
outside the unit circle, and $d_{1}$ is fixed or pre-estimated. They proposed
to test $H_{0}:\phi=0$, in which case $X_{t}\in I(1)$, against $H_{1}:\phi<0$,
in which case it is claimed that $X_{t}\in I(d_{1})$. Indeed, $X_{t}\in I(0)$
if $d_{1}=0$, this being the well-known result from the standard Dickey-Fuller
approach. However, in general, determining the integration order of $X_{t}$
under $H_{1}$ would require conditions on the parameters under which all roots
of the polynomial $\pi(z)=\alpha\left(  z\right)  \left(  1-z\right)
^{1-d_{1}}-\phi z$ are outside the unit circle. The conditions given by
Dolado, Gonzalo, and Mayoral (2002) are insufficient, as shown by the
counter-example in footnote~3 of Johansen and Nielsen (2010). Indeed, general
conditions appear impossible to derive (see Johansen, 2008), so the
integration order of $X_{t}$ under the alternative is unknown. In this sense
the model (\ref{DGM model}) is ill-posed. Similar issues arise in a related
model that was proposed and analyzed by Lobato and Velasco (2006).

Finally, Johansen and Nielsen (2010) introduced a fractional autoregressive
model where a unit root test can be implemented. The idea behind their model
is the following. Consider the usual autoregressive model in error correction
mechanism form%
\begin{equation}
\Delta Y_{t}=\pi Y_{t-1}+\sum_{i=1}^{k}\phi_{i}\Delta Y_{t-i}+\varepsilon_{t},
\label{a29}%
\end{equation}
where $\varepsilon_{t}$ is an i.i.d.\ sequence with $E(\varepsilon_{t})=0$ and
$\operatorname{Var}(\varepsilon_{t})=\sigma^{2}$. Replace in (\ref{a29}) the
usual difference and lag operators, $\Delta$ and $L=1-\Delta$, by the
corresponding fractional operators $\Delta^{b}$ and $L_{b}=1-\Delta^{b}$,
respectively, obtaining%
\begin{equation}
\Delta^{b}Y_{t}=\pi L_{b}Y_{t}+\sum_{i=1}^{k}\phi_{i}\Delta^{b}L_{b}^{i}%
Y_{t}+\varepsilon_{t}. \label{a30}%
\end{equation}
Defining $X_{t}=\Delta^{b-d}Y_{t}$, it is immediate to obtain%
\begin{equation}
\Delta^{d}X_{t}=\pi L_{b}\Delta^{d-b}X_{t}+\sum_{i=1}^{k}\phi_{i}L_{b}%
^{i}\Delta^{d}X_{t}+\varepsilon_{t}, \label{a31}%
\end{equation}
which is the fractional autoregressive model employed by Johansen and Nielsen (2010).

Model (\ref{a31}) has interesting properties that are very similar in essence
to those of the standard autoregressive process (\ref{a29}). Note that
(\ref{a31}) can be expressed as%
\[
\zeta(L_{b})\Delta^{d-b}X_{t}=\varepsilon_{t},
\]
where%
\begin{equation}
\zeta(z)=1-z-\pi z-\sum_{i=1}^{k}\phi_{i}z^{i}(1-z), \label{a32}%
\end{equation}
so $\zeta(z)$ is the usual $(k+1)$-order autoregressive polynomial from
(\ref{a29}). Thus, by performing the change of variables $z=1-(1-v)^{b}$, the
properties of the process (\ref{a31}) depend on the roots of $\zeta(z)$. The
main representation result for this model is that if all roots of $\zeta(z)$
are outside a modified unit circle (the modification having to do with the
change of variables), then $X_{t}\in I(d-b)$. Alternatively, if $\pi=0$ and
all remaining roots are outside the modified unit circle, then $X_{t}\in
I(d)$. Consequently, this model can be the basis for a unit root test, where
$H_{0}:\pi=0$, in which case $X_{t}\in I(d)$, versus $H_{1}:\pi\neq0$, in
which case $X_{t}\in I(d-b)$. This generalizes the usual unit root tests,
which are obtained in the special case $d=b=1$. As in Sowell (1990) or Ling
and Li (2001), $d$ is not restricted to be~1, but unlike earlier papers, the
gap $b$ between integration orders under $H_{0}$ and $H_{1}$ is also not
restricted to be~1.

Johansen and Nielsen (2010, 2012a) proved consistency of the MLE of
$(d,b,\pi,\phi_{1},\dots,\phi_{k})$. They also showed that, when $b_{0}<1/2$,
the estimators are jointly asymptotically Gaussian, and when $b_{0}>1/2$, the
estimator of $(d,b,\phi_{1},\dots,\phi_{k})$ are asymptotically Gaussian
whereas that of $\pi$ is a functional of Type~II fractional Brownian motion
depending on $b_{0}$ and is asymptotically independent of the other estimator.
Finally, they derived the asymptotic distribution of the likelihood ratio
(unit root) test for $\pi=0$, which is another functional of Type~II
fractional Brownian motion that depends only the fractional gap $b_{0}$, and
not $d_{0}$. Thus, this distribution is not pivotal, but a plug-in approach
can be used to derive critical values using the estimator of $b_{0}$ (see
MacKinnon and Nielsen, 2014).

\section{Fractional cointegration}

\label{sec:cointegration}

Once the concept of fractional integration has been introduced, that of
fractional cointegration is a natural step forward. In fact, the definition of
cointegration given in Definition~\ref{def cointegration} can be directly
applied to a fractional situation. In the fractional context, however, there
are many more possibilities of cointegration allowed under
Definitions~\ref{def vector zero} and~\ref{def cointegration} compared with
the $I(1)/I(0)$ paradigm, especially when the observables have different
orders of fractional integration.

For example, denoting by $d_{\max}$ the highest integration order of the
observables, Definitions~\ref{def vector zero} and~\ref{def cointegration}
imply that there is cointegration whenever a linear combination of the
observables has integration order smaller than $d_{\max}$. Thus, if the $i$'th
variable has fractional integration order smaller than $d_{\max}$, then
cointegration arises trivially with the unit vector, $e_{i}$, as a
cointegration vector. Consequently, Fl\^{o}res and Szafarz (1996) strengthened
slightly this idea, requiring also that the linear combination must involve
nontrivially an $I(d_{\max})$ observable. An alternative (and much stronger)
definition is given by Robinson and Marinucci (2003), which required the
linear combination of the observables to have an integration order smaller
that $d_{\min}$, where $d_{\min}$ is the smallest integration order of the
observables. Robinson and Yajima (2002) grouped the vector of observables into
subvectors of variables with the same integration order, and define
cointegration whenever there exists at least one of these subvectors that is
cointegrated in Engle and Granger's (1987) sense. They also exemplified the
differences among the various definitions. In a bivariate situation these
definitions are equivalent. For any of these alternative definitions of
cointegration, a crucial concept is the cointegrating rank; that is, the
number of linearly independent cointegrating vectors. The space spanned by
these cointegrating vectors is the cointegrating space.

Based on the different definitions of (fractional) cointegration, many
different models displaying cointegration have been proposed. To demonstrate
these ideas, we now discuss a rather general model. Let $u_{t}$, $t\in%
\mathbb{Z}
$, be a $p$-dimensional covariance-stationary unobservable process with zero
mean and spectral density, $f_{u}(\lambda)$, given by%
\begin{equation}
E(u_{0}u_{j}^{\prime})=\int_{-\pi}^{\pi}e^{ij\lambda}f_{u}(\lambda
)\mathsf{d}\lambda, \label{1.5}%
\end{equation}
which is assumed to be nonsingular and continuous at all frequencies,
c.f.\ (\ref{def zero}). As will be clarified below, the treatment of
$f_{u}(\lambda)$ is what distinguishes semiparametric and parametric inference
methods. In the former, $f_{u}\left(  \lambda\right)  $ is considered a
nonparametric function, whereas in the latter, $f_{u}(\lambda)$ is assumed
known up to a finite-dimensional vector of unknown parameters. Also, for real
numbers $d_{i}$, $i=1,\dots,p$, such that%
\begin{equation}
d_{1}\leq d_{2}\leq\ldots\leq d_{p}, \label{zz1}%
\end{equation}
and a $p\times p$ nonsingular matrix $\Upsilon$, we define the $p$-dimensional
vector of observables $X_{t}$, $t\in%
\mathbb{Z}
$, as%
\begin{equation}
\Upsilon X_{t}=\operatorname*{diag}\left(  \Delta_{+}^{-d_{1}},\Delta
_{+}^{-d_{2}},\ldots,\Delta_{+}^{-d_{p}}\right)  u_{t}, \label{zz2}%
\end{equation}
noting that there is no loss of generality in (\ref{zz1}). The cointegrating
properties in the general system (\ref{zz2}) depend on $\Upsilon$ and the
possible presence of strict inequalities in (\ref{zz1}). For example, if all
elements in the last column of $\Upsilon^{-1}$ are non-zero, then all
individual components of $X_{t}$ are $I(d_{p})$. This is standard in the
traditional cointegrating setting where $d_{p}=1$. If, for some $1\leq r<p$,
we set%
\[
d_{r}<d_{r+1}=\ldots=d_{p},
\]
then (\ref{zz2}) is a system with cointegrating rank $r$, where the first $r$
rows of $\Upsilon$ are the cointegrating vectors. More general cointegrating
possibilities can be captured by (\ref{zz2}), and a general specification for
$\Upsilon$ and (\ref{zz1}) which allows for the possibility of multiple
cointegrating subspaces is given in Hualde and Robinson (2010). These
subspaces appear naturally in a fractional setting and are characterized by
particular directions in the cointegrating space that lead to larger
reductions in the integration orders.

To illustrate some interesting characteristics of a cointegrated system, we
next discuss a special case of (\ref{zz1})--(\ref{zz2}). For $r<p$ we set%
\begin{equation}
\Upsilon=\left(
\begin{array}
[c]{cc}%
I_{r} & -\beta^{\prime}\\
0_{p-r,r} & I_{p-r}%
\end{array}
\right)  \label{zz3}%
\end{equation}
and%
\begin{equation}
d_{1}=\ldots=d_{r}<d_{r+1}=\ldots=d_{p}, \label{zz4}%
\end{equation}
where $I_{s}$ is the $s$-dimensional identity matrix, $0_{i,j}$ is a $i\times
j$ matrix of zeros and $\beta$ is a $\left(  p-r\right)  \times r$
unrestricted matrix. This particular case has been heavily stressed in theory
and practice and, as explained in Hualde and Robinson (2010), identification
restrictions such as those imposed in (\ref{zz3})--(\ref{zz4}) are always
possible whenever the cointegrating rank among the observables is $r$ and
there is only one cointegrating subspace. However, given this situation, the
ordering of the variables in $X_{t}$ is not innocuous, although this ordering
can be inferred from data (Hualde, 2008). Letting $X_{t}=(X_{1t}^{\prime
},X_{2t}^{\prime})^{\prime}$ and $u_{t}=(u_{1t}^{\prime},u_{2t}^{\prime
})^{\prime}$, where $X_{1t}$ and $u_{1t}$ are $r\times1$ vectors while
$X_{2t}$ and $u_{2t}$ are $\left(  p-r\right)  \times1$ vectors,
(\ref{zz2})--(\ref{zz4}) represent a fractional extension of Phillips' (1991a)
triangular system,%
\begin{align}
X_{1t}  &  =\beta^{\prime}X_{2t}+\Delta_{+}^{-\left(  d-b\right)  }%
u_{1t},\label{zz5}\\
X_{2t}  &  =\Delta_{+}^{-d}u_{2t}, \label{zz6}%
\end{align}
where for some $b>0$ we use the simplifying notation $d-b=d_{1}$ and
$d=d_{r+1}$. In (\ref{zz5})--(\ref{zz6}) all individual components of $X_{2t}$
are $I(d)$, whereas all cointegrating errors are $I(d-b)$. The individual
components of $X_{1t}$ are in general also $I(d)$, although they could be
$I(d-b)$ if the corresponding row of $\beta^{\prime}$ only contains zeros. In
the latter case, such a cointegrating relation will be denoted as trivial and
just indicates that a particular observable has an integration order smaller
than $d$, so that there exists a unit cointegrating vector. Note also that,
apart from the restriction in (\ref{zz4}), both integration orders are
unrestricted, so many different cointegrating possibilities are allowed by
(\ref{zz5})--(\ref{zz6}), including equilibrium relations among nonstationary
observables that are not $I\left(  1\right)  $ or even among stationary ones
(the so-called stationary cointegration). It also permits a slower convergence
to equilibrium due to long memory cointegrating errors. As denoted by Phillips
and Loretan (1991), (\ref{zz5})--(\ref{zz6}) with $d=b=1$, represents
\textquotedblleft a typical cointegrated system\textquotedblright\ in
structural form. In particular, (\ref{zz5}) could be regarded as a stochastic
version of the partial equilibrium relationships $X_{1t}-\beta^{\prime}X_{2t}%
$, with $\Delta_{+}^{-\left(  d-b\right)  }u_{1t}$ representing deviations
from this equilibrium, whereas (\ref{zz6}) is a reduced form equation.

The Type~II nature of (\ref{zz5})--(\ref{zz6}) accommodates integration orders
in both stationary or nonstationary ranges, but it implies that, irrespective
of the values taken by $d$ and $b$, $X_{t}$ is nonstationary as is the
differenced process $\Delta_{+}^{d}X_{t}$ (although it is asymptotically
stationary). It is interesting to describe some of the properties of the
stationary version of $\Delta_{+}^{d}X_{t}$, that is $x_{t}=(u_{2t}^{\prime
}\beta+\Delta^{b}u_{1t}^{\prime},u_{2t}^{\prime})^{\prime}$, because they
illustrate some key features of cointegrated models that form the basis of
many inferential procedures. Noting (\ref{1.5}), the spectral density of
$x_{t}$ is
\[
f_{x}(\lambda)=\left(
\begin{array}
[c]{cc}%
(1-e^{i\lambda})^{b}I_{r} & \beta^{\prime}\\
0_{p-r,r} & I_{p-r}%
\end{array}
\right)  f_{u}(\lambda)\left(
\begin{array}
[c]{cc}%
(1-e^{-i\lambda})^{b}I_{r} & 0_{r,p-r}\\
\beta & I_{p-r}%
\end{array}
\right)  .
\]
Using the approximations $f_{u}(\lambda)=(2\pi)^{-1}\Omega(1+O(\lambda^{2}))$
and $|1-e^{i\lambda}|^{2b}=2^{b}(1-\cos\lambda)^{b}=O(\lambda^{2b})$ as
$\lambda\rightarrow0$, it holds that%
\begin{equation}
f_{x}(\lambda)=\left(
\begin{array}
[c]{cc}%
\beta^{\prime}\Omega_{22}\beta & \beta^{\prime}\Omega_{22}\\
\Omega_{22}\beta & \Omega_{22}%
\end{array}
\right)  (1+O(\lambda^{b})+O(\lambda^{2}))\text{ as }\lambda\rightarrow0,
\label{a41}%
\end{equation}
where $\Omega_{22}$ is the $\left(  p-r\right)  \times\left(  p-r\right)  $
lower right block of $2\pi f_{u}\left(  0\right)  $, i.e.\ the so-called
long-run variance of $u_{2t}$. For $d<1/2$, (\ref{a41}) motivates a
multivariate version of the local approximation (\ref{a40}) given by
\begin{equation}
f_{x}(\lambda)\sim G\lambda^{-2d}\text{ as }\lambda\rightarrow0, \label{a43}%
\end{equation}
where \textquotedblleft$\sim$\textquotedblright\ is taken elementwise for real
and imaginary parts separately and%
\begin{equation}
G=\left(
\begin{array}
[c]{cc}%
\beta^{\prime}\Omega_{22}\beta & \beta^{\prime}\Omega_{22}\\
\Omega_{22}\beta & \Omega_{22}%
\end{array}
\right)  =(\beta,I_{p-r})^{\prime}\Omega_{22}(\beta,I_{p-r}) \label{G matrix}%
\end{equation}
is a $p\times p$ matrix with rank $p-r$. Additionally, it can be shown that,
as $\lambda\rightarrow0$, the $r$ smallest eigenvalues of $f_{x}(\lambda)$ are
$O(\lambda^{b})$, so they approach zero, whereas the $p-r$ largest eigenvalues
are bounded away from zero in the limit; see also Velasco (2003a) and Nielsen
(2004b) for further details.

In the following subsections we will present some testing and estimation
procedures for fractional cointegration. These methods can be classified in
several different ways. One possibility is to distinguish between
semiparametric and parametric methods, depending on whether the parametric
structure of the $I(0)$ error input process that generates the observables and
cointegrating errors is known or unknown. As previously done in
Section~\ref{sec:inference}, we will focus on this method classification, but
within each category we also distinguish between what Jeganathan (1997)
denotes as first and second stage procedures.

According to Jeganathan (1997), limiting distributions of first stage
procedures are nonstandard and unsuitable for use in statistical inference,
whereas second stage procedures imply estimators belonging to the locally
asymptotically mixed normal family. This class of estimators enjoy several
attractive features. They are symmetrically distributed, median unbiased, and
an optimal theory of inference applies under Gaussian assumptions (Phillips,
1991a, and Saikkonen, 1991). Also, they lead to Wald and likelihood ratio test
statistics with standard $\chi^{2}$ null limit distributions. In the context
of standard cointegration with unit root observables and $I(0)$ cointegrating
errors, Jeganathan (1997) suggested that first stage procedures could be used
to test for the presence of unit roots in a given model, and then, by second
stage methods one could estimate cointegrating relationships on the model
where the unit roots tested in the first stage are imposed. Thus, second stage
methods incorporate in the estimation this type of information, allowing
implementation of endogeneity corrections that leads to the desirable
asymptotic properties. As a practical consequence, traditionally, the main
difference between the two types of procedures is that first stage methods do
not require knowledge of the integration orders involved, whereas second stage
methods do (see, e.g., Phillips, 1991a).

In the context of fractional time series models, assuming knowledge of the
integration orders involved is very unrealistic in general, even after
pretesting. As will be seen below, we will present estimation and inference
methods which share in many cases the optimal asymptotic properties of the
second stage procedures without assuming knowledge of the integration orders.

\subsection{Tests for fractional cointegration (rank)}

\label{sec:CI-tests}

As mentioned before, a fair number of tests have been proposed, including
parametric, semiparametric, and nonparametric approaches. We will illustrate
with some detail one method belonging to each category.

First, in a parametric setting, Breitung and Hassler (2002) generalized the
univariate score test against fractional alternatives (see, e.g., Robinson,
1991, 1994b, Agiakloglou and Newbold, 1994, Tanaka, 1999) to a score-type test
for fractional cointegration; see also the score test for cointegration in
Nielsen\ (2004c) and the multivariate score tests in Nielsen (2004d, 2005a).
Let $Y_{t}$ be a $p$-dimensional vector of observables and suppose that, for
$0\leq r<p$, there exists a full rank orthonormal $p\times p$ matrix
$R=(R_{p-r},R_{r})$, where $R_{p-r}$ and $R_{r}$ have $p-r$ and $r$ columns,
respectively, such that%
\begin{equation}
R^{\prime}X_{t}=\left(
\begin{array}
[c]{cc}%
\Delta_{+}^{-d}I_{p-r} & 0\\
0 & \Delta_{+}^{-\left(  d-b\right)  }I_{r}%
\end{array}
\right)  U_{t} \label{a35}%
\end{equation}
for some $d>1/2$ and $b>0$ and a $p$-dimensional process $U_{t}\in I(0)$. If
$r=0$ then $R=R_{p}$ and (\ref{a35}) becomes $R^{\prime}X_{t}=\Delta_{+}%
^{-d}U_{t}$. Under (\ref{a35}), $X_{t}$ is clearly cointegrated with rank~$r$.
We present Breitung and Hassler's (2002) proposal for the white noise case,
where the $p$-dimensional vector $(\Delta_{+}^{d}X_{t}^{\prime}R_{p-r}%
,\Delta_{+}^{d-b}X_{t}^{\prime}R_{r})^{\prime}$ (or $\Delta_{+}^{d}X_{t}$ if
$r=0$) is a zero mean i.i.d.\ sequence with finite variance. Inspired by
Johansen (1995), the test on cointegration rank is based on the eigenvalues
derived as the solutions to%
\begin{equation}
\left\vert \lambda\widehat{\Sigma}-S_{10}^{\prime}S_{11}^{-1}S_{10}\right\vert
=0, \label{a39}%
\end{equation}
where $\widehat{\Sigma}=T^{-1}\sum_{t=1}^{T}\Delta_{+}^{d}X_{t}\Delta_{+}%
^{d}X_{t}^{\prime}$, $S_{11}=T^{-1}\sum_{t=1}^{T}\sum_{j=1}^{t-1}j^{-1}%
\Delta_{+}^{d}X_{t-j}\sum_{k=1}^{t-1}k^{-1}\Delta_{+}^{d}X_{t-k}^{\prime}$,
and $S_{10}=T^{-1}\sum_{t=1}^{T}\sum_{j=1}^{t-1}j^{-1}\Delta_{+}^{d}%
X_{t-j}\Delta_{+}^{d}X_{t}^{\prime}$. Here, the partial sum in Johansen's
setting (where $d=1$), $X_{t-1}=\sum_{j=1}^{t-1}\Delta_{+}X_{t-j}$, is
replaced by the weighted sum $\sum_{j=1}^{t-1}j^{-1}\Delta_{+}^{d}X_{t-j}$.
The latter originates from evaluation of the derivative of $\Delta_{+}%
^{d}X_{t}$ with respect to $d$ at $d=0$ (e.g.\ Tanaka, 1999), noting that for
$j\geq1$, $\left.  \partial\pi_{j}\left(  d\right)  /\partial d\right\vert
_{d=0}=j^{-1}$; see (\ref{a4}). Ordering the eigenvalues derived from
(\ref{a39}) as $\widehat{\lambda}_{1}\leq\widehat{\lambda}_{2}\leq\ldots
\leq\widehat{\lambda}_{p}$, the statistic to test $H_{0}:r=r_{0}$ proposed by
Breitung and Hassler (2002) is%
\[
\Lambda_{r_{0}}\left(  d\right)  =\sum_{j=1}^{p-r_{0}}\widehat{\lambda}_{j},
\]
which is asymptotically $\chi^{2}$ distributed with $\left(  p-r_{0}\right)
^{2}$ degrees of freedom under $H_{0}$. Additionally, under the alternative
$H_{1}:r>r_{0}$, the authors claim that $\Lambda_{r_{0}}\left(  d\right)  $
diverges to $\infty$ at rate $T$. The test can be applied sequentially to
estimate $r$ by considering $r_{0}=0,1,\ldots$, and denoting by $\widehat{r}$
the first non-rejected null value. Short-run dynamics and/or deterministic
terms can be accommodated by replacing observables by appropriate residuals.
Neatly, the results do not depend on the cointegrating gap $b$ (although $b$
presumably affects power), and in fact a unique cointegrating gap $b$ is not
even required. Importantly, however, the procedure relies on knowledge of $d$
and this is a serious issue because replacing this parameter by an estimator
affects the asymptotic null distribution. Equally importantly, there appears
to be no way to verify whether the assumptions underlying the asymptotic
distribution results are likely satisfied (i.e., to ensure that the lag
structure is sufficiently rich); c.f.\ the discussion in Section~\ref{sec:lm}.

Also in a parametric framework, in Subsection~\ref{sec:fcvar} below we will
present a more detailed discussion of the fractionally cointegrated VAR model,
which generalizes the cointegrated VAR model of Johansen (1995) to the
fractional context. Among other things, this of course includes testing for
the cointegrating rank.

In a semiparametric setting, Robinson and Yajima (2002) introduced a procedure
to determine the cointegrating rank in a fractionally cointegrated system.
Their model is quite general, allowing the components of the $p$-dimensional
vector of observables $X_{t}$ to have distinct integration orders, although
belonging to the stationary and invertible region. Then, cointegration is
inferred within blocks of variables sharing the same order. For this reason,
without loss of generality, we exemplify their approach for the case where all
observables have the same integration order, $d$, with $|d|<1/2$. We assume
that the spectral density matrix of $X_{t}$ behaves locally as in (\ref{a43})
but for a generic matrix $G$ which is finite, non-negative definite, and with
non-zero diagonal elements. Note that $G$ is positive definite if and only if
$X_{t}$ is not cointegrated, see (\ref{G matrix}). The cointegrating rank is
$r=p-\operatorname{rank}\left(  G\right)  $ for $0\leq r<p$, which is the
basis of the procedure. Under regularity conditions, with an appropriate
bandwidth $m$ satisfying at least $m/T\rightarrow0$ and a particular estimator
$\widehat{d}$ of $d$, a consistent estimator of $G$ is%
\[
\widehat{G}(\widehat{d})=\frac{1}{m}%
{\displaystyle\sum\limits_{j=1}^{m}}
\lambda_{j}^{2\widehat{d}}\operatorname{Re}(I_{X}(\lambda_{j})).
\]
Assuming the $p-r$ nonzero eigenvalues of $G$ are distinct, let $\lambda_{i}$
(resp.$\ \widehat{\lambda}_{i}$) be the $i$'th eigenvalue of $G$
(resp.$\ \widehat{G}(\widehat{d})$), $i=1,\ldots,p$, ordered such that
$\lambda_{1}>\lambda_{2}>\ldots>\lambda_{p-r}>0$, with $\lambda_{p-r+1}%
=\ldots=\lambda_{p}$ if $r>0$, and $\widehat{\lambda}_{1}\geq\widehat{\lambda
}_{2}\geq\ldots\geq\widehat{\lambda}_{p}$. The main result in Robinson and
Yajima (2002) is that, under regularity conditions, $m^{1/2}(\widehat{\lambda
}_{i}-\lambda_{i})$ are asymptotically independent for $i=1,...,p$, converge
in distribution to $N(0,\lambda_{i}^{2})$ for $i=1,\ldots,p-r$, and are
$o_{p}(1)$ for $i=p-r+1,\ldots,p$ if $r>0$. Defining, for a user chosen number
$v(T)>0$ which tends to zero as $T\rightarrow\infty$,
\[
L(u)=v(T)(p-u)-%
{\displaystyle\sum\limits_{i=1}^{p-u}}
\widehat{\lambda}_{i},
\]
the cointegrating rank is estimated consistently by%
\[
\widehat{r}=\arg\min_{u=1,\ldots,p-1}L(u).
\]
A related procedure can be found in Chen and Hurvich (2003), see also
Section~\ref{sec:CI-est-semi}, while Nielsen and Shimotsu (2007) extended
Robinson and Yajima's (2002) approach to (asymptotically) stationary and
nonstationary Type~II fractionally integrated processes.

In a nonparametric setting, Nielsen (2010) proposed a variance ratio testing
approach for fractional cointegration motivated by the following observation
for an univariate series. Let $u_{t}\in I(0)$ and $x_{t}=\Delta_{+}^{-d}u_{t}$
with $d>1/2$. For $d_{1}>0$, construct the partial sum $\widetilde{x}%
_{t}=\Delta_{+}^{-d_{1}}x_{t}$. Under regularity conditions on $u_{t}$, it can
be shown that, for $r\in\left[  0,1\right]  $,%
\[
T^{1/2-d}x_{\left\lfloor rT\right\rfloor }\Rightarrow\sigma_{u}W_{d-1}%
(r)\text{ and }T^{1/2-d-d_{1}}\widetilde{x}_{\left\lfloor rT\right\rfloor
}\Rightarrow\sigma_{u}W_{d+d_{1}-1}(r),
\]
see (\ref{a14}), where $\sigma_{u}^{2}$ is the long-run variance of $u_{t}$.
Then, the (univariate) variance ratio statistic is%
\begin{equation}
\rho(d_{1})=T^{2d_{1}}\frac{\sum_{t=1}^{T}x_{t}^{2}}{\sum_{t=1}^{T}%
\widetilde{x}_{t}^{2}}\rightarrow_{d}\frac{\int_{0}^{1}W_{d-1}^{2}%
(r)\mathsf{d}r}{\int_{0}^{1}W_{d+d_{1}-1}^{2}(r)\mathsf{d}r}, \label{a34}%
\end{equation}
where the limit follows by simple application of the continuous mapping
theorem. For $d=d_{1}=1$, $\rho(1)$ was proposed by Breitung (2002) as the
basis of a unit root test, but $\rho(1)$ is related to many well known
statistics like the R/S, V/S, KPSS, and DW statistics; see also Giraitis,
Kokoszka, Leipus, and Teyssiere (2003). Note that no parametric assumptions
are made on $u_{t}$, so there is no risk of misspecification. Additionally,
unlike classical nonparametric approaches, a bandwidth (or, alternatively, a
lag length choice) is not needed. Finally, $\rho(d_{1})$ does not depend on
nuisance parameters because $\sigma_{u}^{2}$ is cancelled out by the ratio in
(\ref{a34}). Indeed, this cancellation is the main reason behind the design of
the statistic because estimation of nuisance parameters, and in particular the
long-run variance, is unnecessary. In principle, $d_{1}$ is a user-chosen
parameter, but the choice is reflected in the asymptotic distribution of
$\rho(d_{1})$, and consequently it can be chosen so that the asymptotic local
power is maximized (Nielsen, 2009, showed that $d_{1}=0.1$ appears to be a
good choice). Deterministic components can be dealt with by using residuals
from the corresponding regression specifications.

Nielsen (2010) exploited this idea to propose a cointegration rank test. Let
$X_{t}$ be a $p$-dimensional observable process generated by (\ref{a35}) with
$d>1/2$ and $d-b<1/2$, and define $\widetilde{X}_{t}=\Delta_{+}^{-d_{1}}X_{t}$
for $d_{1}>0$. Defining $A_{T}=\sum_{t=1}^{T}X_{t}X_{t}^{\prime}$, $B_{T}%
=\sum_{t=1}^{T}\widetilde{X}_{t}\widetilde{X}_{t}^{\prime}$, consider the
ordered eigenvalues $\lambda_{1}\leq\lambda_{2}\leq\ldots\leq\lambda_{p}$ of
$R_{T}(d_{1})=A_{T}B_{T}^{-1}$ as the solutions to
\[
|\lambda B_{T}-A_{T}|=0,
\]
and the statistics%
\[
\Lambda_{p,r}(d_{1})=T^{2d_{1}}\sum_{j=1}^{p-r}\lambda_{j},\quad
r=0,\ldots,p-1.
\]
Nielsen (2010) showed that, for $r=0,\ldots,p-1$,%
\[
\Lambda_{p,r}(d_{1})\rightarrow_{d}U_{p-r}(d,d_{1})
\]
for a random variable $U_{p-r}(d,d_{1})$ which depends only on $p-r$, $d$, and
$d_{1}$. As in the univariate case, $\Lambda_{p,r}(d_{1})$ does not depend on
tuning parameters or on the short-run structure of the input process $U_{t}$,
and the long-run covariance matrix does not need to be estimated. The
statistic and its asymptotic null distribution also do not depend on the
cointegration gap, $b$. The cointegrating rank $r$ can be estimated by a
sequential test procedure identical to that presented in relation to Breitung
and Hassler's (2002) approach, but using $\Lambda_{p,r_{0}}(d_{1})$ for
$r_{0}=0,\ldots,p-1$, instead. Nielsen's (2010) results ensure that the test
is consistent, and, as in the univariate setting, $X_{t}$ can be replaced by
corresponding residuals where deterministic components have been removed.
Unlike Breitung and Hassler's (2002) approach, the tests based on
$\Lambda_{p,r_{0}}(d_{1})$ do not require knowledge or estimation of $d$. In
addition to providing an estimator of the cointegrating rank, Nielsen (2010)
also proposed an estimator of the cointegrating space.

To conclude this subsection we comment briefly on some alternative approaches.
Marinucci and Robinson (2001) proposed a Hausman-type procedure based on the
comparison of two different estimators of the integration order of the
observables, one of which is consistent under both the null of no
cointegration and the alternative of cointegration (although relatively
inefficient under the null), while the other is efficient under the null and
inconsistent under the alternative. Robinson (2008a) formalized this idea for
stationary and nonstationary settings. In a semiparametric setting and for
nonstationary observables, Marmol and Velasco (2004) test the null of no
cointegration (against the alternative of cointegration) by comparing OLS and
generalized least squares (GLS) estimators of the projection or fundamental
vector (i.e., the cointegrating vector, if cointegration exists). Again, their
approach is based on a Hausman-type idea, but here the estimators have
opposite consistency properties under the competing hypotheses. In a similar
fashion, Hualde and Velasco (2008) introduced a procedure based on correctly
orthogonalized residuals (obtained from estimating the projection or
fundamental vector) under the null of no cointegration which, unlike Marmol
and Velasco (2004), enjoys a standard asymptotic null distribution. Their
approach can be employed in parametric and semiparametric settings, covers
stationary and nonstationary ranges, and deals effectively with observables
with distinct integration orders. Finally, Hassler and Breitung (2006)
introduced a modified score test for the null of no cointegration in a
nonstationary time series. The test is applied to single equation regression
residuals from a first-step regression with differenced variables and in a
second step an endogeneity correction is implemented so the statistic obtains
a standard asymptotic null distribution. However, like the procedure of
Breitung and Hassler (2002), this procedure requires knowledge of the
integration order of the observables, which is unrealistic in practice, and
replacing this parameter by an estimator affects the asymptotic null distribution.

\subsection{Semiparametric estimation of fractional cointegration}

\label{sec:CI-est-semi}

Within the topic of estimation of fractional cointegration, we present first
several semiparametric approaches, where $f_{u}(\lambda)$ is considered an
unknown nonparametric function. Initially, we will illustrate some estimators
for the particular specification (\ref{zz5})--(\ref{zz6}) with $r=1$ known.
The case where $r=1$ is known has been routinely employed in both theoretical
and empirical work, and an extension to allow $r>1$ in this particular setup
is non-trivial. On the other hand, an extension allowing the individual
components of $X_{2t}$ in (\ref{zz6}) to have distinct integration orders is straightforward.

The most basic estimation approach of $\beta$ in (\ref{zz5}) is the OLS
estimator, which is a first stage procedure. This estimator is%
\begin{equation}
\widehat{\beta}_{OLS}=\left(
{\textstyle\sum\nolimits_{t=1}^{T}}
X_{2t}X_{2t}^{\prime}\right)  ^{-1}%
{\textstyle\sum\nolimits_{t=1}^{T}}
X_{2t}X_{1t}. \label{zz7}%
\end{equation}
For the particular $d=b=1$ case, Phillips and Durlauf (1986) analyzed the
asymptotic properties of $\widehat{\beta}_{OLS}$ for general conditions on the
error input series $u_{t}$. Specifically, they established its $T$-consistency
and derived its limiting distribution, which in general is nonstandard and
therefore unsuitable for inference. In fractional circumstances, the
properties of the OLS estimate (\ref{zz7}) can be very different from those in
the traditional $d=b=1$ situation. Robinson (1994c) showed the inconsistency
of $\widehat{\beta}_{OLS}$ in the model (\ref{zz5})--(\ref{zz6}) when the
observable $X_{t}$ was a covariance stationary long-memory process, i.e.\ when
$d<1/2$. In this framework, the inconsistency of the OLS estimator is due to
correlation between the stationary regressor and cointegrating error.

Robinson and Marinucci (2001, 2003) provided the asymptotic distribution of
the OLS estimator (with or without intercept) for the case $d\geq1/2$ and
$d\geq b$ in a model very similar to (\ref{zz5})--(\ref{zz6}), but where the
processes belonged to a class closely related to, but wider than, the Type~II
fractionally integrated. They showed that the rate of convergence of the OLS
estimator is $T^{\min\left(  2d-1,b\right)  }$, except for the case where
$d>b$ and $2d-b=1$, where the OLS is $T^{b}/\log T$-consistent. In all cases,
the OLS estimator has nonstandard limiting distributions which, as mentioned
before, complicates statistical inference.

A first stage alternative to OLS is the narrow-band least squares (NBLS)
estimator. For $l=0,1$ and an integer bandwidth $m$, with $l\leq m\leq T/2$,
we can estimate $\beta$ in (\ref{zz5}) by
\begin{equation}
\widetilde{\beta}_{l}(m)=\widehat{F}_{X_{2}X_{2}}^{-1}(l,m)\widehat{F}%
_{X_{2}X_{1}}(l,m), \label{nbls}%
\end{equation}
where, given (perhaps identical) scalar or vector sequences $a_{t}$ and
$b_{t}$, $t=1,\ldots,T$,%
\[
\widehat{F}_{ab}(l,m)=2\operatorname{Re}\left\{  \frac{2\pi}{T}\sum
\limits_{j=l}^{m}I_{ab}(\lambda_{j})\right\}  -\frac{2\pi}{T}I_{ab}%
(\pi)\mathbb{I}(m=T/2)
\]
is called the averaged (cross-)periodogram; see (\ref{a23}). Note that
\[
\widehat{F}_{ab}(1,m)=\widehat{F}_{ab}(0,m)-\overline{a}\overline{b},
\]
where $\overline{a}$ and $\overline{b}$ are the corresponding sample averages
of $a_{t}$ and $b_{t}$, such that omission of the zero frequency implies
sample-mean correction. Under the bandwidth condition
\begin{equation}
\frac{1}{m}+\frac{m}{T}\rightarrow0\text{ as }T\rightarrow\infty, \label{zz8}%
\end{equation}
the averaged (cross-)periodograms are based on a degenerating band of
frequencies around~0, so that (\ref{nbls}) only considers low-frequency
components of the series. In this situation, $\widetilde{\beta}_{l}(m)$ is the
NBLS estimator of $\beta$. In fact, since cointegration defines a long-run
relationship, avoiding high-frequency components that could be both distortive
and uninformative seems sensible. Note also that, from the orthogonality
properties of the complex exponential, $\widetilde{\beta}_{0}(\left\lfloor
T/2\right\rfloor )=\widehat{\beta}_{OLS}$, whereas $\widetilde{\beta}%
_{1}(\left\lfloor T/2\right\rfloor )$ is the OLS estimator in a model like
(\ref{zz5}) that includes an intercept.

The NBLS estimator was proposed by Robinson (1994c). It is related to the
band-spectrum estimator proposed by Hannan (1963), developed later by Engle
(1974) and analyzed by Phillips (1991b) in the context of standard $d=b=1$
cointegration, with the fundamental difference that the band-spectrum
estimator focuses on a nondegenerate band of frequencies, so (\ref{zz8}) does
not hold. Due to (\ref{zz8}), NBLS resembles nonparametric spectral
estimation, where now the focus is the parameter $\beta$ instead of a spectral
density at a given fixed frequency.

The NBLS estimator is particularly interesting in the case of so-called
\textquotedblleft stationary cointegration\textquotedblright, i.e.\ with
$d\in(0,1/2)$ and $d-b\in\lbrack0,1/2)$ in the model (\ref{zz5})--(\ref{zz6}),
where, as mentioned before, OLS is inconsistent. Somewhat surprisingly,
Robinson (1994c) proved consistency of the NBLS estimator in this case.
Intuitively, consistency of NBLS is obtained because focusing on a slowly
degenerating band of low frequencies reduces the bias due to the endogeneity
of $X_{2t}$. Robinson and Marinucci (2003) gave a rate of convergence (which
they conjectured as sharp) for the NBLS estimator under stationary
cointegration. Christensen and Nielsen (2006) improved on this result by
providing a better rate than that of Robinson and Marinucci (2003) and showing
that the NBLS estimator has an asymptotic normal distribution. This was under
the additional conditions that the collective memory satisfies $2d-b<1/2$ and
that the coherency between the weakly dependent processes $u_{1t}$ and
$u_{2t}$ is zero at frequency zero. The latter amounts to a type of long-run
exogeneity that can be a strong condition in some contexts.

For the nonstationary case, Robinson and Marinucci (2001, 2003) also exploited
the bias reduction achieved by focusing on a degenerating band of frequencies
around zero, and showed that in case $2d-b<1$ or $2d-b=1$ with $d>b$, the
rates of convergence previously given for the OLS can in fact be improved
upon. For NBLS, these rates are $T^{b}m^{2d-b-1}$ if $2d-b<1$, $T^{b}/\log m$
if $2d-b=1$ with $d>b$, and $T^{b}$ otherwise, noting (\ref{zz8}). Like OLS,
NBLS in general has nonstandard limiting distributions. For $d=b=1$, the
convergence rates of $\widetilde{\beta}_{l}(m)$ with (\ref{zz8}) and
$\widetilde{\beta}_{l}(\left\lfloor T/2\right\rfloor )$ are identical, but
$\widetilde{\beta}_{1}(m)$ eliminates the \textquotedblleft second-order
bias\textquotedblright\ present in the asymptotic distribution of
$\widetilde{\beta}_{1}(\left\lfloor T/2\right\rfloor )$. The superiority of
NBLS over OLS does not appear when comparing $\widetilde{\beta}_{0}(m)$ and
$\widetilde{\beta}_{0}(\left\lfloor T/2\right\rfloor )$, however, for the
standard $d=b=1$ case.

Focusing also on a narrow-band approach, Nielsen (2005b) proposed a
semiparametric version of the weighted least squares estimator of Robinson and
Hidalgo (1997). In the context of stationary cointegration in the model
(\ref{zz5})--(\ref{zz6}), Nielsen's (2005b) estimator is
\begin{equation}
\widehat{\beta}_{\delta,m}=\left(  \sum\limits_{j=1}^{m}\lambda_{j}^{2\delta
}I_{X_{2}X_{2}}(\lambda_{j})\right)  ^{-1}\sum\limits_{j=1}^{m}\lambda
_{j}^{2\delta}I_{X_{2}X_{1}}(\lambda_{j}), \label{a45}%
\end{equation}
where $\delta$ is a user-chosen number. Clearly, if $\delta=0$ (and $m<T/2$
satisfies (\ref{zz8})) then $\widehat{\beta}_{0,m}$ is the NBLS estimator of
$\beta$. However, the appeal of (\ref{a45}) is mainly for the infeasible
choice $\delta=d-b$, in which case $\widehat{\beta}_{d-b,m}$ was denoted the
narrow-band GLS estimator. The estimator $\widehat{\beta}_{d-b,m}$ also
maximizes the local Whittle likelihood (see (\ref{a46})) constructed for
(\ref{zz5}). Nielsen (2005b) showed that, under regularity conditions that
include the zero coherence condition (as in Christensen and Nielsen, 2006),
$\widehat{\beta}_{\delta,m}$ has a Gaussian limiting distribution. In
particular, for the $p=2$ case, he showed that%
\begin{equation}
\sqrt{m}\lambda_{m}^{-b}(\widehat{\beta}_{d-b,m}-\beta)\rightarrow_{d}N\left(
0,\frac{f_{u}^{(1,1)}(0)}{f_{u}^{(2,2)}(0)}\left(  \frac{1}{2}-b\right)
\right)  , \label{a47}%
\end{equation}
where $f_{u}^{(i,j)}(\lambda)$ is the $(i,j)$ element of $f_{u}(\lambda)$. The
result (\ref{a47}) contrasts with that of Christensen and Nielsen (2006),
where%
\[
\sqrt{m}\lambda_{m}^{-b}(\widehat{\beta}_{0,m}-\beta)\rightarrow_{d}N\left(
0,\frac{f_{u}^{(1,1)}(0)}{f_{u}^{(2,2)}(0)}\frac{(1/2-d)^{2}}{(1/2-(2d-b))}%
\right)
\]
with the additional condition that $2d-b<1/2$ (which is not required for
(\ref{a47})). The ratio between the respective asymptotic variances of
$\widehat{\beta}_{0,m}$ and $\widehat{\beta}_{d-b,m}$ is%
\begin{equation}
(1/2-d)^{2}/((1/2-d)^{2}-(d-b)^{2})\geq1, \label{a48}%
\end{equation}
with equality holding if and only if $d-b=0$, so that (\ref{a48}) confirms the
anticipated higher asymptotic efficiency of $\widehat{\beta}_{d-b,m}$ over
$\widehat{\beta}_{0,m}$. As mentioned before, $\widehat{\beta}_{d-b,m}$ is
infeasible. However, Nielsen (2005b) showed that, for a $\log T$-consistent
estimator $\widehat{d-b}$, the feasible estimator $\widehat{\beta
}_{\widehat{d-b},m}$ has the same asymptotic distribution as~$\widehat{\beta
}_{d-b,m}$.

Still in the context of stationary cointegration in the model (\ref{zz5}%
)--(\ref{zz6}), Nielsen (2007) proposed a local Whittle estimator
$\widehat{\theta}$ of $\theta=\left(  d,b,\beta\right)  ^{\prime}$. The main
result, derived again under the local orthogonality condition that
$f_{u}^{(1,2)}(0)=0$, is that the joint asymptotic distribution of the
appropriately normalized and centered $\widehat{\theta}$ is Gaussian with the
following particularities (mainly due to the zero coherence condition). First,
the distribution of the estimators of the integration orders, which are
$\sqrt{m}$-consistent, is identical to that if $\beta$ were a known parameter
(this is especially relevant for the estimator of $d-b$). Second, the result
for the estimator of $\beta$ is the same as (\ref{a47}). Third, the estimators
of the integration orders are asymptotically independent of the estimator of
the cointegrating vector. Velasco (2003b) and Shimotsu (2012) obtained similar
results in a nonstationary bivariate setting using, respectively, tapering in
a two-step approach (based on consistent initial estimators) and the exact
local Whittle approach.

Despite their Gaussian limit in some cases, all previous estimators are first
stage methods, because they do not implement the type of correction needed to
avoid the endogenity created by regressor $X_{2t}$ and which is removed in
some cases by the zero coherence condition. We next present three second stage procedures.

Robinson (2008b) proposed a bivariate version of the model (\ref{zz5}%
)--(\ref{zz6}) with stationary cointegration, where an unknown phase parameter
$\gamma\in(-\pi,\pi]$ between $\Delta^{-(d-b)}u_{1t}$ and $\Delta^{-d}u_{2t}$
at frequency zero is allowed. In the leading ARFIMA case\ $\gamma=b\pi/2$.
Robinson (2008b) studied the properties of a local Whittle estimator,
obtaining corresponding results to Nielsen (2007) but without the
$f_{u}^{(1,2)}(0)=0$ condition. In fact, Robinson (2008b) assumed that
$f_{u}^{(1,2)}(0)\neq0$, such that $\gamma$ is identifiable, but his theory
can be easily extended to cover the case with known $\gamma$ and
$f_{u}^{(1,2)}(0)=0$. As in Nielsen (2007), the estimators have a Gaussian
joint asymptotic distribution and the same convergence rates, including now
$\sqrt{m}$-consistency for the estimator of the phase parameter. However,
unlike Nielsen's (2007) results, the estimators are asymptotically dependent
due to the $f_{u}^{(1,2)}(0)\neq0$ condition.

Considering the more general Type~II fractionally integrated model (\ref{zz2})
with (\ref{zz1}) and $d_{1}\geq0$, Hualde and Robinson (2010) proposed
identification conditions on $\Upsilon$ and on the integration orders such
that (\ref{zz2}) represents a cointegrated process with multiple cointegrating
subspaces. Their framework is very general and notationally involved, so, for
simplicity, we illustrate it for a particular example with $p=4$. Suppose
there is a cointegrating space $S^{(1)}\subset%
\mathbb{R}
^{4}$ of dimension $r_{1}=2$; that is, there exists a full rank $4\times2$
matrix $\beta(1)$ such that $\beta(1)^{\prime}X_{t}\in I(d_{2})$ with
$d_{2}<d_{4}$ (in the sense of Definition~\ref{def vector zero}). Suppose also
that there is another cointegrating subspace $S^{(2)}$ of dimension $r_{2}=1$,
i.e.\ that there exists a $4\times1$ vector $\beta(2)$, which is a linear
combination of the columns of $\beta(1)$, such that $\beta(2)^{\prime}X_{t}\in
I(d_{1})$ with $d_{1}<d_{2}$. Here, $S^{\left(  1\right)  }$ is plane in $%
\mathbb{R}
^{4}$, and all vectors belonging to that plane are cointegrating vectors.
Similarly, $S^{(2)}$ is a line in the plane $S^{(1)}$ that leads to a greater
reduction in the integration orders of the observables, that is from $d_{4}$
to $d_{1}$ (instead of the reduction from $d_{4}$ to $d_{2}$ achieved by the
rest of vectors in the plane $S^{(1)}$ not belonging to $S^{(2)}$). Note that
the order $d_{2}$ is unique in the sense that any other full rank $4\times2$
matrix whose columns are cointegrating vectors leads to the same order
reduction (from $d_{4}$ to $d_{2}$), while $d_{1}$ is unique in the sense that
any vector $\phi\in S^{(2)}$ implies $\phi^{\prime}X_{t}\in I(d_{1})$.

As explained in Hualde and Robinson (2010), data-based procedures can be
applied to determine the number of cointegrating subspaces (two in this
example), and also their dimensions $r_{1}$ and $r_{2}$. Moreover, this
cointegration structure leads, for a particular ordering of the observables in
$X_{t}$ (which can be also inferred from data), to restrictions in $\Upsilon$
and the integration orders given by%
\begin{equation}
\Upsilon=\left(
\begin{array}
[c]{cccc}%
1 & -\beta_{12} & -\beta_{13} & -\beta_{14}\\
0 & 1 & -\beta_{23} & -\beta_{24}\\
0 & 0 & 1 & 0\\
0 & 0 & 0 & 1
\end{array}
\right)  \label{a50}%
\end{equation}
and%
\begin{equation}
d_{1}<d_{2}<d_{3}=d_{4},\label{a51}%
\end{equation}
where the $\beta$'s in (\ref{a50}) are unrestricted parameters. We say that
model (\ref{zz2}) with (\ref{a50}) and (\ref{a51}) is written in generalized
triangular form. In a general and much more complex setting, considering
unknown orders of integration and unknown $f_{u}(\lambda)$ treated as a
nonparametric function, Hualde and Robinson (2010) proposed two frequency
domain GLS (FDGLS) type of estimators for the unrestricted parameters. One
relies on the inverse spectral weighting $f_{u}^{-1}(\lambda_{j})$, while the
other is unweighted, but involved the inverse of $f_{u}(0)$. The motivation
for the second estimator is, apart from its simplicity, that, under regularity
conditions, $f_{u}^{-1}(\lambda)$ and $f_{u}^{-1}(0)$ behave similarly for
$\lambda$ close to zero. Therefore, employing a narrow-band approach in FDGLS
estimation, an estimate of $f_{u}^{-1}(\lambda)$ can be replaced by an
estimate of $f_{u}^{-1}(0)$ without affecting asymptotic properties. Note that
using estimates of $f_{u}^{-1}(\lambda)$ or $f_{u}^{-1}(0)$ corrects the
endogeneity problem caused by correlation between regressors and cointegrating
errors, so both estimators are second stage. Also, they have the advantage of
a closed form representation, although they depend on estimates of the
nuisance parameters ($f_{u}(\lambda)$ and orders of integration), so the
choice of several bandwidth parameters is involved.

In the previous example, the properties of the estimators depend crucially on
whether $d_{4}-d_{2}>1/2$ (so, also, $d_{4}-d_{1}>1/2$), which is described as
\textquotedblleft strong\textquotedblright\ fractional cointegration, or
$d_{4}-d_{1}<1/2$ (so, also, $d_{4}-d_{2}<1/2$), which is the so-called
\textquotedblleft weak\textquotedblright\ fractional cointegration. In the
former case, estimators have a mixed Gaussian asymptotic distribution and the
following convergence rates: estimators of $\beta_{23}$ and $\beta_{24}$ are
$T^{d_{4}-d_{2}}$-consistent, whereas those of $\beta_{12}$, $\beta_{13}$,
$\beta_{14}$ have in general the rate $T^{d_{2}-d_{1}}$, but the estimator of
$\beta_{13}$ has the faster rate $T^{d_{4}-d_{1}}$ if $\beta_{23}=0$, whereas
that of $\beta_{14}$ has the faster rate $T^{d_{4}-d_{1}}$ if $\beta_{24}=0$.
In the \textquotedblleft weak\textquotedblright\ situation, the limiting
distribution of the estimators is Gaussian with the following convergence
rates, where $m$ is a bandwidth parameter such that $m\rightarrow\infty$ as
$T\rightarrow\infty$: estimators of $\beta_{23}$ and $\beta_{24}$ are
$\sqrt{m}\lambda_{m}^{d_{4}-d_{2}}$-consistent, whereas those of $\beta_{12}$,
$\beta_{13}$, $\beta_{14}$ have in general rate $\sqrt{m}\lambda_{m}%
^{d_{2}-d_{1}}$, but the estimator of $\beta_{13}$ has the faster rate
$\sqrt{m}\lambda_{m}^{d_{4}-d_{1}}$ if $\beta_{23}=0$, whereas that of
$\beta_{14}$ has the faster rate $\sqrt{m}\lambda_{m}^{d_{4}-d_{1}}$ if
$\beta_{24}=0$. The case with mixed \textquotedblleft strong\textquotedblright%
\ and \textquotedblleft weak\textquotedblright\ situations is also studied,
where both mixed Gaussian and Gaussian limits co-exist but are mutually
independent. Interestingly, in all cases these asymptotic distribution results
lead to a $\chi^{2}$-limit for Wald statistics based on the proposed estimators.

In a bivariate situation, Hualde and Iacone (2019) derived the asymptotic
theory of related semiparametric estimators when the bandwidth $m$ is kept
fixed in the limit. They showed that this alternative limit theory provides a
more accurate approximation to the sampling distribution of the classical Wald
statistics, which are typically oversized when confronted with the standard
limit theory.

A different type of endogeneity correction is proposed by Nielsen and
Frederiksen (2011) based on a fully-modified approach to NBLS estimation. In
particular, they derive an expression for the asymptotic bias of the NBLS
estimator, which, in terms of model (\ref{zz2}) depends on $d_{1},\ldots
,d_{p}$ and $f_{u}(0)$. This asymptotic bias is consistently estimated, and
based on this an appropriate correction of the NBLS estimator is implemented,
thus obtaining the fully-modified NBLS estimator, which, again, is a second
stage procedure. Their results apply to the \textquotedblleft
weak\textquotedblright\ fractional cointegration case, where there is a unique
nontrivial cointegrating relation, and their estimator enjoys a Gaussian limit
with identical convergence rates to those in Hualde and Robinson (2010).

Finally, a different but related modeling approach of fractional cointegration
can be found in Chen and Hurvich (2003). This is based on a common components
representation and their focus is estimating the cointegrating space using a
semiparametric narrow-band approach. Assuming a known cointegrating rank $r$,
this space is estimated by the eigenvectors corresponding to the $r$ smallest
eigenvalues of the averaged tapered periodogram matrix of the observables,
which are assumed to have identical memory. Chen and Hurvich (2006) extended
this setting so that observables, which again are assumed to have a common
memory, are modeled as a linear combination of unobservable components with
potentially different memories. This prompts consideration of the issue of
different cointegrating subspaces, and Chen and Hurvich's (2006) approach
leads to estimates of the whole cointegration structure; that is, the
cointegrating space and possible subspaces.

\subsection{Parametric estimation of fractional cointegration}

\label{sec:CI-est-par}

We present some parametric approaches where $f_{u}(\lambda)$ is considered a
known parametric function up to a vector of unknown parameters. Early research
(e.g., Cheung and Lai, 1993, or Baillie and Bollerslev, 1994a,b) focused on a
three-step approach assuming a model like (\ref{zz5})--(\ref{zz6}): first, $d$
is estimated or assumed to take a particular value, e.g., $d=1$ (perhaps as a
result of the outcome of a unit root test); then $\beta$ is estimated using a
first stage procedure like OLS; and finally $d-b$ is estimated from the
corresponding residuals of the potentially cointegrated relation. Testing for
(fractional) cointegration, for example, then amounts to finding statistical
evidence in favor of $b>0$. These steps were combined by Dueker and Startz
(1998) who proposed joint maximum likelihood estimation of $d$, $b$, and
$\beta$ in a Type~I version of (\ref{zz5})--(\ref{zz6}) where $p=2$, $r=1$ (so
$X_{1t}$ and $X_{2t}$ are scalars) and $d<1/2$, assuming also that $u_{t}$ is
a vector ARMA process. Their proposal relied on previous unpublished work by
Sowell (1986), but no asymptotic results were derived. Jeganathan (1999)
proposed a MLE of $\beta$ in a Type~I version of (\ref{zz5})--(\ref{zz6}),
focusing on the case where $u_{t}$ is white noise with known distribution and
assuming known $d$ and $b$. He showed that this estimator has a mixed Gaussian
asymptotic distribution and provided some discussion about a feasible
alternative in which $d$ and $b$ are estimated.

Robinson and Hualde (2003) proposed two parametric GLS-type of estimators of
$\beta$ in model (\ref{zz5})--(\ref{zz6}) with $p=2$, $r=1$, and $d\geq
b>1/2$, which captures the \textquotedblleft strong\textquotedblright%
\ fractionally cointegrated case. The motivation behind their estimators can
be illustrated by focusing first on the traditional $d=b=1$ case. In that
case, Phillips (1991a) justified that the second-order simultaneity bias due
to correlation between the error terms in (\ref{zz5}) and (\ref{zz6}) can be
eliminated and the optimal properties of a second stage estimator, including a
mixed Gaussian limiting distribution, can be recovered. This is achieved by
pseudo maximum likelihood (PML) estimation of $\beta$ in the model
(\ref{zz5})--(\ref{zz6}), which is equivalent to OLS in the augmented
regression model%
\begin{equation}
X_{1t}=\beta X_{2t}+\rho\Delta X_{2t}+u_{1.2t},\label{zz62}%
\end{equation}
where $\rho=\omega_{12}/\omega_{22}$, $\omega_{12}=\operatorname{Cov}%
(u_{1t},u_{2t})$, $\omega_{22}=\operatorname{Var}(u_{2t})$, and $u_{1.2t}%
=u_{1t}-\rho u_{2t}$, noting that by construction $\operatorname{Cov}%
(u_{1.2t},u_{2t})=0$.

This idea was extended to the fractional setting by Robinson and Hualde
(2003), where, focusing again on the white noise case for illustration, the
equivalent to (\ref{zz62}) is%
\begin{equation}
\Delta^{d-b}X_{1t}=\beta\Delta^{d-b}X_{2t}+\rho\Delta^{d}X_{2t}+u_{1.2t}%
,\label{zz63}%
\end{equation}
so it could be expected that OLS estimation in (\ref{zz63}) would lead to
optimality properties (see also Nielsen, 2004c, for an identical approach).
Robinson and Hualde (2003) confirmed this result and showed that the feasible
version of this estimator where $d$ and $b$ are replaced by corresponding
estimates (with convergence rates $T^{\phi}$ for $\phi>\max\{0,1-b\}$), has an
identical limiting distribution to the infeasible OLS estimator in
(\ref{zz63}). Furthermore, their results apply for a general $I(0)$ error
$u_{t}$ with parametric spectral density. Based on this parametric structure,
they proposed time domain (based on an autoregressive transformation) and
frequency domain estimators which approximate GLS (and in fact PML).\ Both
estimators have the same mixed Gaussian asymptotic distribution, leading to
Wald test statistics with $\chi^{2}$ null limit distributions for known or
unknown integration orders.

Corresponding results for the \textquotedblleft weak\textquotedblright%
\ cointegration case where $b<1/2$ were provided by Hualde and Robinson
(2007). This case is more difficult to handle because of the smaller
cointegration gap, $b$, which could potentially be very small so the strength
of the cointegrating relation is weaker compared to the \textquotedblleft
strong\textquotedblright\ case. Hualde and Robinson's (2007) estimator, again
for the scalar $p=2$, $r=1$ case, is designed for the VAR$\left(  k\right)  $
situation, where in model (\ref{zz5})--(\ref{zz6}), the error term is%
\begin{equation}
u_{t}=\sum_{j=1}^{k}B_{j}u_{t-j}+\varepsilon_{t},\label{zz67}%
\end{equation}
where $\varepsilon_{t}=(\varepsilon_{1t},\varepsilon_{2t})^{\prime}$ is a zero
mean white noise process and all zeros of $\det\{I_{2}-\sum_{j=1}^{k}%
B_{j}z^{j}\}$ lie outside the unit circle. Defining $\zeta=(1,0)^{\prime}$ and
letting $B_{ij}$ be the $i$'th row of $B_{j}$, the orthogonalization
equivalent to that in (\ref{zz63}) is, for $t>k$,
\begin{align}
\Delta^{d-b}X_{1t} &  =\beta\Delta^{d-b}X_{2t}+\rho\Delta^{d}X_{2t}+%
{\textstyle\sum\limits_{j=1}^{k}}
(B_{1j}-\rho B_{2j})(\Delta^{d-b}X_{1,t-j},\Delta^{d}X_{2,t-j})^{\prime
}\nonumber\\
&  \quad-\beta%
{\textstyle\sum\limits_{j=1}^{k}}
(B_{1j}-\rho B_{2j})\zeta\Delta^{d-b}X_{2,t-j}+\varepsilon_{1.2t},\label{zz68}%
\end{align}
where $\varepsilon_{1.2t}=\varepsilon_{1t}-\rho\varepsilon_{2t}$ and
$\rho=E(\varepsilon_{1t}\varepsilon_{2t})/E(\varepsilon_{2t}^{2})$. It is
shown that the OLS estimator of $\beta$ in (\ref{zz68}) is $\sqrt{T}%
$-consistent and asymptotically normal, but is infeasible because it depends
on $d$ and $b$. Moreover, a feasible estimator of $\beta$ with the same
properties as the infeasible estimator is unachievable in the
\textquotedblleft weak\textquotedblright\ cointegration case because
estimators of the integration orders are at most $\sqrt{T}$-consistent.
Instead, Hualde and Robinson (2007) proposed convenient estimators of the
nuisance parameters $d$ and $b$ based on simple univariate optimizations. The
feasible estimator of $\beta$ based on these estimators is inefficient
relative to the PML estimator, but has an optimal rate of convergence, is
asymptotically normal, and is computationally much simpler than the PML estimator.

\subsection{Fractionally cointegrated VAR model}

\label{sec:fcvar}

The fractionally cointegrated VAR (FCVAR) model can be seen as a
generalization of the CVAR model (Johansen, 1995) to fractional time series,
and also as a generalization of the fractional AR model (Johansen and Nielsen,
2010) discussed in Section~\ref{sec:DF} to multivariate time series. For a
$p$-dimensional time series $X_{t}$, the FCVAR model can be derived from the
usual CVAR\ model by the same steps as in (\ref{a29})--(\ref{a31}) to arrive
at%
\begin{equation}
\Delta^{d}X_{t}=\alpha\beta^{\prime}L_{b}\Delta^{d-b}X_{t}+\sum_{i=1}%
^{k}\Gamma_{i}L_{b}^{i}\Delta^{d}X_{t}+\varepsilon_{t},\label{FCVAR}%
\end{equation}
where $\varepsilon_{t}$ is assumed i.i.d.\ with mean zero and variance~$\Omega
$. An alternative derivation starts with (\ref{zz5})--(\ref{zz6}), though
using $\gamma$ instead of $\beta$ in (\ref{zz5}) because $\beta$ has a
standard meaning in (\ref{FCVAR}). From there it follows (Johansen, 2008) that
$\Delta^{d}X_{t}=\alpha\beta^{\prime}L_{b}\Delta^{d-b}X_{t}+v_{t}$ with
$v_{t}=\alpha(\beta^{\prime}\alpha)^{-1}u_{1t}+\beta_{\perp}(\alpha_{\perp
}^{\prime}\beta_{\perp})^{-1}u_{2t}$ and the parameters are $\beta^{\prime
}=(I_{r},\gamma^{\prime})$ and $\alpha_{\perp}^{\prime}=(0_{p-r,r},I_{p-r})$.
Here, for a matrix $A$, the matrix $A_{\perp}$ is such that $A^{\prime
}A_{\perp}=A_{\perp}^{\prime}A=0$.

The FCVAR model (\ref{FCVAR}) was proposed by Granger (1986) and Johansen
(2008), although Granger had an additional lag in the error-correction term,
$L_{b}\Delta^{d-b}X_{t-1}$, which would appear to be a typo. The FCVAR model
(\ref{FCVAR}) was analyzed by Johansen and Nielsen (2012a, 2018, 2019) who
list (at least) five desirable properties for a fractional cointegration model
that are all achieved by the FCVAR model: (i)~the parameters should have
interesting interpretations, (ii)~there exists testable criteria on the
parameters for when the solution is cointegrated (and for the cointegration
rank), (iii)~estimation is relatively simple, (iv)~asymptotic theory is
straightforward to apply, and (v)~generic computer programs can be made to
estimate the model and test the most interesting hypotheses.

Model (\ref{FCVAR}) includes the Johansen (1995) CVAR model as the special
case $d=b=1$. Some of the parameters are well-known from the CVAR model and
they have the usual interpretations in the FCVAR model. The most important of
these are the long-run parameters $\alpha$ and $\beta$, which are $p\times r$
matrices with $0\leq r\leq p$ and $r$ being the cointegration, or
cofractional, rank. The columns of $\beta$ constitute the $r$ cointegration
vectors such that $\beta^{\prime}X_{t}$ are the long-run equilibrium
relations. The parameters in $\alpha$ are the adjustment or loading
coefficients which represent the speed of adjustment towards equilibrium for
each of the variables. The short-run dynamics of the variables are governed by
the parameters $\Gamma_{1},\ldots,\Gamma_{k}$ in the autoregressive augmentation.

The FCVAR model has two additional parameters compared with the CVAR model,
namely the fractional order of the observables, $d$, and the cointegration
gap, $b$. These parameters are estimated jointly with the remaining
parameters. This model thus has the same main structure as the standard CVAR
model in that it allows for modeling of both cointegration and adjustment
towards equilibrium, but is more general since it accommodates fractional
integration and fractional cointegration.

The representation theory (solution) for the model (\ref{FCVAR}) was derived
by Johansen (2008) and Johansen and Nielsen (2012a), who showed that%
\[
X_{t}=C\Delta_{+}^{-d}\varepsilon_{t}+\Delta_{+}^{-(d-b)}Y_{t}+\mu_{t},
\]
where $C=\beta_{\perp}(\alpha_{\perp}^{\prime}(I_{p}-\sum_{i=1}^{k}\Gamma
_{i})\beta_{\perp})^{-1}\alpha_{\perp}^{\prime}$ such that $\beta^{\prime}%
C=0$, $Y_{t}\in I(0)$, and $\mu_{t}$ is a function of initial values
$X_{-n},n\geq0$. Thus, it is clear that $X_{t}\in I(d)$ while $\beta^{\prime
}X_{t}\in I(d-b)$. Furthermore, the rank of $\alpha\beta^{\prime}(=\Pi)$ is
the cointegration rank, which is testable. The initial values term, $\mu_{t}$,
is most often assumed to be zero (and known) in the literature, but Johansen
and Nielsen (2012a, 2016) analyze more general conditions under which $\mu
_{t}$ induces a bias in finite samples that can be alleviated by either
conditioning the analysis on the first part of the sample or by including a
so-called \textquotedblleft level parameter\textquotedblright. The latter is
related to the notions of restricted and unrestricted constant terms (Johansen
and Nielsen, 2012a, 2016, and Dolatabadi, Nielsen, and Xu, 2015).

The parameters $(d,b,\alpha,\beta,\Gamma_{1},\ldots,\Gamma_{k},\Omega)$ of
model (\ref{FCVAR}), along with possible deterministic terms or level
parameters, can be estimated jointly by maximum likelihood. For fixed $(d,b)$,
maximum likelihood is reduced rank regression, which leaves a numerical
optimization problem over the two parameters $(d,b)$.

The asymptotic distribution theory for the MLEs is justified by Johansen and
Nielsen (2012a, 2018, 2019). They show that, when $b<1/2$, all estimators are
jointly asymptotically Gaussian. On the other hand, when $b>1/2$, $\hat{\beta
}$ is asymptotically mixed normal while the remaining estimators are jointly
asymptotically Gaussian and independent of~$\hat{\beta}$. The important
consequence, of course, is that likelihood ratio tests of hypotheses on any of
the parameters are $\chi^{2}$-distributed, regardless of the value of $b$.
Furthermore, the likelihood ratio test for cointegration rank is
asymptotically $\chi^{2}$-distributed when $b<1/2$, but is a functional of
fractional Brownian motion when $b>1/2$. This dependence on the real-valued
nuisance parameter $b$ complicates tabulation, but a plug-in approach can be
used to compute $p$-values in practice (MacKinnon and Nielsen, 2014).

Thus, estimation is quite simple and asymptotic theory is straightforward to
apply, and computer programs are available in Matlab and R. For details on
practical implementation and computation, see Nielsen and Popiel (2018) and
Morin, Nielsen, and Popiel (2021). We note that the asymptotic analysis in
Johansen and Nielsen (2012a) assumed that $d$ and $b$ are such that $0\leq
d-b<1/2$, i.e.\ that the long-run equilibrium errors have non-negative memory
and are (asymptotically) stationary. The former condition is restrictive
because it implies that the usual CVAR model with $d=b=1$ lies on the boundary
of the parameter space, which complicates testing this restriction. The latter
condition is likely restrictive in practice, where the possibility of
nonstationary cointegration with $d-b\geq1/2$ may be desired. However, these
restrictions were removed in Johansen and Nielsen (2018, 2019), while relaxing
at the same time a strong moment condition in Johansen and Nielsen (2012a).

\section{Applications of fractional integration and cointegration}

\label{sec:applications}

Many empirical researchers have applied the previous methodological
contributions to describe and illustrate features of real data, especially
since the early 1990s. Some of the earlier works are nicely summarized in
Baillie (1996) or Henry and Zaffaroni (2003), but, nevertheless, we first
provide a brief overview of some early results and subsequently focus on some
applications that we find particularly interesting.

In one of the earliest empirical investigations involving long memory behavior
in economics or finance, Diebold and Rudebusch (1989) analyzed aggregate
output and found evidence of fractional integration through estimation of
ARFIMA models. In particular, their estimates of $d$ typically range from 0.5
to 0.90, depending on the particular series, and with quarterly series being
apparently more persistent than annual ones. This indicates that macroeconomic
growth shocks appear to be less persistent than under the classical $I\left(
1\right)  $ prescription, although the results need to be taken with caution
due to relatively wide confidence intervals. In a similar fashion, Sowell
(1992) used an an ARFIMA process to model the first differences of US
quarterly real GNP. Sowell (1992) estimated an ARFIMA$\left(  3,d,2\right)  $
model (selected by AIC) by maximum likelihood and obtained $\widehat{d}=-0.59$
($d$ being the integration order of the first differences of the quarterly
real GNP). Neither of the hypotheses $H_{0}:d=0$ (GNP is difference
stationary) or $H_{0}:d=-1$ (GNP is trend stationary) were rejected, so the
data appear to be consistent with both rival models.

Extending the analysis to other macroeconomic variables, Crato and Rothman
(1994a) applied ARFIMA models to an extended version of the data set used by
Nelson and Plosser (1982) and found evidence supporting the difference
stationary model for many macroeconomic time series. In a similar vein,
Gil-Alana and Robinson (1997) applied Robinson's (1994b) parametric frequency
domain Lagrange multiplier test to the extended Nelson and Plosser dataset and
concluded that prices and money stock are the most nonstationary series, while
unemployment rate and industrial production are the closest to stationarity.

Focusing on inflation data, Hassler and Wolters (1995) examined the question
of whether inflation has a unit root, which is an important issue in monetary
policy. Using ARFIMA modeling and semiparametric estimation methods they
concluded that, for all considered countries (United States, United Kingdom,
France, Germany, Italy), estimates are significantly different from both 0
and~1, with estimated integration orders typically around the
stationary/nonstationary boundary of~0.5. A similar result was obtained by
Baillie, Chung and Tieslau (1996) who used an approximate maximum likelihood
estimator in an ARFIMA-GARCH setting. Using monthly post-World War II data for
ten countries, they found clear evidence of long memory with mean reverting
behaviour (so $0<d<1$) except for Japan, for which the hypothesis that $d=0$
could not be rejected.

There are numerous different applications to exchange rates and the analysis
of the purchasing power parity (PPP). This is a key equilibrium condition in
international economics where nominal exchange rates and prices adjust, so
real exchange rates revert to a parity value. Thus, this condition represents
a cointegrating relation and, in the fractional context, it was first analyzed
by Diebold, Husted, and Rush (1991). They constructed a long data set for 16
real exchange rates covering approximately a century of the gold standard
period and estimated the corresponding integration orders by maximum
likelihod. In contrast to the assumed $I(1)$ condition for nominal exchange
rates, their memory estimates for real exchange rates were in all cases
significantly less than~1, and in some other cases also significantly
different from zero. In their view, these results justify that the PPP holds
in the long run. Similar analyses can be found in Cheung and Lai (1993),
Baillie and Bollerslev (1994a), Crato and Rothman (1994b), Nielsen (2004c),
and Gil-Alana and Hualde (2009).

Long memory in interest rates have been intensively studied (e.g., Shea, 1991,
Lai, 1997, Tsay, 2000, Meade and Maier, 2003, Nielsen, 2004d, 2005a), and
there are also some studies in the fractional cointegration literature.
Actually, the methodological contribution of Dueker and Startz (1998)
mentioned in Section~\ref{sec:CI-est-par} was illustrated by analyzing the
potential cointegration between US and Canadian bond rates. Their results
implied that the observables were nonstationary but mean reverting processes,
whereas the cointegrating error was stationary long memory, supporting the
possibility of a \textquotedblleft weak\textquotedblright\ cointegration
relation. The term structure of interest rates is particularly interesting in
the context of (fractional) cointegration because the unbiasedness hypothesis
in this model implies a single common stochastic trend. Chen and Hurvich
(2003) analyzed fractional cointegration among daily US interest rates with
eight different maturities ranging from 3 months to 30 years as an
illustration of their semiparametric inference procedure. They found a clear
evidence of cointegration with an estimated cointegrating rank of
$\widehat{r}=6$ or $\widehat{r}=7$, and with some of the cointegrating
relationships being stronger than others. Nielsen (2010) addressed a similar
phenomenon by analyzing four daily US Treasury Bill interest rates with
maturities ranging from 3 months to 2 years. His results indicated that each
series was $I(1)$ and found evidence of three cointegrating relations with a
single common stochastic trend. Nielsen (2010) also applied the parametric
procedures of Johansen (1995) and Breitung and Hassler (2002) and conjectured
why they both fail to detect the correct rank.

Another important strand of empirical research has focused on analyzing the
volatility of financial time series. Baillie, Bollerslev, and Mikkelsen (1996)
introduced the fractionally integrated generalized autoregressive
conditionally heteroskedastic (FIGARCH) model. Here, the main novelty with
respect to the popular GARCH model is that the conditional variance is
fractionally integrated. This model was further extended by Bollerslev and
Mikkelsen (1996) to include the exponential (asymmetric) feature. Both these
papers included empirical studies of stock market volatility and found strong
evidence of fractional integration. Andersen, Bollerslev, Diebold, and Ebens
(2001) studied \textquotedblleft realized\textquotedblright\ daily equity
return volatilities and correlations from high-frequency intraday transaction
prices on some individual stocks. They showed that the behaviour of these
statistics is coherent with long memory and also found comovements on
volatilities and correlations across assets. Similarly, Andersen, Bollerslev,
Diebold, and Labys (2001) analyzed daily exchange rate volatility and
correlation showing long-memory dynamics in their behavior.

Beltratti and Morana (2006) analyzed the linkages between stock market and
macroeconomic volatility. They found that, after accounting for some common
structural breaks, the series appeared to be long memory and displayed
fractional cointegration. Specifically, they justified three cointegrating
relationships among stock market, money growth, inflation, the Federal funds
rate, and output growth volatilities.

Following this general overview of some relevant empirical works, we next
present in more details applications of fractional integration and
cointegration that we find particularly interesting and relevant.

\subsection{The Deaton paradox}

This paradox refers to an apparently contradictory result related to the
permanent income hypothesis (PIH) under rational expectations; in particular
the variability of consumption changes versus that of income innovations.
Following Diebold and Rudebusch (1991b) and letting $C_{t}$ and $Y_{t}$ be
consumption and income, respectively, the PIH implies that
\begin{equation}
\Delta C_{t}=r\sum_{i=0}^{\infty}\beta^{i}(E_{t}Y_{t+i}-E_{t-1}Y_{t+i}),
\label{zz69}%
\end{equation}
where $r$ is the real interest rate, $\beta=r/(1+r)$ is a discount factor, and
$E_{t}$ is the conditional expectation formed at time $t$. Thus, (\ref{zz69})
implies that the behavior of consumption depends on the specification of the
income process. It is typically assumed that $Y_{t}$ is $I(1)$, so letting%
\begin{equation}
\Delta Y_{t}=\gamma+a(L)\epsilon_{t}, \label{zz70}%
\end{equation}
where $\epsilon_{t}$ is a zero mean white noise and $a(z)=1+a_{1}z+a_{2}%
z^{2}+\ldots$ has all roots outside the unit circle, it can be shown that%
\[
\Delta C_{t}=c_{\infty}^{\beta}\epsilon_{t},
\]
where the multiplier is $c_{\infty}^{\beta}=1+\sum_{i=1}^{\infty}\beta
^{i}a_{i}$. For many ARIMA specifications in (\ref{zz70}) and realistic
assumptions about $r$, $c_{\infty}^{\beta}$ is substantially above~1.

Noting that%
\[
\text{std}(\Delta C_{t})=c_{\infty}^{\beta}\text{std}(\epsilon_{t}),
\]
the ARIMA specification in (\ref{zz70}) implies that the variability of
consumption changes should be greater than the variability of income
innovations when $c_{\infty}^{\beta}>1$. However, in real data the opposite is
observed, i.e.\ consumption is too smooth. This contradictory result has been
named the \textquotedblleft Deaton paradox.\textquotedblright

Diebold and Rudebusch (1991b) proposed to model $Y_{t}$ as an ARFIMA$(p,d,q)$
process instead of (\ref{zz70}). This alternative model is important in the
context of the Deaton paradox\ because the multiplier $c_{\infty}^{\beta}$
depends critically on the impact of past innovations to $\Delta Y_{t}$, which
is driven by the chosen specification for $Y_{t}$. Specifically, Diebold and
Rudebusch (1991b) show how the observed excess smoothness of consumption is
theoretically supported by values of $d$ less than one. Unfortunately,
estimates of $d$ are typically close to one, with confidence intervals
including one, so it cannot be assessed whether their results are a departure
from the predictions of the PIH. Either way, a key message of the paper is
that the introduction of a more flexible specification (ARFIMA vs.\ ARIMA)
shows that departures from the classical PIH, in this case excess smoothness
of consumption, could be due to misspecification of the time series properties
of the data.

\subsection{Long memory inflation uncertainty and the term structure of
interest rates}

Backus and Zin (1993) examined the apparent contradiction between the
predictions of theory and observed data in the relationship of short-term
interest rates and long-term yields (or, equivalently, long forward rates).
Specifically, if the short rate process is a stationary ARMA$(p,q)$ process,
then theory implies that the expected long-term yield converges to a constant
while its variance tends to zero exponentially. This result contrasts with
observed data, where, typically, the mean yield curve flattens out for longer
maturities and the variance of the long yields declines with maturity although
far from exponentially. The former is coherent with theory, but the latter is
not. An alternative $I(1)$ specification of the short rate implies that
expected long-term yield is linear in maturity while the variance is constant,
which is also at odds with observed data.

A possible explanation of this contradictory behavior was proposed by Backus
and Zin (1993) who suggested modeling the short rate by a stationary
fractionally integrated noise. They showed that this alternative model implies
that the variability of yields decreases with maturity, but at a much slower
rate than for the stationary ARMA process. In particular, for values of $d$
around 0.3, the theoretically implied shape of the yield curve, both in terms
of expected yield and variance of the yield, is clearly superior to the
alternative ARMA and $I(1)$ specifications. Unfortunately, empirical support
for a stationary fractionally integrated short rate process is weak. Backus
and Zin (1993) reported stronger evidence of fractional behavior for inflation
and money growth, which is relevant because they conjecture that the behavior
of short rates is driven to some extent by those variables.

\subsection{Fractional beta-convergence}

As noted by Michelacci and Zaffaroni (2000), there are three stylized facts in
modern empirical macroeconomics which seem to be inconsistent: unit root in
output per capita, output per capita of different economies converging to
their long-run steady state value at a uniform exponential rate of 2\% per
year (beta-convergence), and, finally, that the steady state output could well
be represented by a smooth linear trend. The first stylized fact implies that
output is not mean reverting. This does not necessarily contradict
beta-convergence because current and steady state outputs may be cointegrated.
However, if the smooth trend representation for steady state output is correct
and current output has a unit root, testing beta-convergence would be
equivalent to testing trend stationarity in output which is contradictory.

However, Michelacci and Zaffaroni (2000) proposed an alternative model for
current output which reconciles these findings. Their solution is to model the
de-trended current per capita output as a fractionally integrated process with
memory $d$ such that $1/2<d<1$. This is a nonstationary but mean reverting
process, which is consistent with beta-convergence (to a smooth trend steady
state), and which could easily lead to non-rejections of the unit root
hypothesis. In addition, the authors found conditions under which long memory
in output arises (by aggregation of heterogeneous units as in
Section~\ref{sec:sources}) and provided an extensive empirical analysis with
evidence in favor of nonstationary and mean reverting behavior of per capita
output and also of unconditional convergence across OECD countries with
similar rates of convergence.

\subsection{The forward premium anomaly}

This is a puzzle in the international finance literature which refers to the
inability of forward exchange rates to forecast future spot rates. Denote by
$s_{t}$ the log-spot exchange rate at time $t$ and $f_{t,k}$ the log-forward
rate with $k$ being the length of the forward contract. According to the
forward rate unbiasedness hypothesis, the forward rate is an unbiased
forecaster of the future spot rate. That is, $E_{t}s_{t+k}=f_{t,k}$, where
$E_{t}$ is the conditional expectation formed at time $t$. This hypothesis
implies market efficiency, rational expectations, and risk neutrality. Tests
of unbiasedness have commonly been implemented in a regression of the spot
return, $s_{t+k}-s_{t}$, on the forward premium, $f_{t,k}-s_{t}$, i.e.,
\begin{equation}
s_{t+k}-s_{t}=\alpha+\beta(f_{t,k}-s_{t})+\varepsilon_{t+k}, \label{zz71}%
\end{equation}
where the null hypothesis of unbiasedness would be $H_{0}:\alpha=0,\beta=1$ or
sometimes just $H_{0}:\beta=1$ to allow for a constant risk premium.
Surprisingly, estimates are typically significantly smaller than one and even
negative in some cases (see Ballie and Bollerslev, 2000, for references).

Ballie and Bollerslev (2000) described this anomaly as a statistical problem
related to the intrinsic dependence structures of the regressor and dependent
variable in (\ref{zz71}). There is strong evidence in favour of $s_{t}$ and
$f_{t,k}$ both being $I(1)$, implying that $s_{t+k}-s_{t}$ is $I(0)$, whereas
$f_{t,k}$ and $s_{t}$ appear to be cointegrated such that the forward premium
$f_{t,k}-s_{t}$ is $I(d)$ with most likely $1/2<d<1$. This implies an
imbalance in the regression (\ref{zz71}) caused by the different integration
orders of the regressor and regressand. Maynard and Phillips (2001) provided
formal theoretical justification of the consequences of this imbalance when
the forward premium is nonstationary but mean reverting. Specifically, the OLS
estimator in (\ref{zz71}) converges in probability to zero at rate $T^{1-2d}$,
so that the appropriately normalized estimator tends to a non-standard random
variable whose sign depends on the one-sided long-run covariance, say
$\lambda$, between the spot return and the forward premium. A non-standard
behaviour is also reported for the $t$-statistic, which, even if the
OLS\ estimator of $\beta$ converges to zero, diverges to $+\infty$ if
$\lambda>0$ or to $-\infty$ if $\lambda<0$, at rate $T^{1-d}$. To overcome
this problem, Maynard, Smallwood, and Wohar (2013) provided a two-step
procedure to rebalance the regression by fractionally differencing the
regressor with an appropriate integration order obtained in the first step,
and use their procedure to test the forward rate unbiasedness hypothesis.

\subsection{The implied-realized volatility relation}

A similar unbiasedness hypothesis as in the previous subsection can be found
in other contexts. For example, under market efficiency and rationality,
option prices should reflect all available information about expected future
return volatility of the underlying asset. Early work (e.g., Christensen and
Prabhala, 1998, and the references therein) considered regression (in logs) of
realized volatility of the underlying asset, $\sigma_{RV,t}$, on the
volatility implied by option prices, $\sigma_{IV,t}$, i.e.,%
\begin{equation}
\log\sigma_{RV,t}=\alpha+\beta\log\sigma_{IV,t}+\varepsilon_{t}, \label{zz72}%
\end{equation}
where the unbiased hypothesis of interest is $H_{0}:\beta=1$ allowing for a
constant risk premium. The regression (\ref{zz72}) was estimated by OLS with
estimated values of $\beta$ being significantly less than one.

Bandi and Perron (2006) and Christensen and Nielsen (2006) analyzed this
problem from a stationary fractional cointegration perspective, noting that
log-volatilities appear to be stationary long memory processes. In this case,
as discussed in Section~\ref{sec:CI-est-semi}, OLS is inconsistent and this
could explain the results obtained by OLS estimation in earlier work.
Christensen and Nielsen (2006) estimated the integration orders of $\log
\sigma_{RV,t}$ and $\log\sigma_{IV,t}$ using the local Whittle procedure, and
for different bandwidth choices obtained values in the range 0.35--0.48. They
then estimated (\ref{zz72}) by NBLS obtaining estimated $\beta$'s in the range
0.84--0.89 (again, for different bandwidth choices) and insignificantly
different from $\beta=1$. Finally, estimates of the integration order of
$\varepsilon_{t}$, obtained by local Whittle applied to the NBLS residuals,
were obtained in the range 0.09--0.11 and thus supported the existence of
fractional cointegration with weakly dependent cointegrating errors. These
results are consistent with those in Bandi and Perron (2006) who, in
independent work, applied a narrow-band approach with subsampling and provided
further support to the $\beta=1$ hypothesis. Nielsen (2007) revisited this
issue using instead a local Whittle quasi-maximum likelihood estimation
approach, where the integration orders of the regressors and error and the
cointegration parameter in (\ref{zz72}) are jointly estimated. This
alternative methodology allows for testing interesting joint hypotheses like
$H_{0}:d_{\varepsilon}=0,\beta=1$, where $d_{\varepsilon}$ is the integration
order of $\varepsilon_{t}$. This particular hypothesis is rejected for some
bandwidth choices, though not for others, but in any case the existence of
stationary fractional cointegration is supported by Nielsen's (2007) analysis.

\subsection{Political science}

Several political science questions have been addressed by fractional
integration and cointegration techniques. The appeal of this methodology is
that, in many circumstances, data in political studies are formed through
aggregation of heterogeneous dynamic behaviour at individual level, which, as
explained in Section~\ref{sec:sources}, is one of the sources of long memory.
Using the aggregation argument and a model of voter behavior,
Box-Steffensmeier and Smith (1996), Byers, Davidson, and Peel (1997), and
Dolado, Gonzalo, and Mayoral (2002) showed that aggregate opinion poll data
may be best modeled using fractional time series models, and empirical
estimates of integration orders around~0.8 were found.

The relationship between measures of political support and economic
indicators, a phenomenon known as economic voting, has been investigated by
Box-Steffensmeier and Tomlinson (2000) and Davidson (2003) in the context of
fractional integration and cointegration. Their results are inconclusive about
the existence of fractional cointegration between the two types of variables.
Based on a fractionally integrated vector error correction model, Davidson,
Byers, and Peel (2006) analyzed the relationship between the approval of the
performances of prime ministers and governments in the UK, providing some
evidence supporting the existence of fractional cointegration. A similar
analysis was conducted by Jones, Nielsen, and Popiel (2014) using the FCVAR
model (Section~\ref{sec:fcvar}) to study the potential cointegration linkages
between economic performance and political support both in Canadian terms and
also relative to US performance. Interestingly, they found that support to the
Progressive Conservative and Liberal parties depend substantially on the
economic situation: periods of high interest rates and low unemployment
benefit the conservatives, while the opposite economic conditions lead to
higher liberal support. Additionally, their results indicate that US economic
performance does not appear to have an effect on Canadian political support.

In a similar fashion the working paper version of MacKinnon and Nielsen (2014)
analyzed the possibility of fractional cointegration between the support for
the Conservative and Labour parties in UK using monthly Gallup opinion poll
data from 1951 to 2000. They applied the FCVAR methodology and provided
evidence in favour of the nonstationary mean reverting behaviour of both
observable series, which is in line with previous studies cited above, but
found no evidence that the two series are cointegrated. Furthermore, they paid
particular attention to the important issue of initial values. In a
nonstationary setting, due to the truncation inherent in
Definition~\ref{def type II} of fractional models, it is necessary to
condition on some observed (but not modeled) initial values. The conditioning
argument was proposed and rigorously justified by Johansen and Nielsen (2010,
2012a, 2016), while Hualde and Robinson (2010) provided an heuristic solution
based on omitting observations at the beginning of the sample. In addition to
theoretical arguments, Johansen and Nielsen (2016) gave a detailed analysis of
the polling data and the consequences of alternative specifications of the
initial conditions.

A forecasting analysis based on polling data of political support in the UK
for 2010--2015 was provided by Nielsen and Shibaev (2018). They compared the
FCVAR model with relevant competing models, and their results showed evidence
of superior performance of the former model with the relative forecast
improvement being higher at longer forecast horizons, which is expected.

\subsection{Risk-return relationship in asset pricing}

An important research topic in the finance literature is the risk-return
trade-off. Simple CAPM and factor models suggest a strong relation between the
conditional mean and variance of returns. This relation has been extensively
studied in the empirical finance literature, but the evidence is mixed and in
particular there does not appear to be any agreement in the literature on the
sign of the risk-return relation. Christensen and Nielsen (2007) noted that,
even though the stationary long memory properties of volatility have been
extensively documented in the literature, previous empirical and theoretical
work on the risk-return relationship has omitted this important feature.
Moreover, standard specifications that incorporate volatility in returns in a
linear fashion would imply long memory in returns, which is not empirically warranted.

Consequently, Christensen and Nielsen (2007) specify and estimate new vector
ARFIMA models for the joint dynamics of stock returns and volatility that
allow for long memory in volatility without imposing this property on returns.
They also show how asset pricing theory implies testable cross-equation
restrictions on the system, and these are not rejected in their preferred
specifications. The latter include a strong financial leverage effect, a
positive risk-return trade-off, long memory in volatility, but a small and
short lived effect of volatility shocks on stock prices.

\subsection{The purchasing power parity revisited}

We have already mentioned several studies of the PPP in a fractional context.
The PPP involves the relationship between the domestic log-price index
($p_{t}$), foreign log-price index ($p_{t}^{\ast}$), and log-exchange rate
($e_{t}$). One approach would be to analyze the complete cointegration
structure of the vector $z_{t}=(p_{t},p_{t}^{\ast},e_{t})^{\prime}$ and check
whether the condition $p_{t}=p_{t}^{\ast}-e_{t}$, which is the absolute
version of the PPP, is supported the data. This approach is further
complicated by the possibility that the individual integration orders of the
observables in $z_{t}$ might be distinct.

Hualde and Robinson (2010) used their semiparametric methodology (see
Section~\ref{sec:CI-est-semi}) and the step-wise procedure described in Hualde
(2008) on US/UK data and\ concluded that the cointegrating rank is $r_{1}=2$
with a further cointegrating subspace of dimension $r_{2}=1$. That is, there
exists a two-dimensional cointegration space spanned by the two linearly
independent cointegrating vectors $\gamma_{1}=(\beta_{21},1,0)^{\prime}$ with
$\beta_{21}\neq0$ and $\gamma_{2}=(0,0,1)^{\prime}$.\ The first vector
reflects that $p_{t}$ and $p_{t}^{\ast}$ have the same integration order and
are cointegrated; the second simply reflects that $e_{t}$ has a smaller
integration order than $p_{t}$ and $p_{t}^{\ast}$, so there exists a trivial
cointegrating relation. Additionally, the existence of a cointegrating
subspace means that the order-reducing linear combination of $p_{t}$ and
$p_{t}^{\ast}$ cointegrates with $e_{t}$ to get a linear combination involving
the three variables with a further reduced order. Hualde and Robinson (2010)
thus described the cointegrating structure as%
\begin{align}
\beta_{11}p_{t}+\beta_{12}p_{t}^{\ast}+e_{t}  &  =\alpha_{1}+\Delta
_{+}^{-d_{1}}u_{1t},\label{zz74}\\
\beta_{21}p_{t}+p_{t}^{\ast}  &  =\alpha_{2}+\Delta_{+}^{-d_{2}}%
u_{2t},\label{zz75}\\
p_{t}  &  =\alpha_{3}+\Delta_{+}^{-d_{3}}u_{3t}, \label{zz76}%
\end{align}
where $d_{1}<d_{2}<d_{3}$; see also (\ref{zz2})--(\ref{zz3}). They obtained
estimates of $d_{1}$, $d_{2}$, and $d_{3}$ of about 1.5--1.8, around~1.3, and
around~1, respectively. Based on their estimates of $\beta_{11}$, $\beta_{12}%
$, and $\beta_{21}$, they were unable to reject the hypothesis $\beta
_{11}=1,\beta_{12}=-1$, which represents the absolute version of the PPP, but
their results do not clarify the question of whether the deviations from
equilibrium in (\ref{zz74}) are mean reverting or not.

\subsection{Empirical applications of the FCVAR model}

Given the enormous impact of Johansen's (1995) CVAR methodology as a tool to
analyze the standard $I(1)/I(0)$ cointegration setting (as well as $I(2)$
situations), the development of extensions to cover fractional time series is
a very relevant research area. We have addressed the main theoretical
achievements, and in particular the FCVAR model, in Section~\ref{sec:fcvar}.
There are now a large number of empirical applications of this methodology
facilitated by freely available software packages in both Matlab~(Nielsen and
Popiel, 2018) and R~(Morin, Nielsen, and Popiel, 2021). In addition to the
applications involving political science and opinion polls mentioned
previously, we will now mention a small selection of empirical applications of
the FCVAR model in a variety of areas.

In an early application, Osterrieder and Schotman (2011) analyzed the joint
behaviour of real-estate returns and the rent-to-price ratio. Using a long
time series of 355 years of real-estate returns and rent-to-price ratios they
estimated an FCVAR model and compared the predictive performance of this model
to that of a triangular specification. Their results support the superiority
of the FCVAR model. Bollerslev, Osterrieder, Sizova, and Tauchen (2013)
estimated an FCVAR model using high-frequency data of futures contracts for
the S\&P500 and the corresponding VIX volatility index. Specifically, they
proposed a trivariate model to explain the joint evolution of two
log-volatility measures and the returns. Their main conclusion is that both
volatilities appear to be $I(d)$ (with estimated $d$ about 0.4) and
cointegrated with an $I(0)$ cointegration error, while the returns appear to
be $I(0)$ as expected. Interestingly, the cointegrating relation between the
volatilities is related to the variance risk premium (which is linked to
aggregate economic uncertainty), and the joint modeling provides nontrivial
return predictability. In another application to volatilities, Rossi and De
Magistris (2013) estimated a bivariate FCVAR model for two volatility
measures, the futures and spot log-daily ranges, $\log\sigma_{t;F}$ and
$\log\sigma_{t;S}$, respectively. The model is estimated assuming the
cointegrating relation $\log\sigma_{t;F}+\beta\log\sigma_{t;S}$. Here, the
value $\beta=-1$ represents the non-arbitrage condition and the authors
provide evidence supporting it. Moreover, while the observables appear to be
stationary long memory processes (with estimated memory close to but below
0.5), the cointegration error appears to be $I(0)$. Furthermore, the
out-of-sample forecast superiority of the proposed FCVAR model over other
model specifications is illustrated.\ Related to the volatility applications
and the risk-return applications mentioned earlier, Chen, Chiang, and
H\"{a}rdle (2018) analyze down-side risk, volatility spill-overs, and
comovement of stock returns in an FCVAR model. Their results indicate, for
example, that the downside risk for each G7 market is cointegrated with that
of the world market.

Finally, Dolatabadi, Nielsen, and Xu (2015, 2016) analyze equilibrium
relations between spot and futures prices of several commodities and
implications for price discovery. Specifically, for each commodity they
estimated a bivariate FCVAR model and in each case found evidence of
cointegration. They compared their results with the standard non-fractional
CVAR and showed that the cointegration errors appear to be stationary but
fractionally integrated, so that the fractional model is more appropriate than
the standard CVAR model. In a similar fashion, Dolatabadi, Narayan, Nielsen,
and Xu (2018) analyzed the relationship between spot and futures prices for
many commodities and demonstrated the superiority of the FCVAR model against
the CVAR model in terms of in-sample fit and out-of-sample forecasting, and
further illustrated this by means of \textquotedblleft economic
significance.\textquotedblright

\section{Concluding remarks and further readings}

\label{sec:concluding}

We have presented a general overview of the literature on fractional
integration and cointegration, both from theoretical and empirical
perspectives. Our aim has been to provide a good introduction to most of the
relevant issues in these research areas, but, undoubtedly, some important
topics have been omitted or not covered in detail. To conclude our review we
briefly outline a few such topics.

A prominent example relates to stationary ARCH-type models that have been
introduced to capture long memory features of asset return volatility. Some of
the most relevant ideas regarding this general topic are covered in the survey
of Giraitis, Leipus, and Surgailis (2009). The most well-known of these models
appears to be the integrated FIGARCH process introduced by Baillie,
Bollerslev, and Mikkelsen (1996) briefly mentioned in
Section~\ref{sec:applications} in the context of empirical applications, but
this is only one particular ARCH-type model with long memory. A general model
that may exhibit power law decay in the autocovariances of squared returns is
the ARCH$\left(  \infty\right)  $ model introduced by Robinson (1991), which
is a nonparametric generalization of the GARCH$\left(  p,q\right)  $ model.
Other process which display long memory in squares are the Linear ARCH of
Giraitis, Robinson, and Surgailis (2000, 2004), the bilinear model of Giraitis
and Surgailis (2002), the HYGARCH model of Davidson (2004), and the long
memory stochastic volatility model introduced by Breidt, Crato, and de Lima
(1998) and Harvey (1998); see also Robinson and Zaffaroni (1998) and Robinson (2001).

Another important class of models that we have not covered so far are those
where the singularity of the spectral density of the process does not occur at
the zero frequency as in (\ref{a7}), but instead at a seasonal or cyclical
frequency. This type of process is relevant for the treatment of seasonally
unadjusted time series when the dependence among seasonal (or cyclical)
observations display the typical slow decay of standard long memory processes.
Properties of these models and relevant estimation methods have been analyzed
by, among others, Gray, Zhang, and Woodward (1989), Robinson (1994b), Giraitis
and Leipus (1995), Arteche and Robinson (1999, 2000), Arteche (2002, 2020),
Nielsen (2004a), and Arteche and Velasco (2005). Within this literature, a
particularly interesting inferential problem appears when, in addition to the
exponent of the spectral singularity (the memory parameter), the location of
the pole is also unknown and needs to be estimated from the data. The location
of the pole is known for seasonal processes, but in cases where the interest
is to measure the length of a cycle (e.g., when modeling macroeconomic
observables), it is realistic to assume that the location of the pole is
unknown. In the latter case, inference is substantially more complicated.
Estimation of both the location of the pole and the related memory parameter
is analyzed by Giraitis, Hidalgo, and Robinson (2001), Hidalgo and Soulier
(2004), and Hidalgo (2005).

Related to fractional cointegration and focusing on a bivariate case, Hualde
(2006) considered a situation where the integration orders of the two
observables are different, but their corresponding balanced versions are
cointegrated in the usual sense. The balanced versions are obtained by
fractionally differencing one series with the appropriate parameter, termed
the imbalance parameter, such that both series have identical integration
orders. Hualde (2006) termed this concept unbalanced cointegration, and it is
a particular case of polynomial (fractional) cointegration. In the context of
the FCVAR model, Johansen (2008) and Franchi (2010) gave conditions under
which polynomial cointegration can arise. Interestingly, in this situation
both the cointegration parameter and the imbalance parameter drive the
long-run linkages between the observables. Hualde (2014) proposed a model of
unbalanced cointegration and a semiparametric estimator of the cointegration
and imbalance parameters, and described their limiting properties. Johansen
and Nielsen (2021) proposed a generalization of the FCVAR model in which each
observable has its own memory parameter. They showed that the regression-based
unbalanced cointegration model analyzed by Hualde (2006, 2014) arises as a
special case of their model, and further discussed the concept of unbalanced
cointegration, model properties, as well as maximum likelihood estimators and
their asymptotic properties.

\section*{References}%

\setlength{\parindent}{-6pt}%
%

\hspace*{-6pt}%
Abadir, K.M., Distaso, W., \& Giraitis, L.\ (2007). Nonstationarity-extended
local Whittle estimation. \emph{Journal of Econometrics}, 141, 1353--1384.

Agiakloglou, C., \& Newbold, P.\ (1994). Lagrange multiplier tests for
fractional difference. \emph{Journal of Time Series Analysis}, 15, 253--262.

Agiakloglou, C., Newbold, P., \& Wohar, M.\ (1993). Bias in an estimator of
the fractional difference parameter. \emph{Journal of Time Series Analysis},
14, 235--246.

Akonom, J., \& Gourieroux, C.\ (1987). A functional central limit theorem for
fractional processes. Technical report 8801, Paris: CEPREMAP.

Andersen, T.G., Bollerslev, T., Diebold, F.X., \& Ebens, H.\ (2001). The
distribution of realized stock return volatility. \emph{Journal of Financial
Economics}, 61, 43--76.

Andersen, T.G., Bollerslev, T., Diebold, F.X., \& Labys, P.\ (2001). The
distribution of realized exchange rate volatility. \emph{Journal of the
American Statistical Association}, 96, 42--55.

Andrews, D.W.K., \& Guggenberger, P.\ (2003). A bias-reduced log-periodogram
regression estimator for the long-memory parameter. \emph{Econometrica}, 71, 675--712.

Andrews, D.W.K., \& Sun, Y.\ (2004). Adaptive local polynomial Whittle
estimation of long-range dependence. \emph{Econometrica}, 72, 569--614.

Arteche, J.\ (2002). Semiparametric robust tests on seasonal or cyclical long
memory time series. \emph{Journal of Time Series Analysis}, 23, 251--285.

Arteche, J.\ (2004). Gaussian semiparametric estimation in long memory in
stochastic volatility and signal plus noise models. \emph{Journal of
Econometrics}, 119, 131--154.

Arteche, J.\ (2006). Semiparametric estimation in perturbed long memory
series. \emph{Computational Statistics \&\ Data Analysis}, 51, 2118--2141.

Arteche, J.\ (2020). Exact local Whittle estimation in long memory time series
with multiple poles. \emph{Econometric Theory}, 36, 1064--1098.

Arteche, J., \&\ Robinson, P.M.\ (1999). Seasonal and cyclic long memory. In
\emph{Asymptotics, Nonparametrics and Time Series:\ A Tribute to Madan Lal
Puri} (Ghosh, S.,\ ed.), 115--145, Marcel Dekker.

Arteche, J., \& Robinson, P.M.\ (2000). Semiparametric inference in seasonal
and cyclical long memory processes. \emph{Journal of Time Series Analysis},
21, 1--25.

Arteche, J., \& Velasco, C.\ (2005). Trimming and tapering semi-parametric
estimates in asymmetric long memory time series. \emph{Journal of Time Series
Analysis}, 26, 581--611.

Backus, D.K., \& Zin, S.E.\ (1993). Long-memory inflation
uncertainty:\ evidence from the term structure of interest rates.
\emph{Journal of Money, Credit, and Banking}, 25, 681--700.

Baillie, R.T.\ (1996). Long memory processes and fractional integration in
econometrics. \emph{Journal of Econometrics}, 73, 5--59.

Baillie, R.T.,\ \& Bollerslev, T.\ (1994a). Cointegration, fractional
cointegration and exchange rate dynamics. \emph{Journal of Finance}, 49, 737--745.

Baillie, R.T., \& Bollerslev, T.\ (1994b). The long memory of the forward
premium. \emph{Journal of International Money and Finance}, 13, 565--571.

Baillie, R.T., \&\ Bollerslev, T.\ (2000). The forward premium anomaly is not
as bad as you think. \emph{Journal of International Money and Finance}, 19, 471--488.

Baillie, R.T., Bollerslev, T., \& Mikkelsen, H.-O.\ (1996). Fractionally
integrated generalized autoregressive conditional heteroskedasticity.
\emph{Journal of Econometrics}, 74, 3--30.

Baillie, R.T., Chung, C.-F., \& Tieslau, M.A.\ (1996). Analysing inflation by
the fractionally integrated ARFIMA-GARCH model. \emph{Journal of Applied
Econometrics}, 11, 23--40.

Bandi, F.M., \& Perron, B.\ (2006). Long memory and the relation between
implied and realized volatility. \emph{Journal of Financial Econometrics}, 4, 636--670.

Beltratti, A., \& Morana, C.\ (2006). Breaks and persistency: macroeconomic
causes of stock market volatility. \emph{Journal of Econometrics}, 131, 151--177.

Beran, J.\ (1995). Maximum likelihood estimation of the differencing parameter
for invertible short and long memory autoregressive integrated moving average
models. \emph{Journal of the Royal Statistical Society: Series B}, 57, 659--672.

Bloomfield, P.\ (1973). An exponential model for the spectrum of a scalar time
series. \emph{Biometrika}, 60, 217--226.

Bollerslev, T., \& Mikkelsen, H.-O.\ (1996). Modeling and pricing long memory
in stock market volatility. \emph{Journal of Econometrics}, 73, 151--184.

Bollerslev, T., Osterrieder, D., Sizova, N., \& Tauchen, G.\ (2013). Risk and
return: long-run relations, fractional cointegration, and return
predictability. \emph{Journal of Financial Economics}, 108, 409--424.

Box, G.E.P., \& Jenkins, G.M.\ (1970). \emph{Time Series Analysis,
Forecasting, and Control}. San Francisco: Holden-Day.

Box-Steffensmeier, J.M., \& Smith, R.M.\ (1996). The dynamics of aggregate
partisanship. \emph{American Political Science Review}, 90, 567--580.

Box-Steffensmeier, J.M., \& Tomlinson, A.R.\ (2000). Fractional integration
methods in political science. \emph{Electoral Studies}, 19, 63--76.

Breidt, F.J., Crato, N., \& de Lima, P.\ (1998). The detection and estimation
of long memory in stochastic volatility. \emph{Journal of Econometrics}, 83, 325--348.

Breitung, J.\ (2002). Nonparametric tests for unit roots and cointegration.
\emph{Journal of Econometrics}, 108, 343--363.

Breitung, J., \& Hassler, U.\ (2002). Inference on the cointegration rank in
fractionally integrated processes. \emph{Journal of Econometrics}, 110, 167--185.

Brockwell, P.J., \& Davis, R.A.\ (1991). \emph{Time Series:\ Theory and
Methods}. New York:\ Springer.

Byers, D., Davidson, J., \& Peel, D.\ (1997). Modeling political popularity:
an analysis of long range dependence in opinion poll series. \emph{Journal of
the Royal Statistical Society:\ Series A}, 160, 471--490.

Byers, D., Davidson, J., \& Peel, D.\ (2007). The long memory model of
political support: some further results. \emph{Applied Economics}, 39, 2547--2552.

Cavaliere, G., Nielsen, M.\O ., \& Taylor, A.M.R.\ (2015). Bootstrap score
tests for fractional integration in heteroskedastic ARFIMA models, with an
application to price dynamics in commodity spot and futures markets.
\emph{Journal of Econometrics}, 187, 557--579.

Cavaliere, G., Nielsen, M.\O ., \& Taylor, A.M.R.\ (2017). Quasi-maximum
likelihood estimation and bootstrap inference in fractional time series models
with heteroskedasticity of unknown form. \emph{Journal of Econometrics}, 198,\ 165--188.

Cavaliere, G., Nielsen, M.\O ., \& Taylor, A.M.R.\ (2020). Adaptive inference
in heteroscedastic fractional time series models. Forthcoming in \emph{Journal
of Business and Economic Statistics}.

Chan, N.H., \& Terrin, N.\ (1995). Inference for unstable long-memory
processes with applications to fractional unit root autoregressions.
\emph{Annals of Statistics}, 23, 1662--1683.

Chen, C.Y.-H., Chiang, T.C., \& H\"{a}rdle, W.K.\ (2018). Downside risk and
stock returns in the G7 countries: an empirical analysis of their long-run and
short-run dynamics. \emph{Journal of Banking \&\ Finance}, 93, 21--32.

Chen, W.W.,\ \& Hurvich, C.M.\ (2003). Semiparametric estimation of
multivariate fractional cointegration. \emph{Journal of the American
Statistical Association}, 98, 629--642.

Chen, W.W., \& Hurvich, C.M.\ (2006). Semiparametric estimation of fractional
cointegrating subspaces. \emph{Annals of Statistics}, 34, 2939--2979.

Cheung, Y.-W., \& Lai, K.\ (1993). A fractional cointegration analysis of
purchasing power parity. \emph{Journal of Business and Economic Statistics},
11, 103--112.

Chevillon, G., Hecq, A., \& Laurent, S.\ (2018). Generating univariate
fractional integration within a large VAR(1). \emph{Journal of Econometrics},
204, 54--65.

Chevillon, G., \& Mavroeidis, S.\ (2017). Learning can generate long memory.
\emph{Journal of Econometrics}, 198, 1--9.

Christensen, B.J., \& Nielsen, M.\O .\ (2006). Asymptotic normality of
narrow-band least squares in the stationary fractional cointegration model and
volatility forecasting. \emph{Journal of Econometrics}, 133, 343--371.

Christensen, B.J., \& Nielsen, M.\O .\ (2007). The effect of long memory in
volatility on stock market fluctuations. \emph{Review of Economics and
Statistics}, 89, 684--700.

Christensen, B.J., \& Prabhala, N.R.\ (1998). The relation between implied and
realized volatility. \emph{Journal of Financial Economics}, 50, 125--150.

Crato, N., \& Rothman, P.\ (1994a). Fractional integration analysis of
long-run behavior for US macroeconomic time series. \emph{Economics Letters},
45, 287--291.

Crato, N., \&\ Rothman, P.\ (1994b). A reappraisal of parity reversal for UK
real exchange rates. \emph{Applied Economics Letters}, 1, 139--141.

Dahlhaus, R.\ (1989). Efficient parameter estimation for self-similar
processes. \emph{Annals of Statistics}, 17, 1749--1766.

Davidson, J.\ (2003). A model of fractional cointegration, and tests for
cointegration using the bootstrap. \emph{Journal of Econometrics}, 110, 187--212.

Davidson, J.\ (2004). Moment and memory properties of linear conditional
heteroscedasticity models, and a new model. \emph{Journal of Business and
Economic Statistics}, 22, 16--29.

Davidson, J., Byers, D., \& Peel, D.\ (2006). Support for governments and
leaders: fractional cointegration analysis of poll evidence from the UK,
1960-2004. \emph{Studies in Nonlinear Dynamics and Econometrics}, 10, article 1.

Davidson, J., \& Sibbertsen, P.\ (2005). Generating schemes for long memory
processes: regimes, aggregation and linearity. \emph{Journal of Econometrics},
128, 253--282.

Davydov, Y.A.\ (1970). The invariance principle for stationary processes.
\emph{Theory of Probability \& Its Applications}, 15, 487--498.

Deo, R.S., \& Hurvich, C.M.\ (2001). On the log periodogram regression
estimator of the memory parameter in long memory stochastic volatility models.
\emph{Econometric Theory}, 17, 686--710.

Diebold, F.X., Husted, S., \& Rush, M.\ (1991). Real exchange rates under the
gold standard. \emph{Journal of Political Economy}, 99, 1252--1271.

Diebold, F. X., \& Inoue, A.\ (2001). Long memory and regime switching.
\emph{Journal of Econometrics}, 101, 131--159.

Diebold, F.X., \& Rudebusch, G.D.\ (1989). Long memory and persistence in
aggregate output. \emph{Journal of Monetary Economics}, 24, 189--209.

Diebold, F. X., \& Rudebusch, G.D.\ (1991a). On the power of Dickey-Fuller
tests against fractional alternatives. \emph{Economics Letters}, 35, 155--160.

Diebold, F.X., \& Rudebusch, G.D.\ (1991b). Is consumption too smooth?\ Long
memory and the Deaton paradox. \emph{Review of Economics and Statistics}, 74, 1--9.

Dolado, J.J., Gonzalo, J., Mayoral, L.\ (2002). A fractional Dickey-Fuller
test for unit roots. \emph{Econometrica}, 70, 1963--2006.

Dolatabadi, S., Nielsen, M.\O ., \& Xu, K.\ (2015). A fractionally
cointegrated VAR analysis of price discovery in commodity futures markets.
\emph{Journal of Futures Markets}, 35, 339--356.

Dolatabadi, S., Nielsen, M.\O ., \& Xu, K.\ (2016). A fractionally
cointegrated VAR model with deterministic trends and application to commodity
futures markets. \emph{Journal of Empirical Finance}, 38B, 623--639.

Dolatabadi, S., Narayan, P.K., Nielsen, M.\O ., \& Xu, K.\ (2018). Economic
significance of commodity return forecasts from the fractionally cointegrated
VAR model. \emph{Journal of Futures Markets}, 38, 219--242.

Dueker, M., \& Startz, R.\ (1998). Maximum-likelihood estimation of fractional
cointegration with an application to US and Canadian bond rates. \emph{Review
of Economics and Statistics}, 80, 420--426.

Engle, R.F.\ (1974). Band spectrum regression. \emph{International Economic
Review}, 15, 1--11.

Engle, R.F., \& Granger, C.W.J.\ (1987). Co-integration and error correction:
representation, estimation, and testing. \emph{Econometrica}, 55, 251--276.

Fl\^{o}res, R.G., \& Szafarz, A.\ (1996). An enlarged definition of
cointegration. \emph{Economics Letters}, 50, 193--195.

Fox, R., \& Taqqu, M.S.\ (1986). Large-sample properties of parameter
estimates for strongly dependent stationary Gaussian time series. \emph{Annals
of Statistics}, 14, 517--532.

Franchi, M.\ (2010). A representation theory for polynomial cofractionality in
vector autoregressive models. \emph{Econometric Theory}, 26, 1201--1217.

Frederiksen, P., \& Nielsen, M.\O .\ (2008). Bias-reduced estimation of
long-memory stochastic volatility. \emph{Journal of Financial Econometrics},
6, 496--512.

Frederiksen, P., Nielsen, F.S., \& Nielsen, M.\O .\ (2012). Local polynomial
Whittle estimation of perturbed fractional processes. \emph{Journal of
Econometrics}, 167, 426--447.

Gadea, M.D., \& Mayoral, L.\ (2006). The persistence of inflation in OECD
countries:\ a fractionally integrated approach. \emph{International Journal of
Central Banking}, March issue.

Garc\'{\i}a-Enr\'{\i}quez, J., \& Hualde, J.\ (2019). Local Whittle estimation
of long memory: standard versus bias-reducing techniques. \emph{Econometrics
and Statistics}, 12, 66-77.

Geweke, J., \& Porter-Hudak, S.\ (1983). The estimation and application of
long memory time series models. \emph{Journal of Time Series Analysis}, 4, 221--238.

Gil-Alana, L.A., \& Hualde, J.\ (2009). Fractional integration and
cointegration: an overview and an empirical application. In \emph{Palgrave
Handbook of Econometrics}, vol.\ 2 (Patterson, K.\ \& Mills, T.C., eds.). Palgrave-MacMillan.

Gil-Alana, L.A., \& Robinson, P.M.\ (1997). Testing of unit root and other
nonstationary hypotheses in macroeconomic time series. \emph{Journal of
Econometrics}, 80, 241--268.

Giraitis, L., Hidalgo, J., \& Robinson, P.M.\ (2001). Gaussian estimation of
parametric spectral density with unknown pole. \emph{Annals of Statistics},
29, 987--1023.

Giraitis, L., Kokoszka, P., Leipus, R., \&\ Teyssiere, G.\ (2003). Rescaled
variance and related tests for long memory in volatility and levels.
\emph{Journal of Econometrics}, 112, 265--294.

Giraitis, L., \& Leipus, R.\ (1995). A generalized fractionally differencing
approach in long-memory modeling. \emph{Lithuanian Mathematical Journal}, 35, 53--65.

Giraitis, L., Leipus, R., \& Surgailis, D.\ (2009). ARCH($\infty$) models and
long memory properties. In \emph{Handbook of Financial Time Series} (Mikosch,
T., Kreiss, J.P., Davis, R., \& Andersen, T.G., eds.). New York: Springer.

Giraitis, L., Robinson, P.M., \& Samarov, A.\ (1997). Rate optimal
semiparametric estimation of the memory parameter of the Gaussian time series
with long-range dependence. \emph{Journal of Time Series Analysis}, 18, 49--60.

Giraitis, L., Robinson, P.M., \& Surgailis, D.\ (2000). A model for long
memory conditional heteroskedasticity. \emph{Annals of Applied Probability},
10, 1002--1024.

Giraitis, L., Robinson, P.M., \& Surgailis, D.\ (2004). LARCH, leverage, and
long memory. \emph{Journal of Financial Econometrics}, 2, 177--210.

Giraitis, L., \& Surgailis, D.\ (1990). A central limit theorem for quadratic
forms in strongly dependent linear variables and its application to
asymptotical normality of Whittle's estimate. \emph{Probability Theory and
Related Fields}, 86, 87--104.

Giraitis, L., \& Surgailis, D.\ (2002). ARCH-type bilinear models wiht double
long memory. \emph{Stochastic Processes and their Applications}, 100, 275--300.

Gourieroux, C., \& Jasiak, J.\ (2001). Memory and infrequent breaks.
\emph{Economics Letters}, 70, 29--41.

Granger, C.W.J.\ (1980). Long memory relationships and the aggregation of
dynamic models. \emph{Journal of Econometrics}, 14, 227--238.

Granger, C.W.J.\ (1981). Some properties of time series data and their use in
econometric model specification. \emph{Journal of Econometrics}, 16, 121--130.

Granger, C.W.J.\ (1986). Developments in the study of cointegrated economic
variables. \emph{Oxford Bulletin of Economics and Statistics}, 48, 213--228.

Granger, C.W.J., \& Hyung, N.\ (2004). Occasional structural breaks and long
memory withan application to the S\&P 500 absolute stock returns.
\emph{Journal of Empirical Finance}, 11, 399--421.

Granger, C.W.J., \&\ Joyeux, R.\ (1980). An introduction to long-memory time
series models and fractional differencing. \emph{Journal of Time Series
Analysis}, 1, 15--29.

Granger, C.W.J., \& Ter\"{a}svirta, T.\ (1999). A simple nonlinear time series
model with misleading linear properties. \emph{Economics Letters}, 62, 161--165.

Gray, H.L., Zhang, N.-F., \& Woodward, W.A.\ (1989). On generalized fractional
processes. \emph{Journal of Time Series Analysis}, 10, 233--257.

Haldrup, N., \& Nielsen, M.\O .\ (2007). Estimation of fractional integration
in the presence of data noise. \emph{Computational Statistics \& Data
Analysis}, 51, 3100--3114.

Haldrup, N., \& Vera Vald\'{e}s, J.E.\ (2017). Long memory, fractional
integration, and cross-sectional aggregation. \emph{Journal of Econometrics},
199, 1--11.

Hamilton, J.D.\ (1989). A new approach to the economic analysis of
nonstationary time series and the business cycle. \emph{Econometrica}, 57, 357--384.

Hannan, E.J.\ (1963). Regression for time series. In \emph{Proceedings of the
Symposium on Time Series Analysis} (Rosenblatt, M., ed.). New York:\ John
Wiley and Sons.

Hannan, E.J.\ (1973). The asymptotic theory of linear time-series models.
\emph{Journal of Applied Probability}, 10, 130--145.

Harvey, A.C.\ (1998). Long memory in stochastic volatility. In
\emph{Forecasting Volatility in Financial Markets} (Knight, J., \& Satchell,
S., eds.), 307--320. Oxford: Butterworth-Heineman.

Hassler, U.\ (2018). \emph{Time series analysis with long memory in view}. New
York: John Wiley and Sons.

Hassler, U., \& Breitung, J.\ (2006). A residual-based LM-type test against
fractional cointegration. \emph{Econometric Theory}, 22, 1091--1111.

Hassler, U., \& Wolters, J.\ (1995). Long memory in inflation rates:
international evidence. \emph{Journal of Business and\ Economic Statistics},
13, 37--45.

Henry, M., \& Zaffaroni, P.\ (2003). The long-range dependence paradigm for
macroeconomics and finance. In \emph{Theory and applications of long-range
dependence} (Doukhan, P., Oppenheim, G., \& Taqqu, M.S., eds.), 417--438. Birkhauser.

Hidalgo, J.\ (2005). Semiparametric estimation for stationary processes whose
spectra have an unknown pole. \emph{Annals of Statistics}, 33, 1843--1889.

Hidalgo, J., \& Soulier, P.\ (2004). Estimation of the location and exponent
of the spectral singularity of a long memory process. \emph{Journal of Time
Series Analysis}, 25, 55--81.

Hidalgo, J., \& Robinson, P.M.\ (1996). Testing for structural change in a
long-memory environment. \emph{Journal of Econometrics}, 70, 159--174.

Hosking, J.R.M.\ (1981). Fractional differencing. \emph{Biometrika}, 68, 165--176.

Hosoya, Y.\ (1996). The quasi-likelihood approach to statistical inference on
multiple time-series with long-range dependence. \emph{Journal of
Econometrics}, 73, 217--236.

Hou, J., \& Perron, P.\ (2014). Modified local Whittle estimator for long
memory processes in the presence of low frequency (and other) contaminations.
\emph{Journal of Econometrics}, 182, 309--328.

Hualde, J.\ (2006). Unbalanced cointegration. \emph{Econometric Theory}, 22, 765--814.

Hualde, J.\ (2008). Consistent estimation of cointegrating subspaces.
Preprint, Universidad P\'{u}blica de Navarra.

Hualde, J.\ (2014). Estimation of long-run parameters in unbalanced
cointegration. \emph{Journal of Econometrics}, 178, 761--778.

Hualde, J., \& Iacone, F.\ (2019). Fixed bandwidth inference for fractional
cointegration. \emph{Journal of Time Series Analysis}, 40, 544--572.

Hualde, J., \& Nielsen, M.\O .\ (2020). Truncated sum of squares estimation of
fractional time series models with deterministic trends. \emph{Econometric
Theory}, 36, 751--772.

Hualde, J., \&\ Nielsen, M.\O .\ (2021). Truncated sum-of-squares estimation
of fractional time series models with generalized power law trend. Working
paper, Aarhus University.

Hualde, J., \& Robinson, P.M.\ (2007). Root-n-consistent estimation of weak
fractional cointegration. \emph{Journal of Econometrics}, 140, 450--484.

Hualde, J., \&\ Robinson, P.M.\ (2010). Semiparametric inference in
multivariate fractionally cointegrated systems. \emph{Journal of
Econometrics}, 157, 492--511.

Hualde, J., \&\ Robinson, P.M.\ (2011). Gaussian pseudo-maximum likelihood
estimation in fractional time series models. \emph{Annals of Statistics}, 39, 3152--3181.

Hualde, J., \& Velasco, C.\ (2008). Distribution-free tests of fractional
cointegration. \emph{Econometric Theory}, 24, 216--255.

Hurvich, C.M., \& Brodsky, J.\ (2001). Broadband semiparametric estimation of
the memory parameter of a long-memory time series using fractional exponential
models. \emph{Journal of Time Series Analysis}, 22, 221--249.

Hurvich, C.M., Deo, R.S., \& Brodsky, J.\ (1998). The mean squared error of
Geweke and Porter-Hudak's estimator of the memory parameter of a long-memory
time series. \emph{Journal of Time Series Analysis}, 19, 19--46.

Hurvich, C.M., Moulines, E., \& Soulier, P.\ (2005). Estimating long memory in
volatility. \emph{Econometrica}, 73, 1283--1328.

Hurvich, C.M., \& Ray, B.K.\ (1995). Estimation of the memory parameter for
nonstationary or noninvertible fractionally integrated processes.
\emph{Journal of Time Series Analysis}, 16, 17--41.

Hurvich, C.M., \& Ray, B.K.\ (2003). The local Whittle estimator of
long-memory stochastic volatility. \emph{Journal of Financial Econometrics},
1, 445--470.

Iacone, F., Leybourne, S.J., \& Taylor, A.M.R.\ (2019). Testing the order of
fractional integration of a time series in the possible presence of a trend
break at an unknown point. \emph{Econometric Theory}, 35, 1201--1233.

Iacone, F., Nielsen, M.\O ., \& Taylor, A.M.R.\ (2021). Semiparametric tests
for the order of integration in the possible presence of level breaks.
Forthcoming in \emph{Journal of Business and Economic Statistics}.

Jeganathan, P.\ (1997). On asymptotic inference in linear cointegrated time
series systems. \emph{Econometric Theory}, 13, 692--745.

Jeganathan, P.\ (1999). On asymptotic onference in cointegrated time series
with fractionally integrated errors. \emph{Econometric Theory}\textit{, }15,
583--621. (with unpublished correction, 2001).

Jensen, A.N., \& Nielsen, M.\O .\ (2014). A fast fractional difference
algorithm. \emph{Journal of Time Series Analysis}, 35, 428--436.

Johansen, S.\ (1995). \emph{Likelihood-Based Inference in Cointegrated Vector
Autoregressive Models}. Oxford: Oxford University Press.

Johansen, S.\ (2008). A representation theory for a class of vector
autoregressive models for fractional processes. \emph{Econometric Theory}, 24, 651--676.

Johansen, S., \& Juselius, K.\ (1990). Maximum likelihood estimation and
inference on cointegration---with applications to the demand for money.
\emph{Oxford Bulletin of Economics and Statistics}, 52, 169--210.

Johansen, S., \& Nielsen, M.\O .\ (2010). Likelihood inference for a
nonstationary fractional autoregressive model. \emph{Journal of Econometrics},
158, 51--66.

Johansen, S., \& Nielsen, M.\O .\ (2012a). Likelihood inference for a
fractionally cointegrated vector autoregressive model. \emph{Econometrica},
80, 2667--2732.

Johansen, S., \& Nielsen, M.\O .\ (2012b). A necessary moment condition for
the fractional functional central limit theorem. \emph{Econometric Theory},
28, 671--679.

Johansen, S., \& Nielsen, M.\O .\ (2016). The role of initial values in
conditional sum-of-squares estimation of nonstationary fractional time series
models. \emph{Econometric Theory}, 32, 1095--1139.

Johansen, S., \& Nielsen, M.\O .\ (2018). Testing the CVAR in the fractional
CVAR model. \emph{Journal of Time Series Analysis}, 39, 836--849.

Johansen, S., \& Nielsen, M.\O .\ (2019). Nonstationary cointegration in the
fractionally cointegrated VAR model. \emph{Journal of Time Series Analysis},
40, 519--543.

Johansen, S., \& Nielsen, M.\O .\ (2021). Statistical inference in the
multifractional cointegrated VAR model. Working paper, Aarhus University.

Jones, M.E.C., Nielsen, M.\O ., \& Popiel, M.K.\ (2014). A fractionally
cointegrated VAR analysis of economic voting and political support.
\emph{Canadian Journal of Economics}, 47, 1078--1130.

Klemes, V.\ (1974). The Hurst phenomenon: a puzzle? \emph{Water Resources
Research}, 10, 675--688.

K\"{u}nsch, H.\ (1987). Statistical aspects of self-similar processes. In
\emph{Proceedings of the First World Congress of the Bernoulli Society},
vol.\ 1 (Prokhorov, Yu., \& Sazanov, V.V., eds.), 67--74. Utrecht: VNU Science Press.

Lai, K.S.\ (1997). Long-term persistence in the real interest rate: some
evidence of a fractional unit root. \emph{International Journal of Finance \&
Economics}, 2, 225--235.

Li, W.K., \& McLeod, A.I.\ (1986). Fractional time series modelling.
\emph{Biometrika}, 73, 217--221.

Ling, S., \& Li, W.K.\ (2001). Asymptotic inference for nonstationary
fractionally integrated autoregressive moving-average models.
\emph{Econometric Theory}, 17, 738--764.

Lobato, I., \& Savin, N.\ (1997). Real and spurious long memory properties of
stock market data. \emph{Journal of Business and Economic Statistics}, 16, 261--283.

Lobato, I., \& Robinson, P.M.\ (1998). A nonparametric test for I(0).
\emph{Review of Economic Studies}, 65, 475--495.

Lobato, I.N., \& Velasco, C.\ (2006). Optimal fractional Dickey--fuller tests.
\emph{Econometrics Journal}, 9, 492--510.

MacKinnon, J.G., \& Nielsen, M.\O . (2014). Numerical distribution functions
of fractional unit root and cointegration tests. \emph{Journal of Applied
Econometrics}, 29, 161--171.

Mandelbrot, B.B., \& Van Ness, J.W.\ (1968). Fractional Brownian motions,
fractional noises and applications. \emph{SIAM Review}, 10, 422--437.

Marinucci, D., \& Robinson, P.M.\ (1999). Alternative forms of fractional
Brownian motion. \emph{Journal of Statistical Planning and Inference}, 80, 111--122.

Marinucci, D., \& Robinson, P.M.\ (2000). Weak convergence of multivariate
fractional processes. \emph{Stochastic Processes and their Applications}, 86, 103--120.

Marinucci, D., \& Robinson, P.M.\ (2001). Semiparametric fractional
cointegration analysis. \emph{Journal of Econometrics}, 105, 225--247.

Marmol, F., \& Velasco, C.\ (2004). Consistent testing of cointegrating
relationships. \emph{Econometrica}, 72, 1809--1844.

Maynard, A., \& Phillips, P.C.B.\ (2001). Rethinking an old empirical puzzle:
econometric evidence on the forward discount anomaly. \emph{Journal of Applied
Econometrics}, 16, 671--708.

Maynard, A., Smallwood, A., \& Wohar, M.E.\ (2013). Long memory regressors and
predictive testing: a two-stage rebalancing approach. \emph{Econometric
Reviews}, 32, 318--360.

McCloskey, A., \& Perron, P.\ (2013). Memory parameter estimation in the
presence of level shifts and deterministic trends. \emph{Econometric Theory},
29, 1196--1237.

Meade, N., \& Maier, M.R.\ (2003). Evidence of long memory in short-term
interest rates. \emph{Journal of Forecasting}, 22, 553--568.

Michelacci, C., \& Zaffaroni, P.\ (2000). (Fractional) beta convergence.
\emph{Journal of Monetary Economics}, 45, 129--153.

Morin, L., Nielsen, M.\O ., Popiel, M.K.\ (2021). FCVAR: an R package for the
fractionally cointegrated vector autoregressive model. Working paper, Aarhus University.

Moulines, E., \& Soulier, P.\ (1999). Broadband log-periodogram regression of
time series with long-range dependence. \emph{Annals of Statistics}, 27, 1415--1439.

Nelson, C.R., \& Plosser, C.I.\ (1982). Trends and random walks in
macroeconmic time series: some evidence and implications. \emph{Journal of
Monetary Economics}, 10, 139--162.

Nielsen, M.\O .\ (2004a) Efficient likelihood inference in nonstationary
univariate models. \emph{Econometric Theory}, 20, 116--146.

Nielsen, M.\O .\ (2004b) Spectral analysis of fractionally cointegrated
systems. \emph{Economics Letters}, 83, 225--231.

Nielsen, M.\O .\ (2004c) Optimal residual-based tests for fractional
cointegration and exchange rate dynamics. \emph{Journal of Business and
Economic Statistics}, 22, 331--345.

Nielsen, M.\O .\ (2004d) Efficient inference in multivariate fractionally
integrated time series models. \emph{Econometrics Journal}, 7, 63--97.

Nielsen, M.\O .\ (2005a) Multivariate Lagrange multiplier tests for fractional
integration. \emph{Journal of Financial Econometrics}\ 3, 372--398.

Nielsen, M.\O .\ (2005b) Semiparametric estimation in time-series regression
with long-range dependence. \emph{Journal of Time Series Analysis}, 26, 279--304.

Nielsen, M.\O .\ (2007) Local Whittle analysis of stationary fractional
cointegration and the implied-realized volatility relation. \emph{Journal of
Business and Economic Statistics}, 25, 427--446.

Nielsen, M.\O .\ (2009) A powerful test of the autoregressive unit root
hypothesis based on a tuning parameter free statistic. \emph{Econometric
Theory}, 25, 1515--1544.

Nielsen, M.\O .\ (2010) Nonparametric cointegration analysis of fractional
systems with unknown integration orders. \emph{Journal of Econometrics}, 155, 170--187.

Nielsen, M.\O .\ (2015) Asymptotics for the conditional-sum-of-squares
estimator in multivariate fractional time-series models. \emph{Journal of Time
Series Analysis}, 36, 154--188.

Nielsen, M.\O ., \& Frederiksen, P.H.\ (2005) Finite sample comparison of
parametric, semiparametric, and wavelet estimators of fractional integration.
\emph{Econometric Reviews}, 24, 405--443.

Nielsen, M.\O ., \& Frederiksen, P.\ (2011) Fully modified narrow-band least
squares estimation of weak fractional cointegration. \emph{Econometrics
Journal}, 14, 77--120.

Nielsen, M.\O ., \& Hualde, J.\ (2019) Special issue of the Journal of Time
Series Analysis in honour of the 35th anniversary of the publication of Geweke
and Porter-Hudak (1983): Guest editors' introduction. \emph{Journal of Time
Series Analysis}, 40, 386--387.

Nielsen, M.\O ., \& Popiel, M.K.\ (2018). A Matlab program and user's guide
for the fractionally cointegrated VAR model. QED working paper 1330, Queen's University.

Nielsen, M.\O ., \& Shibaev, S.\ (2018) Forecasting daily political opinion
polls using the fractionally cointegrated vector auto-regressive model.
\emph{Journal of the Royal Statistical Society: Series A}, 181, 3--33.

Nielsen, M.\O ., \& Shimotsu, K.\ (2007) Determining the cointegrating rank in
nonstationary fractional systems by the exact local Whittle approach.
\emph{Journal of Econometrics}, 141, 574--596.

Ohanissian, A., Russell, J.R,\ \& Tsay, R.S.\ (2008). True or spurious memory?
a new test. \emph{Journal of Business and Economic Statistics}, 26, 161--175.

Osterrieder, D., \& Schotman, P.C.\ (2011). Predicting returns with a
co-fractional VAR model. Working paper, Maastricht University.

Parke, W.R.\ (1999). What is fractional integration?. \emph{Review of
Economics and Statistics}, 81, 632--638.

Perron, P.\ (1989). The great crash, the oil price shock and the unit root
hypothesis. \emph{Econometrica}, 57, 1361--1401.

Perron, P., \& Qu, Z.\ (2010). Long-memory and level shifts in the volatility
of stock market return indices. \emph{Journal of Business and Economic
Statistics}, 28, 275--290.

Phillips, P.C.B.\ (1991a). Optimal inference in cointegrated systems.
\emph{Econometrica}, 59, 283--306.

Phillips, P.C.B.\ (1991b). Spectral regression for cointegrated time series.
In \emph{Nonparametric and Semiparametric Methods in Economics and Statistics}
(Barnett, W.\ ed.). Cambridge: Cambridge University Press.

Phillips, P.C.B., \& Durlauf, S.N.\ (1986). Multiple time series regression
with integrated processes. \emph{Review of Economic Studies}, 53, 473--495.

Phillips, P.C.B., \& Loretan, M.\ (1991). Estimating long-run economic
equilibria. \emph{Review of Economic Studies}, 58, 407--436.

Qu, Z.\ (2011). A test against spurious long memory. \emph{Journal of Business
and Economic Statistics}, 29, 423--438.

Robinson, P.M.\ (1978). Statistical inference for a random coefficient
autoregressive model. \emph{Scandinavian Journal of Statistics}, 5, 163--168.

Robinson, P.M.\ (1991). Testing for strong serial correlation and dynamic
conditional heteroskedasticity in multiple regression. \emph{Journal of
Econometrics}, 47, 67--84.

Robinson, P.M.\ (1993). Highly insignificant F-ratios. \emph{Econometrica},
61, 687--696.

Robinson, P.M.\ (1994a). Time series with strong dependence. In \emph{Advances
in Econometrics: Sixth World Congress}, vol.\ 1 (Sims, C.A., ed.), 47--96.
Cambridge: Cambridge University Press.

Robinson, P.M.\ (1994b). Efficient tests of nonstationary hypotheses.
\emph{Journal of the American Statistical Association}, 89, 1420--1437.

Robinson, P.M.\ (1994c). Semiparametric analysis of long memory time series.
\emph{Annals of Statistics}, 22, 515--539.

Robinson, P.M.\ (1995a). Log-periodogram regression of time series with long
range dependence. \emph{Annals of Statistics}, 23, 1048--1072.

Robinson, P.M.\ (1995b). Gaussian semiparametric estimation of long range
dependence. \emph{Annals of Statistics}, 23, 1630--1661.

Robinson, P.M.\ (2001). The memory of stochastic volatility models.
\emph{Journal of Econometrics}, 101, 195--218.

Robinson, P.M.\ (2003). Long memory time series. In \emph{Time Series With
Long Memory} (Robinson, P.M., ed.), 1--48. Oxford: Oxford University Press.

Robinson, P.M.\ (2005a) The distance between rival nonstationary fractional
processes. \emph{Journal of Econometrics}, 128, 283--300.

Robinson, P.M.\ (2005b). Efficiency improvements in inference on stationary
and nonstationary fractional time series. \emph{Annals of Statistics}, 33, 1800--1842.

Robinson, P.M.\ (2006). Conditional-sum-of-squares estimation of models for
stationary time series with long memory. In \emph{Time Series and Related
Topics: In Memory of Ching-Zong Wei} (Ho, H.-C., Ing, C.-K., \& Lai, T.L.,
eds.), 130--137. IMS Lecture Notes Monograph Series, vol.\ 52.

Robinson, P.M.\ (2008a) Diagnostic testing for cointegration. \emph{Journal of
Econometrics}, 143, 206--225.

Robinson, P.M.\ (2008b). Multiple local Whittle estimation in stationary
systems. \emph{Annals of Statistics}, 36, 2508--2530.

Robinson, P.M., \& Henry, M.\ (2003). Higher-order kernel semiparametric
M-estimation of long memory. \emph{Journal of Econometrics}, 114, 1--27.

Robinson, P.M., \& Hidalgo, F.J.\ (1997). Time series regression with long
range dependence. \emph{Annals of Statistics}, 25, 77--104.

Robinson, P.M., \& Hualde, J.\ (2003). Cointegration in fractional systems
with unknown integration orders. \emph{Econometrica}, 71, 1727--1766.

Robinson, P.M., \& Iacone, F.\ (2005). Cointegration in fractional systems
with deterministic trends. \emph{Journal of Econometrics}, 129, 263--298.

Robinson, P.M., \& Marinucci, D.\ (2001). Narrow-band analysis of
nonstationary processes. \emph{Annals of Statistics}, 29, 947--986.

Robinson, P.M., \& Marinucci, D.\ (2003). Semiparametric frequency domain
analysis of fractional cointegration. In \emph{Time Series With Long Memory}
(Robinson, P.M., ed.), 334--373. Oxford: Oxford University Press.

Robinson, P.M., \& Yajima, Y.\ (2002). Determination of cointegrating rank in
fractional systems. \emph{Journal of Econometrics}, 106, 217--241.

Robinson, P. M., \& Zaffaroni, P.\ (1998). Nonlinear time series with long
memory: a model for stochastic volatility. \emph{Journal of Statistical
Planning and Inference}, 68, 359--371.

Rossi, E., \& De Magistris, P.S.\ (2013). Long memory and tail dependence in
trading volume and volatility. \emph{Journal of Empirical Finance}, 22, 94--112.

Saikkonen, P.\ (1991). Asymptotically efficient estimation of cointegration
regressions. \emph{Econometric Theory}, 7, 1--21.

Shea, G.S.\ (1991). Uncertainty and implied variance bounds in long-memory
models of the interest rate term structure. \emph{Empirical Economics}, 16, 287--312.

Shimotsu, K.\ (2010). Exact local Whittle estimation of fractional integration
with unknown mean and time trend. \emph{Econometric Theory}, 26, 501--540.

Shimotsu, K.\ (2012). Exact local Whittle estimation of fractionally
cointegrated systems. \emph{Journal of Econometrics}, 169, 266--278.

Shimotsu, K., \& Phillips, P.C.B.\ (2005). Exact local Whittle estimation of
fractional integration. \emph{Annals of Statistics}, 33, 1890--1933.

Smith, A.\ (2005). Level shifts and the illusion of long memory in economic
time series. \emph{Journal of Business and Economic Statistics}, 23, 321--335.

Sowell, F.\ (1986). \emph{Fractionally Integrated Vector Time Series}\ (PhD
dissertation), Duke University.

Sowell, F.\ (1990). The fractional unit root distribution. \emph{Econometrica}%
, 58, 495--505.

Sowell, F.\ (1992). Maximum likelihood estimation of stationary univariate
fractionally integrated time series models. \emph{Journal of Econometrics},
53, 165--188.

Stock, J.H.\ (1994). Unit roots and trend breaks. In \emph{Handbook of
Econometrics}, vol.\ IV (Engle, R.F., \& McFadden, D., eds.). Amsterdam: North-Holland.

Stock, J.H., \& Watson, M.W.\ (1988). Testing for common trends. \emph{Journal
of the American Statistical Association}, 83, 1097--1107.

Sun, Y., \& Phillips, P.C.B.\ (2003). Nonlinear log-periodogram regression for
perturbed fractional processes. \emph{Journal of Econometrics}, 115, 355--389.

Tanaka, K.\ (1999). The nonstationary fractional unit root. \emph{Econometric
Theory}, 15, 549--582.

Taqqu, M.S.\ (1975). Weak convergence to fractional Brownian motion and to the
Rosenblatt process. \emph{Advances in Applied Probability}, 7, 249--249.

Tsay, W.J.\ (2000). Long memory story of the real interest rate.
\emph{Economics Letters}, 67, 325--330.

Velasco, C.\ (1999a). Gaussian semiparametric estimation of non-stationary
time series. \emph{Journal of Time Series Analysis}, 20, 87--127.

Velasco, C.\ (1999b). Non-stationary log-periodogram regression. \emph{Journal
of Econometrics}, 91, 325--371.

Velasco, C.\ (2000). Non-Gaussian log-periodogram regression.
\emph{Econometric Theory}, 16, 44--79.

Velasco, C.\ (2003a). Nonparametric frequency domain analysis of nonstationary
multivariate time series. \emph{Journal of Statistical Planning and
Inference}, 116, 209--247.

Velasco, C.\ (2003b). Gaussian semi-parametric estimation of fractional
cointegration. \emph{Journal of Time Series Analysis}, 24, 345--378.

Velasco, C.\ (2006). Semiparametric estimation of long-memory models. In
\emph{Palgrave Handbook of Econometrics}, vol.\ 1 (Patterson, K., \& Mills,
T.C., eds.), 353--395. Palgrave-MacMillan.

Velasco, C., \& Robinson, P.M. (2000). Whittle pseudo-maximum likelihood
estimation for nonstationary time series. \emph{Journal of the American
Statistical Association}, 95, 1229--1243.

Whittle, P.\ (1953). Estimation and information in stationary time series.
\emph{Arkiv f\"{o}r Matematik}, 2, 423--434.

Zaffaroni, P.\ (2004). Contemporaneous aggregation of linear dynamic models in
large economies. \emph{Journal of Econometrics}, 120, 75--102.

\end{document}